\documentclass[a4paper, 11pt, sort&compress]{elsarticle}

\usepackage{graphics}
\usepackage{amsmath}
\usepackage{subfig}
\usepackage{float}
\usepackage{algorithm}
\usepackage{setspace}
\usepackage{bm}
\usepackage{comment}
\usepackage{amssymb}
\usepackage{bbm, bbold}
\usepackage{subfig}
\usepackage{xcolor}
\usepackage{multirow}
\usepackage{url}
\usepackage{tcolorbox}
\usepackage{listings}
\usepackage{algorithm}
\usepackage{algpseudocode}
\usepackage{enumerate}
\usepackage{hyperref}

\newcommand{\T}{\mathrm{T}}
\newcommand{\tm}{\, \text{-} \,}

\newcommand{\pderiv}[2]{\frac{\partial #1}{\partial #2}}

\newcommand{\mydashedline}{$\text{-} \, \text{-} \, \text{-} \, \text{-} \, \text{-} \, \text{-}$}
\newcommand{\mysolidline}{\rule[0.5ex]{10mm}{0.25mm}}

\RequirePackage[margin=20mm]{geometry}

\graphicspath{{./figures/}}

\begin{document}
\fontfamily{ptm}\selectfont

\begin{frontmatter}

\title{A simple extrapolated predictor for overcoming the starting and tracking issues in the arc-length method for nonlinear structural mechanics}

\author{Chennakesava Kadapa \corref{cor1}}
\ead{c.kadapa@bolton.ac.uk}

\cortext[cor1]{Corresponding authors}

\address{School of Engineering, University of Bolton, Bolton BL3 5AB, United Kingdom}
\begin{abstract}
This paper presents a simplified implementation of the arc-length method for computing the equilibrium paths of nonlinear structural mechanics problems using the finite element method. In the proposed technique, the predictor is computed by extrapolating the solutions from two previously converged load steps. The extrapolation is a linear combination of the previous solutions; therefore, it is simple and inexpensive. Additionally, the proposed extrapolated predictor also serves as a means for identifying the forward movement along the equilibrium path without the need for any sophisticated techniques commonly employed for explicit tracking. The ability of the proposed technique to successfully compute complex equilibrium paths in static structural mechanics problems is demonstrated using seven numerical examples involving truss, beam-column and shell models. The computed numerical results are in excellent agreement with the reference solutions. The present approach does not require prohibitively small increments for its success.
\end{abstract}

\end{frontmatter}

\section{Introduction}
Nonlinear structures experience complex deformation behaviour beyond the limit points, for example, post-buckling, plastic yielding and damage. For computing the complex nonlinear response of structures, the arc-length method has become the de-facto incremental-iterative numerical technique in computational structural mechanics using the finite element method (FEM). Numerical methods for computing buckling instabilities are becoming even more important in the design and computer modelling of metamaterials \cite{CoulaisPRL2015, LiuPRSA2017, RenMDPIMatl2018, VangelatosEML2019, JanbazMH2019}, soft structures \cite{LiSM2012, BuddayPTRSA2017, SteinSM2019, DortdivanliogluJMPS2019} and additively manufactured components \cite{SeifiES2018, BaroutajiES2021}.

Based on the techniques originally pioneered by Wempner \cite{WempnerIJSS1971}, Riks \cite{RiksJAM1972, RiksIJSS1979} and Crisfield \cite{CrisfieldCandS1981}, numerous flavours of the arc-length method have been proposed for computing complex equilibrium paths in nonlinear static structural mechanics problems, see Bergan et al. \cite{BerganIJNME1978}, Batoz and Dhatt \cite{BatozIJNME1979}, Ramm \cite{RammArcLength1980}, Powell and Simons \cite{PowellIJNME1981}, Fried \cite{FriedCMAME1984}, Gierlinski and Smith \cite{GierlinskiCandS1985}, Schweizerhof and Wriggers  \cite{SchweizerhofCMAME1986}, and Krenk \cite{KrenkIJNME1995}.

In the arc-length method, a constraint equation, called the arc-length equation, is added to the original (discrete) nonlinear equations of the problem. The augmented system of equations thus obtained is solved for the incremental load factor along with the incremental displacements. Depending upon the type of the constraint equation, the arc-length method is known as the spherical arc-length method \cite{WempnerIJSS1971, RiksIJSS1979, CrisfieldCandS1981}, cylindrical arc-length method \cite{CrisfieldCandS1981, RammArcLength1980}, and elliptical arc-length method \cite{ParkIJNME1982}. Linearised arc-length methods in which a linearised form of the arc-length equation is employed instead of the original arc-length equation, are also available \cite{WempnerIJSS1971, RiksJAM1972, CrisfieldIJNME1983}.

Some critical issues encountered in the computer implementation of the arc-length methods are those associated with 
computing the solution of a matrix system that is increased in size; 
finding the solution of a quadratic equation for the load parameter increment, especially the case of complex roots of the quadratic equation; 
computation of the arc-length radius at the beginning of each load step; and 
the determination of the sign of the load parameter increment at the first iteration of each load step to ensure forward movement along the equilibrium path.

Over the years, several researchers have proposed different techniques for overcoming these issues towards computing complex equilibrium paths in static nonlinear structures. These techniques consist of one or a combination of (a) imposing the orthogonality conditions, (b) computing the angle between residuals and displacement increments, (c) imposing certain constraints, and (d) computing the sign of the determinant of the stiffness matrix. The discussion of all of these techniques is beyond the scope and the interest of this article. The reader is referred to Crisfield \cite{CrisfieldCandS1981}, Bergan et al. \cite{BerganIJNME1978}, Krenk \cite{KrenkIJNME1995}, Bellini and Chulya \cite{BelliniCandS1987}, Lam and Morley \cite{LamJSE1992}, Carrera \cite{CarreraCandS1994}, Feng et al. \cite{FengCS1996}, Ritto-Correa and Camotim \cite{RittoCandS2008}, Al-Rasby \cite{AlRasbyCandS1991}, Krenk and Hededal \cite{KrenkCMAME1995}, and Kouhia \cite{KouhiaCMAME2008}, and references cited therein for the details regarding the commonly-encountered issues in the arc-length method, limitations of different flavours of the arc-length method and various techniques proposed for overcoming them.

Despite their tremendous success in computing the complex equilibrium paths in the response of nonlinear structures, the existing techniques for overcoming the starting and tracking in the arc-length are cumbersome and expensive. In this work, a simple extrapolation operator is proposed for overcoming the critical issues related to the starting and tracking in the arc-length method. In the proposed technique, the solution at the predictor step is computed as a linear combination of solutions at the two previously converged load steps. Therefore, the proposed technique is simple, inexpensive and easier to implement. Furthermore, the proposed technique also helps in tracking the solution in the forward direction along the equilibrium path, without the need for any sophisticated techniques commonly employed in the classical arc-length implementations.

The rest of the paper is organised as follows. The governing equations for the arc-length method and the details of the proposed technique are discussed in Section \ref{sec-formulation}. The suitability of the proposed technique in successfully computing complex equilibrium paths is illustrated in Section \ref{sec-examples} using seven benchmark examples involving nonlinear truss, beam-column and shell models. The paper is concluded with Section \ref{sec-conclusion} with a summary of observations made and conclusions drawn from the present work.

\section{The arc-length method} \label{sec-formulation}
By adapting the finite element method (FEM) for computing numerical solutions, the governing discrete system of equations for the nonlinear elasticity problem can be written as,
\begin{align} \label{eqn-force-equilibrium}
\mathbf{R}(\bm{u}, \lambda) = \mathbf{F}^{\mathrm{int}}(\bm{u}) - \lambda \, \mathbf{F}^{\mathrm{ext}} = \bm{0},
\end{align}
where, $\bm{u}$ is the nodal displacement vector, $\lambda$ is the load factor, $\mathbf{F}^{\mathrm{int}}(\bm{u})$ is the internal force vector, $\mathbf{F}^{\mathrm{ext}}$ is the external force vector, and $\mathbf{R}$ is the residual vector.

Often, it is impossible to obtain the numerical solutions of Eq. (\ref{eqn-force-equilibrium}) when the external load is applied all at once. Moreover, the response of the structures, especially in the post-buckling regime, can be quite complex with curves and loops in the load-deflection curves. Therefore, numerical solutions of Eq. (\ref{eqn-force-equilibrium}) are obtained by using an incremental approach, in which the solutions are computed using an iterative technique, for example, the Newton-Raphson scheme, at each increment.

In the incremental approach, the displacement and load factor at the current load step, $\bm{u}_{n+1}$ and $\lambda_{n+1}$, respectively, are computed as increments, $\Delta \bm{u}$ and $\Delta \lambda$, from their respective values at the previously converged load step, $\bm{u}_{n}$ and $\lambda_{n}$, as
\begin{align}
\bm{u}_{n+1} = \bm{u}_{n} + \Delta \bm{u}, \label{eqn-unp1} \\
\lambda_{n+1} = \lambda_{n} + \Delta \lambda. \label{eqn-lnp1}
\end{align}
where, the subscripts $n{+}1$ and $n$, respectively, denote the current and previously converged load steps. Using (\ref{eqn-unp1}) and (\ref{eqn-lnp1}), the residual vector for the current load step can be written as,
\begin{align} \label{eqn-resinp1}
\mathbf{R}(\bm{u}_{n+1}, \lambda_{n+1}) = \mathbf{F}^{\mathrm{int}}(\bm{u}_{n+1}) - \lambda_{n+1} \, \mathbf{F}^{\mathrm{ext}} = \bm{0}.
\end{align}

If the region of interest of the response of the structure under consideration does not include any limit points, then either load control method (LCM) or the displacement control method (DCM) or their variations \cite{YangAIAA1990, LeonMRC2014} can be used. However, if one is interested in tracking the response of the structure beyond the limit points, then the arc-length method (ALM) must be employed, see Leon et al. \cite{LeonAMR2011} and references cited therein for the comprehensive details on various iterative techniques for numerical solutions of nonlinear structures.

In the arc-length method, the system of nonlinear equations in Eq. (\ref{eqn-resinp1}) is solved for both the displacement, $\bm{u}_{n+1}$, as well as the loading parameter, $\lambda_{n+1}$. This approach increases the number of degrees of freedom (DOFs) by one, making Eq. (\ref{eqn-resinp1}) an under-determined system of equations. To overcome this issue, the system of nonlinear equations in Eq. (\ref{eqn-resinp1}) is augmented with an additional equation, called as \emph{the arc-length equation}, see \cite{WempnerIJSS1971, RiksIJSS1979, CrisfieldCandS1981}. The generic form of the arc-length equation is given as
\begin{align} \label{eqn-arclength}
[\Delta \bm{u}]^{\T} \, [\Delta \bm{u}] + \psi \, [\Delta \lambda]^2 \, \mathbf{F}^{\T} \, \mathbf{F} = [\Delta s]^2,
\end{align}
where, $s$ is the arc-length parameter which parametrises the equilibrium path \cite{FengCS1996}, $\Delta s$ is the increment in the arc-length parameter and $\mathbf{F}=\mathbf{F}^{\mathrm{ext}}$. Here, $\psi$ is a scalar parameter which helps to recover different arc-length schemes. For $\psi=0$, the cylindrical arc-length method is recovered \cite{CrisfieldCandS1981, RammArcLength1980}; for $\psi=1$, the method becomes spherical arc length method \cite{WempnerIJSS1971, RiksIJSS1979, CrisfieldCandS1981}; and for $\psi > 1$, the elliptical arc-length method is recovered \cite{ParkIJNME1982}. For the given $\Delta s$, Eqs. (\ref{eqn-resinp1}) and (\ref{eqn-arclength}) are solved together for increments $\Delta \bm{u}$ and $\Delta \lambda$ using the Newton-Raphson scheme.

Starting with an initial guess $\left( \Delta \bm{u}^{(1)}, \Delta \lambda^{(1)} \right)$, the displacement increment, $\Delta \bm{u}$, and the load increment, $\Delta \lambda$, are computed by iterative updates, $\delta \bm{u}$ and $\delta \lambda$, as
\begin{equation} \label{eqn-Dukp1}
\left.
\begin{array}{ll}
 \Delta \bm{u}^{(k+1)}  = \Delta \bm{u}^{(k)} + \delta \bm{u}, \\
 \Delta \lambda^{(k+1)} = \Delta \lambda^{(k)} + \delta \lambda,
\end{array}
\right \}
\qquad \mathrm{for} \quad k=1,2,3,\ldots, k_{\max},
\end{equation}
where, $k$ is the iteration counter and $k_{\max}$ is the maximum number of iterations. Using the expressions in Eq. (\ref{eqn-Dukp1}), $\bm{u}_{n+1}$ and $\lambda_{n+1}$ at the current iteration $k{+}1$ can be written as,
\begin{align} \label{eqn-ukp1}
\left.
\begin{array}{ll}
\bm{u}_{n+1}^{(k+1)} &= \bm{u}_{n} + \Delta \bm{u}^{(k+1)} = \bm{u}_{n} + \Delta \bm{u}^{(k)} + \delta \bm{u} = \bm{u}_{n+1}^{(k)} + \delta \bm{u}, \\
\lambda_{n+1}^{(k+1)} &= \lambda_{n} + \Delta \lambda^{(k+1)} = \lambda_{n} + \Delta \lambda^{(k)} + \delta \lambda = \lambda_{n+1}^{(k)} + \delta \lambda,
\end{array}
\right \}
\qquad \mathrm{for} \quad k=1,2,3,\ldots, k_{\max}.
\end{align}

\begin{figure}[H]
\centering
\includegraphics[clip, scale=0.5]{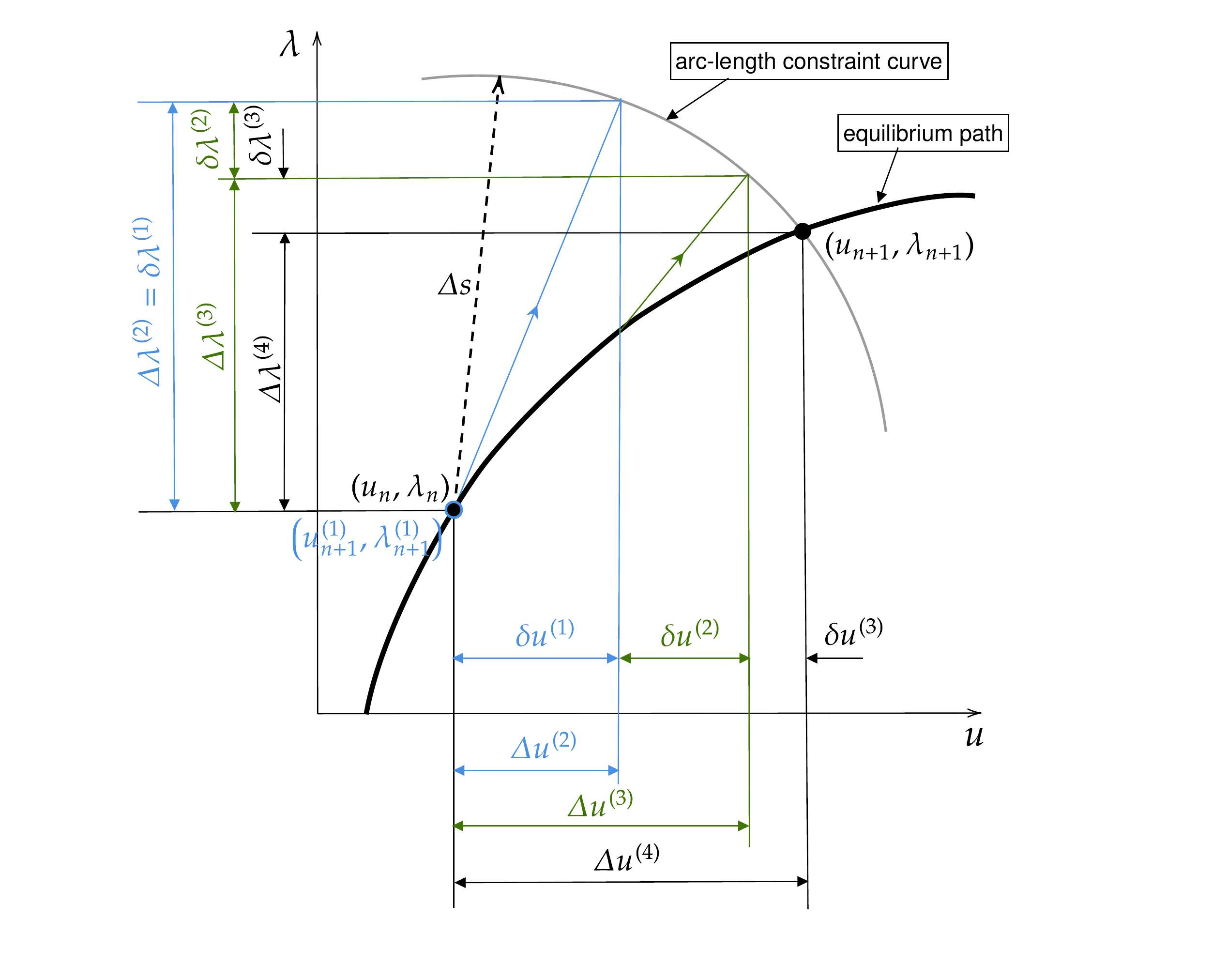}
\caption{A typical evolution of solution increments in the arc-length method.}
\label{fig-arclength-original}
\end{figure}

By applying the Newton-Raphson scheme to solve Eqs. (\ref{eqn-resinp1}) and (\ref{eqn-arclength}), we get the following matrix system,
\renewcommand{\arraystretch}{1.5}
\begin{align} \label{eqn-matsys-arclen}
\begin{bmatrix}
\mathbf{K}(\bm{u}_{n+1}^{(k)})  &  -\mathbf{F} \\
\bm{a}^{\T}  &  b
\end{bmatrix}
\,
\begin{Bmatrix}
\delta \bm{u} \\
\delta \lambda
\end{Bmatrix}
=
- \, 
\begin{Bmatrix}
\mathbf{R}(\bm{u}_{n+1}^{(k)}, \lambda_{n+1}^{(k)}) \\
\mathcal{A}(\Delta \bm{u}^{(k)}, \Delta \lambda^{(k)})
\end{Bmatrix},
\end{align}
\renewcommand{\arraystretch}{1.0}
where,
\begin{align}
\mathbf{K}(\bm{u}_{n+1}^{(k)}) &= \left. \pderiv{\mathbf{F}^{\mathrm{int}}}{\bm{u}} \right \vert_{\bm{u}_{n+1}^{(k)}}, \\
\bm{a} &= 2 \, [\Delta \bm{u}^{(k)}], \\
b &= 2 \, \psi \, [ \Delta \lambda^{(k)} ] \; [\mathbf{F}^{\T} \, \mathbf{F}], \\
\mathcal{A}(\Delta \bm{u}^{(k)}, \Delta \lambda^{(k)}) &= [\Delta \bm{u}^{(k)}]^{\T} \, [\Delta \bm{u}^{(k)}] + \psi \, [\Delta \lambda^{(k)}]^2 \, \mathbf{F}^{\T} \, \mathbf{F} - [\Delta s]^2.
\end{align}
The matrix system in Eq. (\ref{eqn-matsys-arclen}) is solved for $(\delta \bm{u}, \delta \lambda)$ until a predefined convergence criterion is satisfied. Typical evolution of solution increments in the arc-length method is illustrated schematically in Fig. \ref{fig-arclength-original}.

\subsection{Issues associated with the arc-length method}
When solving the coupled matrix system (\ref{eqn-matsys-arclen}), one quickly runs into four main difficulties:
\begin{enumerate}[(i)]
\item solving a matrix system of equations that is increased in size,
\item starting the arc-length algorithm at the first load step,
\item predicting the solution increments at the first iteration for the subsequent load steps, and
\item identifying the correct (forward) direction of evolution along the equilibrium path.
\end{enumerate}

These difficulties, along with the techniques for overcoming them, are discussed below.

\subsubsection{Issue 1: solving a matrix system of equations that is increased in size}
The size of the matrix system given by Eq. (\ref{eqn-matsys-arclen}) is one higher than the stiffness matrix system obtained with the FEM with the load-control method. While this might not be a big issue in in-house codes, it does require modifications to the code to account for the increased matrix size. Moreover, the increased bandwidth due to the off-diagonal terms $\mathbf{F}$ and $\bm{a}^{\T}$ deteriorates the performance of the matrix solver. To overcome this issue, different approaches were proposed in the literature, c.f. Crisfield \cite{CrisfieldCandS1981}, Batoz and Dhatt \cite{BatozIJNME1979}, and Schweizerhof \cite{SchweizerhofCMAME1986}, all of which solve Eq. (\ref{eqn-matsys-arclen}) by splitting it into a suitable form such that the matrix system of the original size can be used.

In the present work, the splitting approach based on the Schur complement of the stiffness matrix ($\mathbf{K}$) \cite{BatozIJNME1979}, which is also valid for the cylindrical arc-length method ($\psi{=}b{=}0$), is adapted. Accordingly, the coupled matrix system given in Eq. (\ref{eqn-matsys-arclen}) is solved for $\delta \lambda$ and $\delta \bm{u}$ as
\begin{align}
\delta \lambda &= \frac{\mathbf{a}^{\T} \, \delta \bm{u}^{II} - \mathcal{A}}{b + \mathbf{a}^{\T} \, \delta \bm{u}^{I}},   \label{eqn-appr2-dl} \\
\delta \bm{u} &= - \, \delta \bm{u}^{II} + \delta \lambda \, \delta \bm{u}^{I}, \label{eqn-appr2-du}
\end{align}
where,
\begin{align}
\delta \bm{u}^{I}  = \mathbf{K}^{-1} \, \mathbf{F}, \qquad
\delta \bm{u}^{II} = \mathbf{K}^{-1} \, \mathbf{R}.
\end{align}

This approach is similar to the one proposed by Crisfield \cite{CrisfieldCandS1981} in which a quadratic equation is solved for $\delta \lambda$. Note that, in this approach, the case of complex roots as well as the ambiguity associated with choosing the correct solution of the quadratic equation, are completely avoided.

\vspace*{5mm}
\noindent
\textbf{Remark I:} The disadvantage of the splitting scheme is that two matrix solves are required for computing $\delta \bm{u}^{I}$ and $\delta \bm{u}^{II}$ at every iteration. The associated cost can be minimised by factorising the matrix once and then using the factorisation for computing the solution of multiple right-hand sides. In fact, many matrix solvers, for example, SuperLU \cite{superlu}, PARDISO \cite{pardiso-6.0a} and MUMPS \cite{mumps1} support the solution of matrix systems for multiple right-hand sides at once. An alternative is to compute $\delta \bm{u}^{I}$ at the first iteration only; this, however, deteriorates the convergence of iterations. In this work, both $\delta \bm{u}^{I}$ and $\delta \bm{u}^{II}$ are solved for at every iteration.

\subsubsection{Issue 2: starting the arc-length algorithm at the first load step}
The lack of information regarding the arc-length increment $\Delta s$ makes it difficult to start the algorithm at the first load step. However, since the limit points are not encountered in the first few load steps, the difficulty of starting the arc-length method can be easily overcome by computing displacement at the first load step using the load control method in which the load increment $\Delta \lambda$ is specified as the user input. Based on the author's experience, \emph{a value of $\Delta \lambda$ that produces a noticeable deformation of the structure from the original configuration}, is sufficient enough.

For the first load step, the values of $b$, $\mathcal{A}$ and $\bm{a}$ in the matrix system  (\ref{eqn-matsys-arclen}) are taken as,
\begin{align}
b=1; \qquad \mathcal{A}=0; \qquad \mathrm{and} \qquad \bm{a} = \bm{0},
\end{align}
so that
\begin{align}
\delta \lambda = 0; \qquad \mathrm{and} \qquad
\delta \bm{u} &= - \, \delta \bm{u}^{II}.
\end{align}

Once $\Delta \bm{u}$ is obtained at the first load step using the load control method, the arc-length increment $\Delta s$ can be computed using Eq. (\ref{eqn-arclength}) as
\begin{align}
\Delta s = \sqrt{[\Delta \bm{u}]^{\T} \, [\Delta \bm{u}] + \psi \, [\Delta \lambda]^2 \, \mathbf{F}^{\T} \, \mathbf{F}}, \qquad \because \Delta s > 0.
\end{align}
The value of $\Delta s$ thus computed is used in the subsequent load steps, either with or without a multiplication factor ($>1$).

\subsubsection{Issue 3: predicting the solution increments at the first iteration for the subsequent load steps}
This issue, together with issue 4, are the most crucial and difficult issues in the successful implementation of the arc-length method. In the load-control method, the solution at the previously converged load step is often employed as the predictor at the current load step. Using this choice, we get,
\begin{subequations} \label{eqn-pred1}
\begin{align}
\Delta \bm{u}^{(1)}  &= \bm{u}_{n+1}^{(1)} - \bm{u}_{n} = \bm{u}_{n} - \bm{u}_{n} = \bm{0}, \label{eqn-pred1-Du} \\
\Delta \lambda^{(1)} &= \lambda_{n+1}^{(1)} - \lambda_{n} = \lambda_{n} - \lambda_{n} = 0. \label{eqn-pred1-Dl}
\end{align}
\end{subequations}

Although the predicted displacement increment in Eq. (\ref{eqn-pred1-Du}) works for the load control method, the choice $\left( \Delta \bm{u}^{(1)}, \Delta \lambda^{(1)}\right) = \left( \bm{0},0 \right)$ results in all kinds of problems in the arc-length method since it yields $\bm{a}=\bm{0}$ and $b=0$, which makes it impossible to compute $\delta \lambda$ from Eq. (\ref{eqn-matsys-arclen}). A variety of techniques have been proposed for computing suitable non-zero values as the predictors at the first iteration for the arc-length methods. These techniques vary in the complexity of implementation, computational cost incurred and their success in computing the complex equilibrium paths.

In the proposed work, an extrapolation operator based on the solutions at the two previously converged load steps is employed for predicting the solution increments at the first iteration in each load step (except the first load step) of the arc-length method. Accordingly,
\begin{equation} \label{eqn-arclen-preds}
\renewcommand{\arraystretch}{2.0}
 \left.
  \begin{array}{ll}
    \bm{u}_{n+1}^{(1)} = [1+\alpha] \, \bm{u}_{n} - \alpha \, \bm{u}_{n-1} \\
    \lambda_{n+1}^{(1)} = [1+\alpha] \, \lambda_{n} - \alpha \, \lambda_{n-1}
  \end{array}
 \right \}
\qquad \mathrm{with} \qquad \alpha = \frac{\Delta s}{\Delta s_{n}} > 0,
\renewcommand{\arraystretch}{1.0}
\end{equation}
where, $\Delta s$ and $\Delta s_{n}$ are the increments of arc-length parameter, respectively, for the current and previously converged load steps. The parameter $\alpha$ accounts for the adaptive load stepping. For uniform load increments, $\alpha=1$.

Using the predictor given by Eq. (\ref{eqn-arclen-preds}), the solution increments at the predictor step become
\begin{subequations} \label{eqn-pred2}
\begin{align}
\Delta \bm{u}^{(1)} &= \bm{u}_{n+1}^{(1)} - \bm{u}_{n} = \alpha \, \left[ \bm{u}_{n} - \bm{u}_{n-1} \right] = \alpha \, \Delta \bm{u}_{n}, \label{eqn-pred2-Du} \\
\Delta \lambda^{(1)} &= \lambda_{n+1}^{(1)} - \lambda_{n} = \alpha \, \left[ \lambda_{n} - \lambda_{n-1} \right] = \alpha \, \Delta \lambda_{n}, \label{eqn-pred2-Dl}
\end{align}
\end{subequations}
where, $\Delta \bm{u}_{n}$ and $\Delta \lambda_{n}$, respectively, are the displacement increment and load factor increment at the previously converged load step.

\begin{figure}[H]
\centering
\includegraphics[clip, scale=0.5]{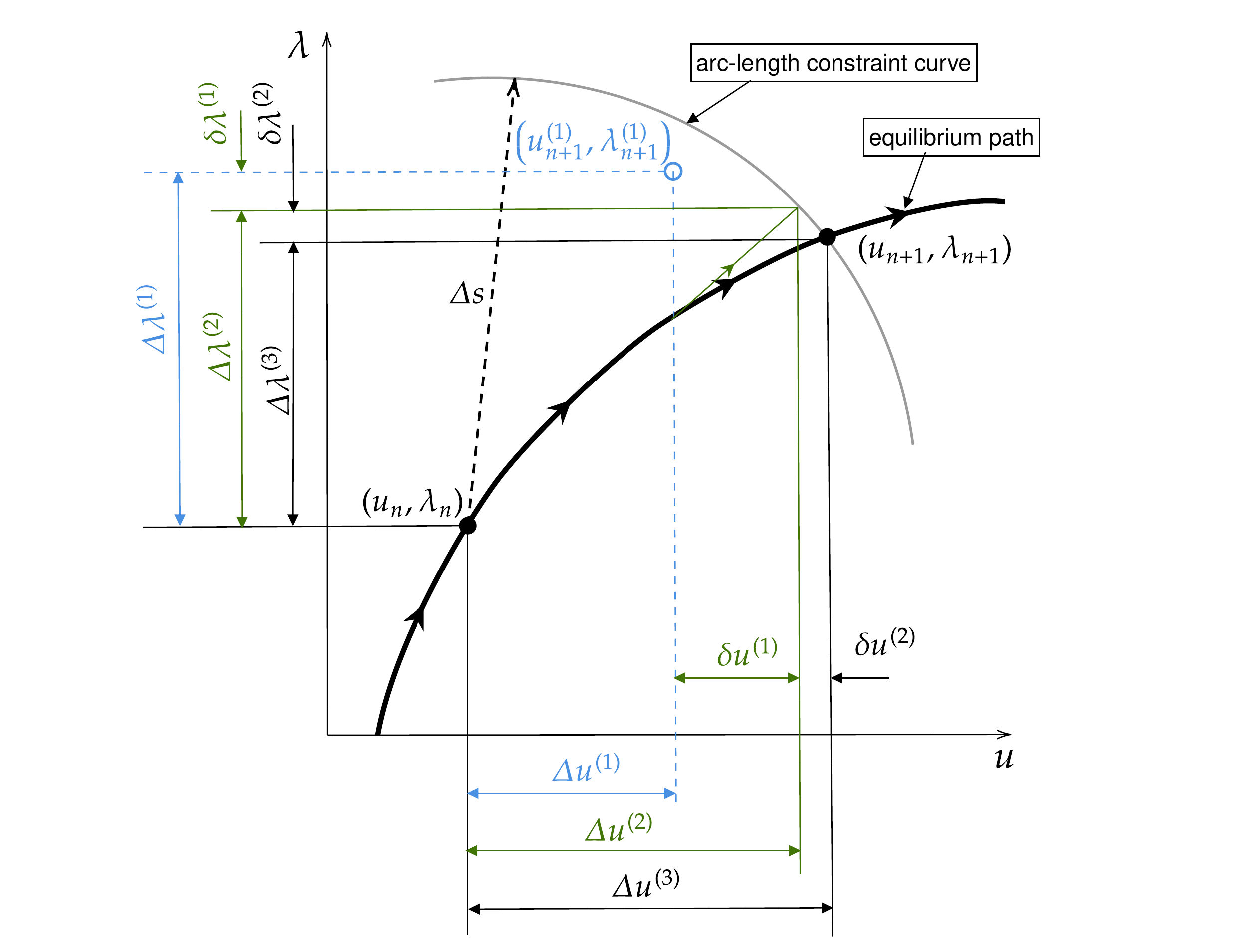}
\caption{Evolution of solution increments in the arc-length method with $(\Delta \bm{u}^{(1)},\Delta \lambda^{(1)}) = \alpha \, (\Delta \bm{u}_{n},\Delta \lambda_{n})$.}
\label{fig-arclength-predicted}
\end{figure}

The above equations indicate that the initial guess for the solution increment at the current load step is a constant multiple of the solution increments at the previously converged load step. Therefore, the predictor given by Eq. (\ref{eqn-pred2}) is, in fact, a better estimate than zero, and it brings the initial guess $\left( \bm{u}_{n+1}^{(1)}, \lambda_{n+1}^{(1)} \right)$ closer to the vicinity of the actual solution $\left( \bm{u}_{n+1}, \lambda_{n+1} \right)$, as illustrated in Fig. \ref{fig-arclength-predicted}.

It is worth pointing that the proposed predictor step is simple and also inexpensive when compared with the techniques employed in the classical arc-length implementations in the literature \cite{RiksIJSS1979, CrisfieldCandS1981, RammArcLength1980, SchweizerhofCMAME1986, KrenkIJNME1995}.

\subsubsection{Issue 4: identifying the correct (forward) direction of evolution along the equilibrium path}
Identification of the correct direction of movement along the equilibrium path is another crucial issue in the successful implementation of the arc-length method. Popular techniques proposed for overcoming this issue consist of sophisticated methods which require comparison of the sign of vector products \cite{BelliniCandS1987, FengCS1996}, evaluation of the sign of the determinant the stiffness matrix \cite{CrisfieldCandS1981, FengCS1996}, enforcement of orthogonal conditions \cite{KrenkIJNME1995, KrenkCMAME1995}, or a combination of these \cite{FengCS1996, LeonAMR2011}. Although these techniques have been proven to be successful in computing the complex equilibrium paths in the response of static nonlinear structures, their computer implementation is quite cumbersome.

\begin{figure}[H]
\centering
\subfloat[$u_{n+1} > u_{n}, \; \lambda_{n+1} > \lambda_{n}$]{\includegraphics[trim=0mm 0mm 0mm 0mm, clip, scale=0.42]{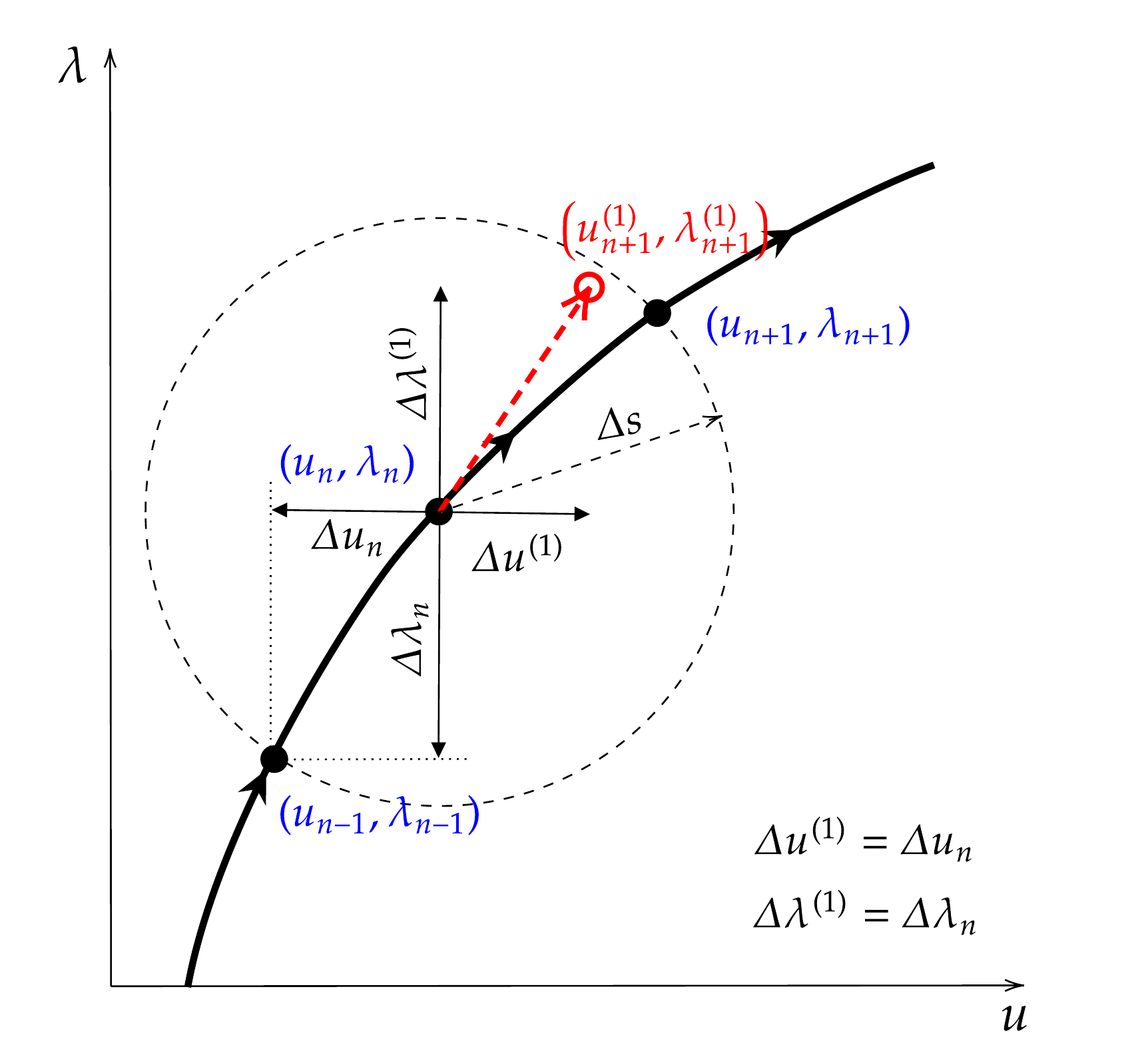} }
\subfloat[$u_{n+1} > u_{n}, \; \lambda_{n+1} < \lambda_{n}$]{\includegraphics[trim=0mm 0mm 0mm 0mm, clip, scale=0.42]{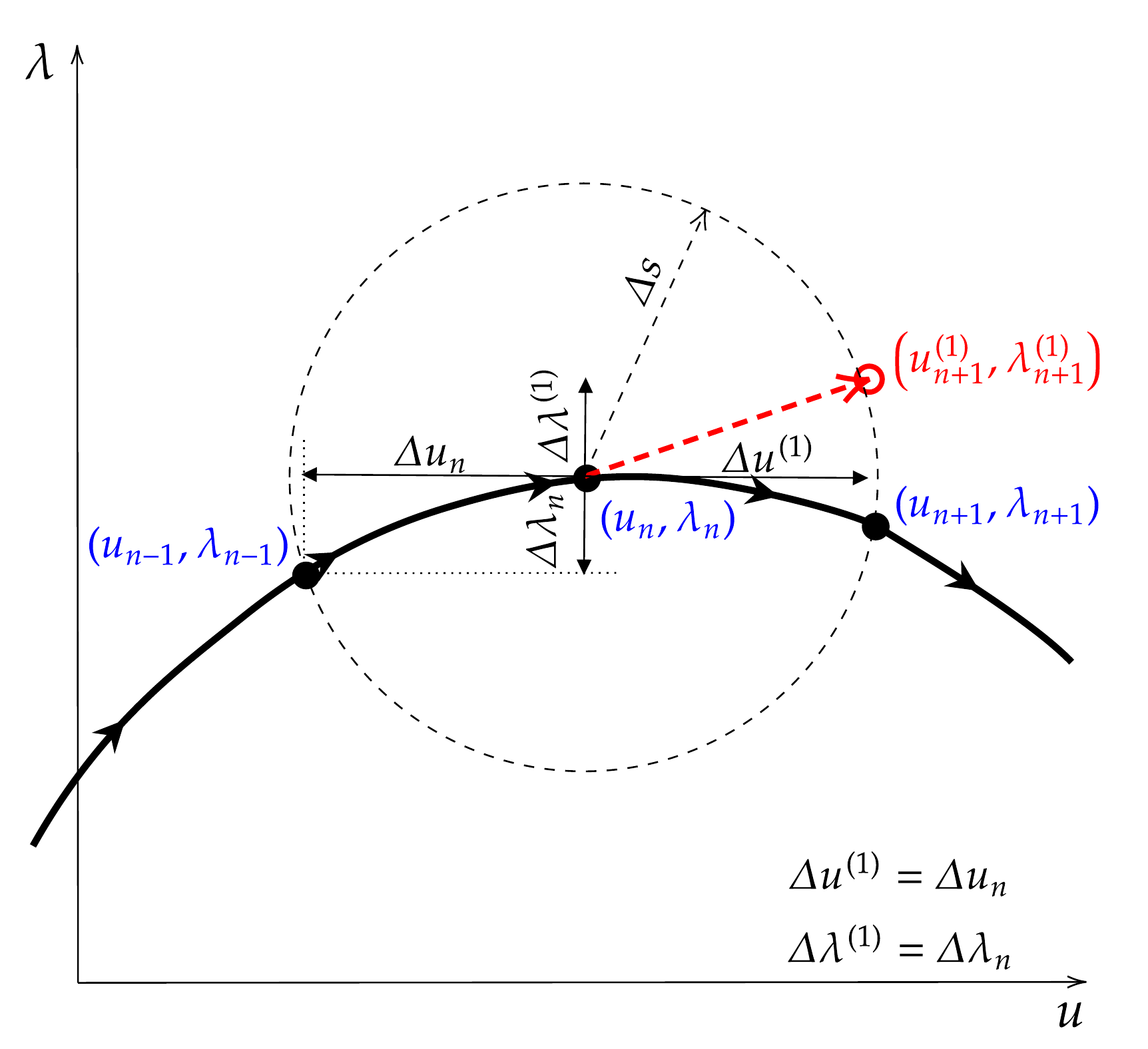} } \\
\subfloat[$u_{n+1} < u_{n}, \; \lambda_{n+1} < \lambda_{n}$]{\includegraphics[trim=0mm 0mm 0mm 0mm, clip, scale=0.42]{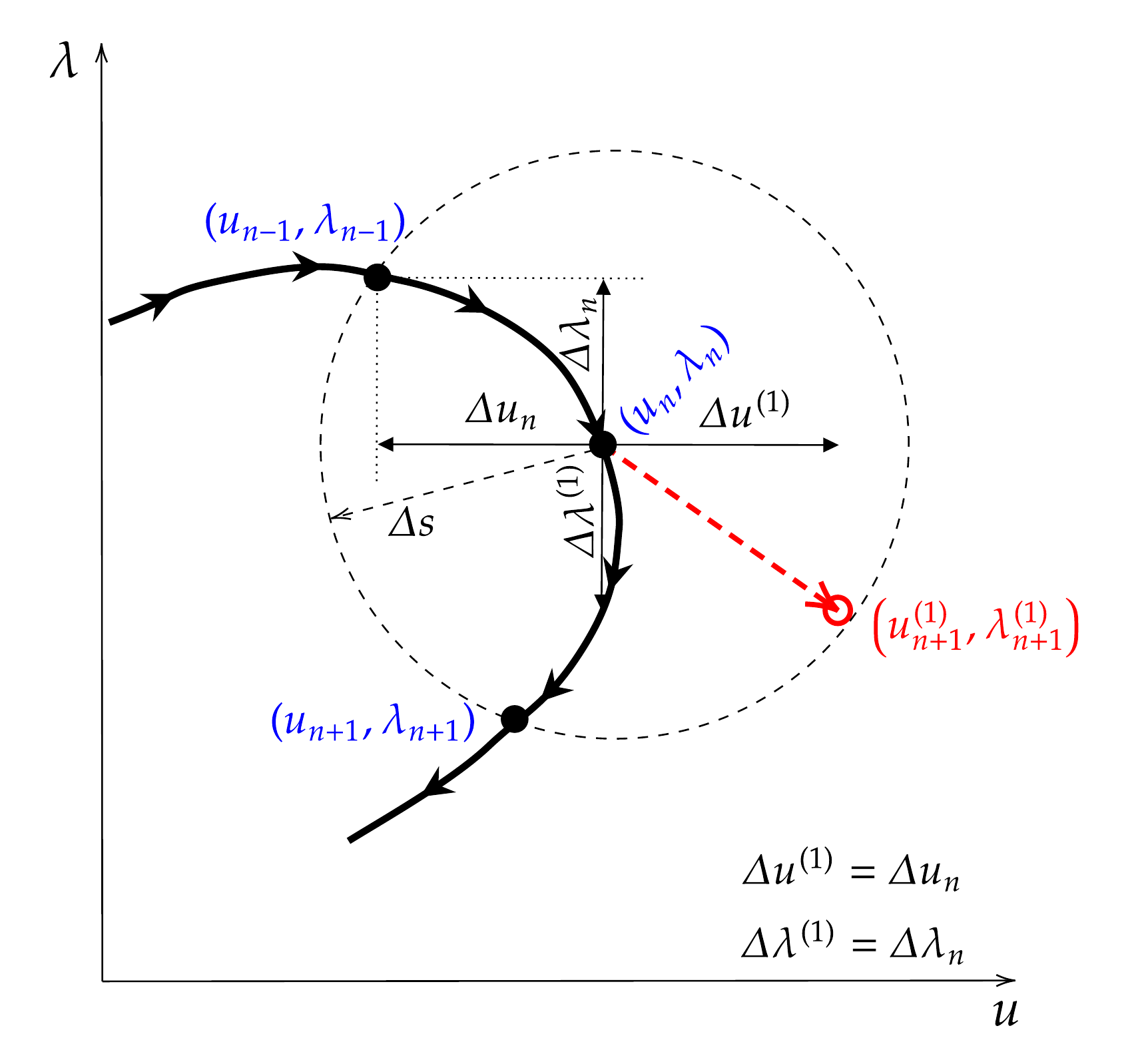} }
\subfloat[$u_{n+1} < u_{n}, \; \lambda_{n+1} > \lambda_{n}$]{\includegraphics[trim=0mm 0mm 0mm 0mm, clip, scale=0.42]{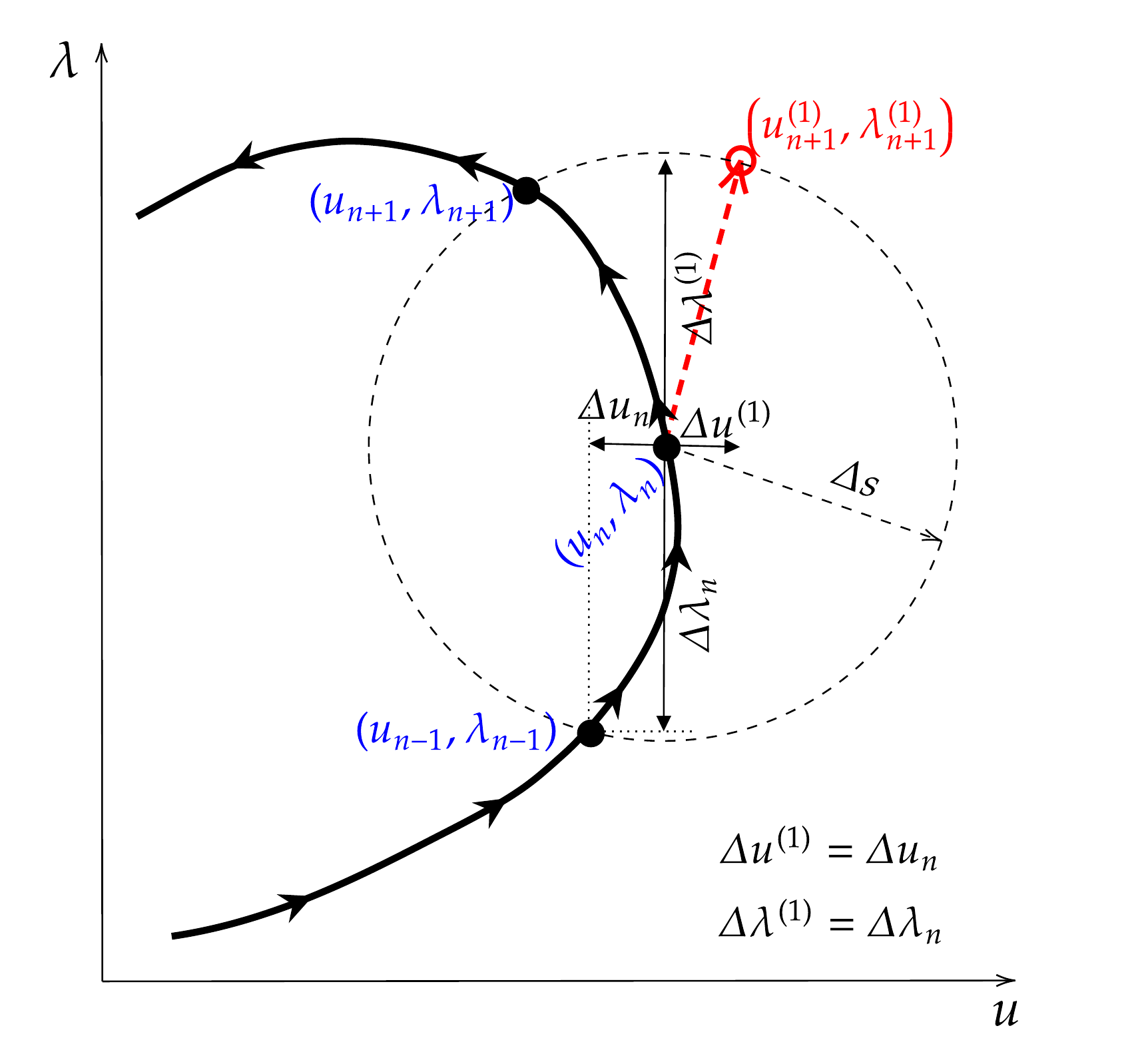} }
\caption{Illustration of predictors for different scenarios along the equilibrium path. $\circ$ and \textbullet \; represent the predicted and actual solutions, respectively. The direction of the predicted solution is denoted with a thick red dashed line.}
\label{fig-arclength-scenrios-1}
\end{figure}

\begin{figure}[H]
\centering
\subfloat[Uniform increment]{\includegraphics[trim=0mm 0mm 0mm 0mm, clip, scale=0.42]{arclength-predictor2.pdf} }
\subfloat[Non-uniform increment (adaptive cutting)]{\includegraphics[trim=0mm 0mm 0mm 0mm, clip, scale=0.42]{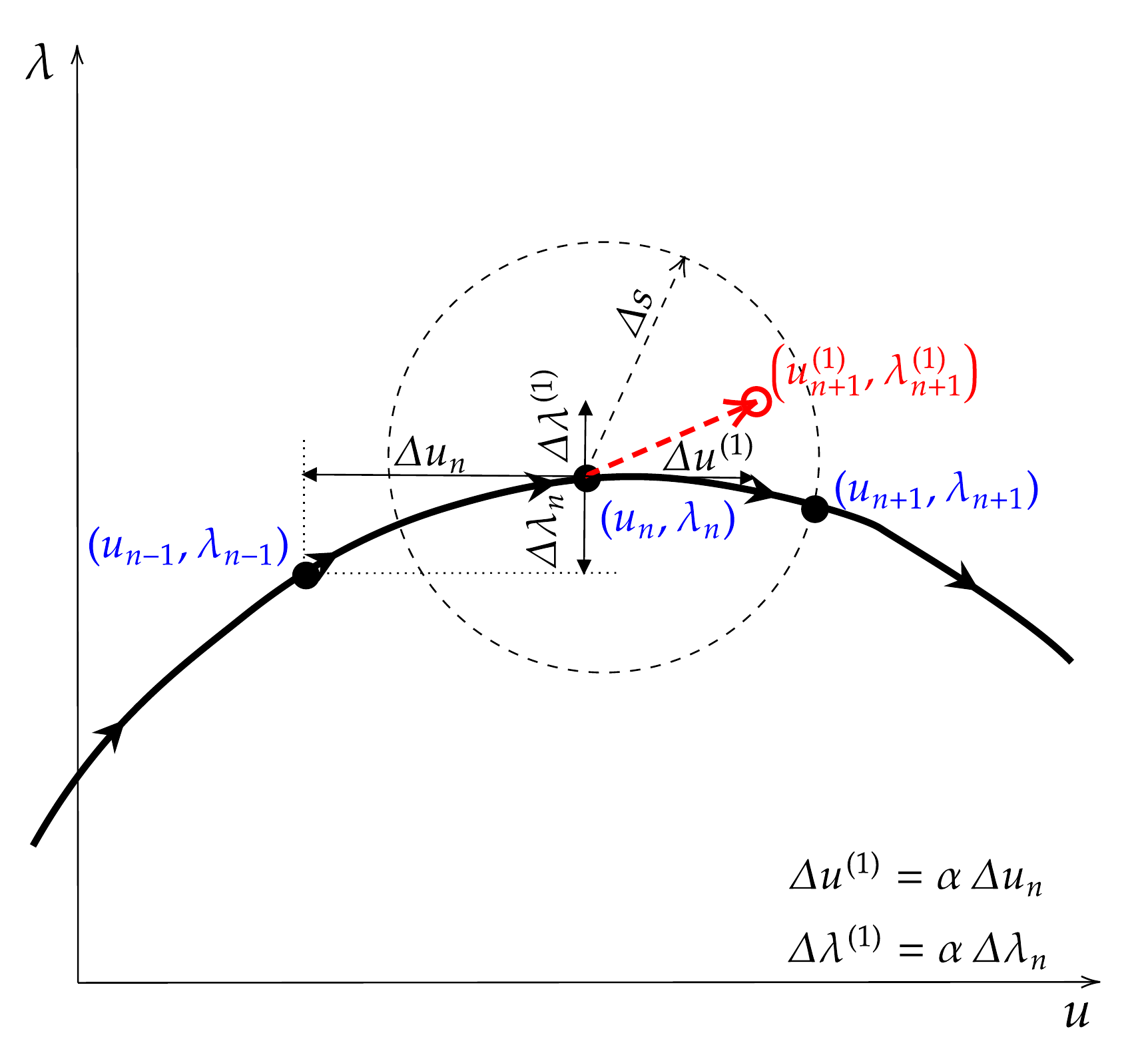} }
\caption{Illustration of predictors with uniform and non-uniform increments in the arc-length parameter.}
\label{fig-arclength-scenrios-2}
\end{figure}

In the present work, no special technique is employed for determining and/or identifying the correct direction of evolution along the equilibrium path. This is taken care by the extrapolation operator used for predicting the solution increments at the first iteration given by Eqs. (\ref{eqn-pred2-Du}) and (\ref{eqn-pred2-Dl}). Towards understanding the reason(s) behind the ability of the proposed technique to successfully compute complex equilibrium paths, the predicted solutions for the 1-DOF problem are illustrated schematically in Fig. \ref{fig-arclength-scenrios-1} for four different scenarios along the equilibrium path for the case with uniform increments of the arc-length parameter. A similar illustration is presented in Fig. \ref{fig-arclength-scenrios-2} for the case with an adaptive cutting. As illustrated in Figs. \ref{fig-arclength-scenrios-1} and \ref{fig-arclength-scenrios-2}, the solution $\left( \bm{u}^{(1)}_{n+1},\lambda^{(1)}_{n+1} \right)$ predicted using the proposed technique is always located on the same side of the actual solution $\left( \bm{u}_{n+1},\lambda_{n+1} \right)$. Another interpretation is that the direction of $\left( \bm{u}^{(1)}_{n+1},\lambda^{(1)}_{n+1} \right) \rightarrow \left( \bm{u}_{n},\lambda_{n} \right)$ is always opposite to that of $\left( \bm{u}_{n},\lambda_{n} \right) \rightarrow \left( \bm{u}_{n-1},\lambda_{n-1} \right)$, thereby forcing the solution increments at the predictor step in the correct direction along the equilibrium path. 

Therefore, the proposed seemingly simple and low-cost extrapolation operator given by Eq. (\ref{eqn-arclen-preds}) not only predicts the solution at the first iteration but also serves to identify the correct direction along the equilibrium path, without the need for any sophisticated techniques. This ability of the proposed technique to successfully compute complex equilibrium paths in nonlinear structural mechanics problems is illustrated with numerical examples in Section \ref{sec-examples}.

\vspace{-10mm}

\section{Numerical examples} \label{sec-examples}
The ability of the proposed arc-length implementation in capturing complex equilibrium paths is illustrated using seven benchmark examples consisting of nonlinear truss, beam-column and shell models. The pseudocode for the arc-length method using the proposed predictor is presented in Algorithm. \ref{algo-arclength}. The nonlinear space truss finite element models are discussed briefly in \ref{sec-appendixA}. For the examples modelled with beams, the geometrically exact beam-column element is used; the reader is referred to Chapter 17 in Zienkiewicz and Taylor \cite{book-fem-ZienkiewiczVol2} for the details. The numerical solutions of the shell problems are computed using continuum finite elements by adapting the mixed displacement-pressure formulation recently proposed in Kadapa and Mokarram \cite{KadapaMAMS2020}.

The spherical arc-length method ($\psi=1$) is used in all the simulations reported in this work. The convergence tolerance ($\epsilon$) is assumed to be $10^{-6}$, and the maximum number of iterations, $k_{\max}$, is set to 10. When the convergence is not achieved according to these two criteria, the arc-length increment ($\Delta s$) is changed adaptively, see lines 42-48 in Algorithm. \ref{algo-arclength}. The specified value of the point load is one, unless stated otherwise explicitly. Note that the numerical solutions computed in all the examples follow only the primary solution branch. For all the numerical examples considered in this work, the simulation time ranges from a few seconds to a few minutes on a personal computer fitted with Intel i7-8750H CPU. The total number of load steps, total number of iterations, average number of iterations and number of restarts for all the examples are tabulated in Table \ref{table-counts}.

\renewcommand{\arraystretch}{1.5}
\begin{table}[H]
\centering
\begin{tabular}{|c|c|c|c|c|}
\hline
Example  & No. of load steps & No. total of iterations  &  No. of average iterations & No. of restarts  \\
\hline
3.1  &   50    & 151, 148, 163, 184  & 3.00, 2.96, 3.26, 3.68  &    0   \\
3.2  &  100    &   344               &        3.44             &    0   \\
3.3  &   50    &   277               &        5.54             &    0   \\
3.4  &  120    &   843               &        7.03             &   16   \\
3.5  &  400, 600    &  2320, 4185               &        5.80, 6.98             &   44, 152   \\
3.6  &   20, 30    &    86, 134               &        4.30, 4.47             &    0, 0   \\
3.7  &   50    &   275               &        5.50             &    0   \\
\hline
\end{tabular}
\caption{Details of load step and iteration counts for the numerical examples. Multiple entries in the columns correspond to the different cases in the order they appear in the respective examples.}
\label{table-counts}
\end{table}
\renewcommand{\arraystretch}{1.0}

\begin{algorithm}[H]
\caption{Algorithm for the arc-length method} \label{algo-arclength}
\begin{algorithmic}[1]
\State Set $\Delta \lambda$, $n_{\mathrm{max}}$, $k_{\mathrm{max}}$ and $\epsilon$. converged=False. $\psi=1$. Initialise variables.
\State Compute $\mathbf{F}$
\For{$n=1$ to $n_{\mathrm{max}}$}
   \State \textsc{\textcolor{blue}{\#1 Predictor step:}}
   \If{$n > 1$}
     \State $\alpha = \Delta s/\Delta s_n$
     \State $\bm{u}_{n+1}^{(1)}  = [1+\alpha] \, \bm{u}_{n}  - \alpha \, \bm{u}_{n-1}$
     \State $\lambda_{n+1}^{(1)} = [1+\alpha] \, \lambda_{n} - \alpha \, \lambda_{n-1}$
   \EndIf
   \State $\Delta \bm{u}^{(1)}  = \bm{u}_{n+1}^{(1)} - \bm{u}_{n}$
   \State $\Delta \lambda^{(1)} = \lambda_{n+1}^{(1)} - \lambda_{n}$
   \State convergedPrev = converged
   \State converged = False
   \State \textsc{\textcolor{blue}{\#2 Corrector step:}}
   \For{$k=1$ to $k_{\mathrm{max}}$}
   \State Compute: $\mathbf{K}(\bm{u}_{n+1}^{(k)})$, $\bm{a}$, $b$, $\mathcal{A}(\bm{u}_{n+1}^{(k)}, \lambda_{n+1}^{(k)})$ and $\mathbf{R}(\bm{u}_{n+1}^{(k)}, \lambda_{n+1}^{(k)})$ in Eq. (\ref{eqn-matsys-arclen})
   \If{$\vert \mathbf{R}(\bm{u}_{n+1}^{(k)}, \lambda_{n+1}^{(k)}) \vert \leq \epsilon $}
   \State converged = True
   \State Exit iteration loop
   \EndIf
   \State Solve: $\delta \lambda$ and $\delta \bm{u}$ from Eqs. (\ref{eqn-appr2-dl}) and (\ref{eqn-appr2-du})
   \State $\Delta \bm{u}^{(k+1)} = \Delta \bm{u}_{n+1}^{(k)} + \delta \bm{u}$
   \State $\Delta \lambda_{n+1}^{(k+1)} = \Delta \lambda_{n+1}^{(k)} + \delta \lambda$
   \State $\bm{u}_{n+1}^{(k+1)} = \bm{u}_{n+1}^{(k)} + \delta \bm{u}$
   \State $\lambda_{n+1}^{(k+1)} = \lambda_{n+1}^{(k)} + \delta \lambda$
   \EndFor
   \State \textsc{\textcolor{blue}{\#3 Solution update:}}
   \If{converged}
      \If{n == 1}
         \State $\Delta s = \sqrt{[\Delta \bm{u}]^{\T} \, [\Delta \bm{u}] + \psi \, [\Delta \lambda]^2 \, \mathbf{F}^{\T} \, \mathbf{F}}$
         \State $\Delta s_{\max} = \Delta s$
         \State $\Delta s_{\min} = \Delta s/1024$
      \EndIf
      \State $\Delta \lambda_{n-1} = \Delta \lambda_{n}$
      \State $\Delta \lambda_{n}   = \Delta \lambda_{n+1}$
      \State $\Delta s_{n} = \Delta s$
      \If{convergedPrev}
        \State $\Delta s = \min\Big(\max(2 \, \Delta s, \Delta s_{\min}), \Delta s_{\max}\Big)$
      \EndIf
   \State $(\bm{u}_{n-1}, \lambda_{n-1}) \gets (\bm{u}_{n},\lambda_{n})$
   \State $(\bm{u}_{n}, \lambda_{n}) \gets (\bm{u}_{n+1},\lambda_{n+1})$
   \Else
      \If{convergedPrev}
        \State $\Delta s = \max(\Delta s/2, \; \Delta s_{\min})$
      \Else
        \State $\Delta s = \max(\Delta s/4, \; \Delta s_{\min})$
      \EndIf
   \EndIf
\EndFor
\end{algorithmic}
\end{algorithm}

\subsection{Plane truss with three members}
\begin{figure}[H]
\centering
\subfloat[]{\includegraphics[trim=10mm 0mm 10mm 0mm, clip, scale=0.44]{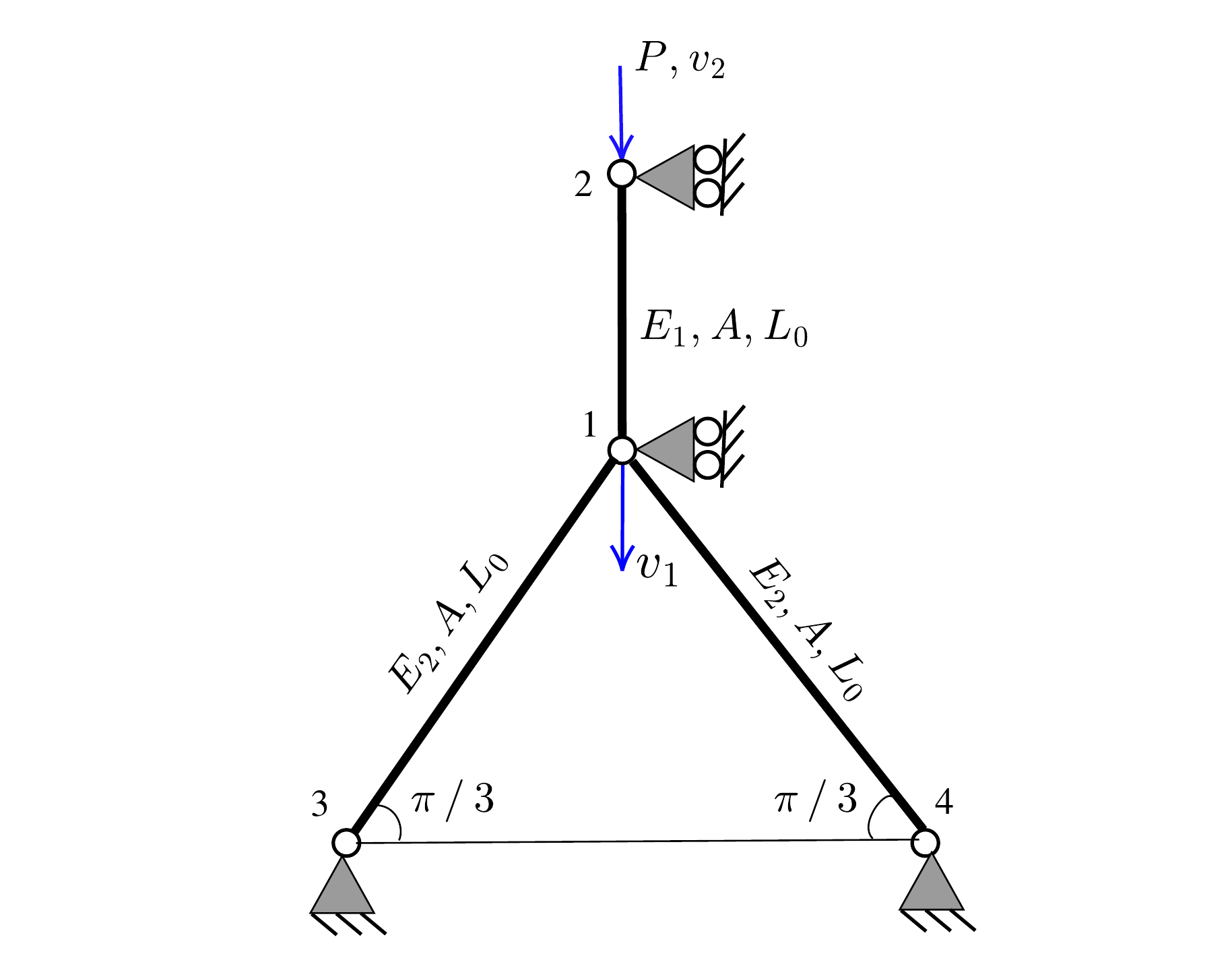} \label{fig-planartruss-geom}}
\subfloat[]{\includegraphics[clip, scale=0.37]{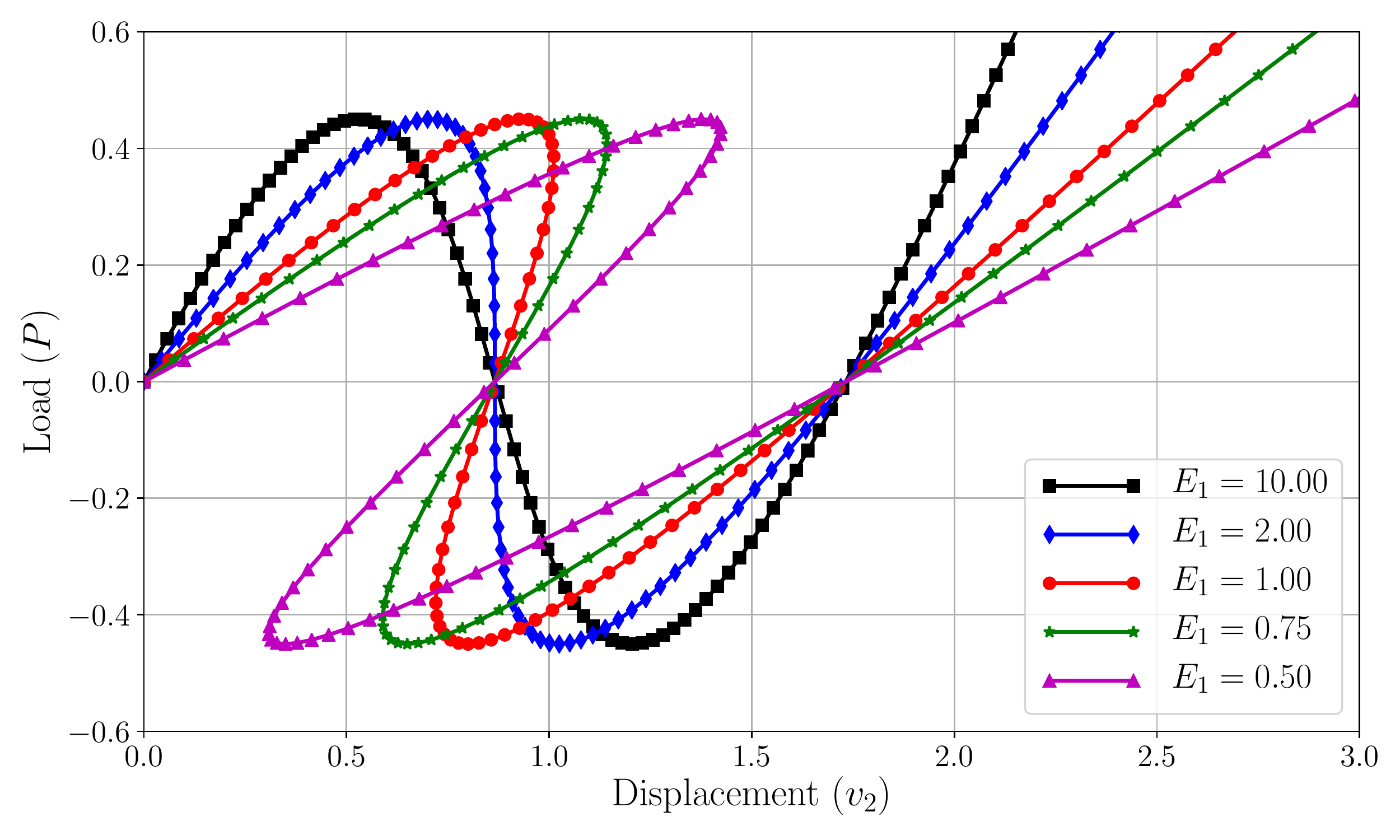} \label{fig-planartruss-graphs1}}
\caption{3-member planar truss: (a) geometry and boundary conditions and (b) analytical load-displacement curves for node 2 for various values of $E_1$ with $L_0=A=E_2=1$.}
\end{figure}

The first numerical example consists of a planar truss with 3 bars which are arranged in the configuration shown in Fig. \ref{fig-planartruss-geom}. For this problem, the truss model based on the engineering strain ($\varepsilon_{E}$) is used, see \ref{sec-appendixA}. This truss structure experiences a highly nonlinear deformation behaviour, as illustrated with load-displacement curves for node 2 in Fig. \ref{fig-planartruss-graphs1}. For $E_1 \gg E_2$, the bar 1-2 acts like a rigid structure and it undergoes significantly less deformation. For this case, the load-control method results in a snap-through behaviour but accurate solutions can be obtained using the displacement-control method. However, for $E_1 < E_2$, the displacement-control method also fails to track the correct equilibrium path as it results in a snap-back response. To accurately compute the response of the structure over a wide range of parameters, the arc-length method is essential.

\begin{figure}[H]
\centering
\subfloat[$E_1=10.0$]{\includegraphics[clip, scale=0.3]{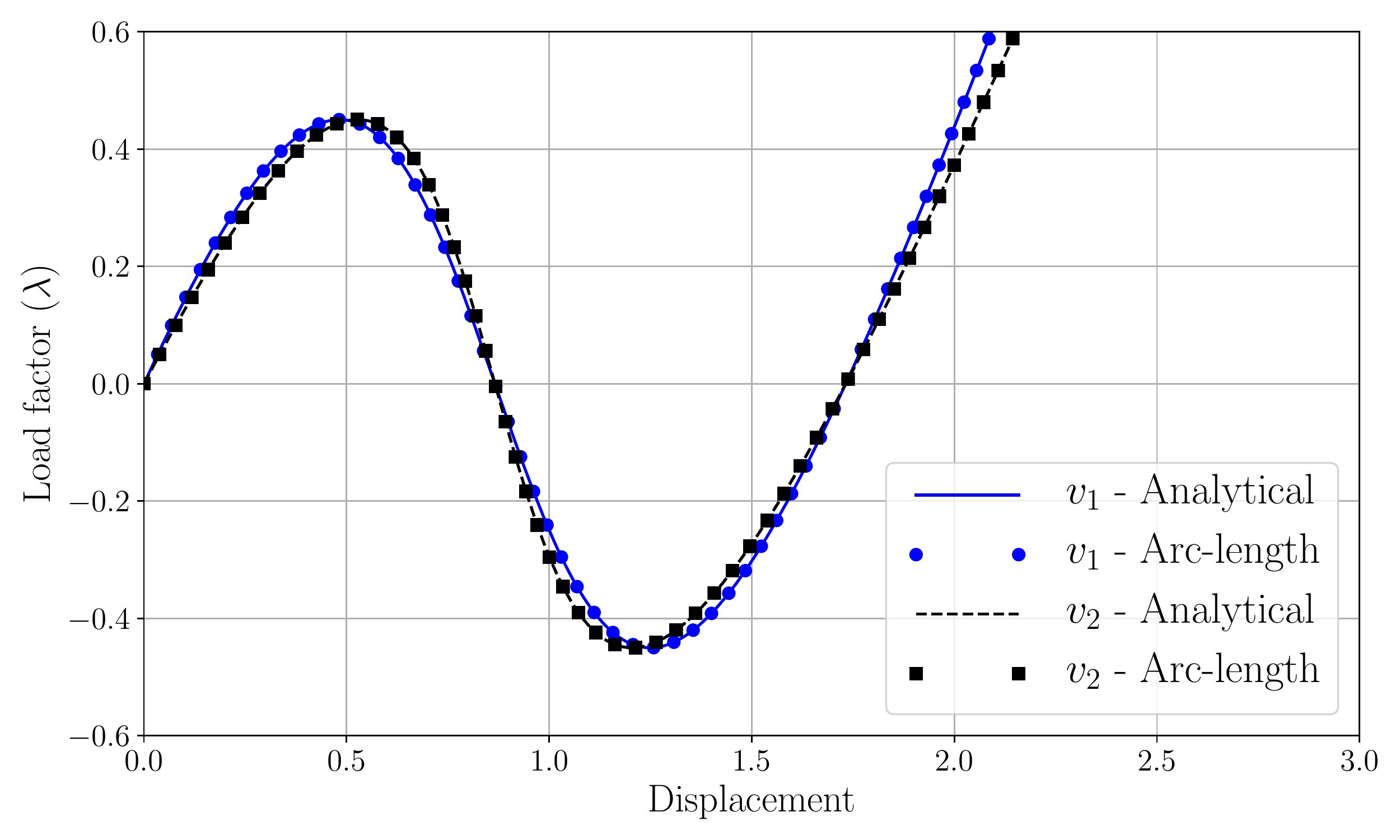}}
\subfloat[$E_1=2.0$]{\includegraphics[clip, scale=0.3]{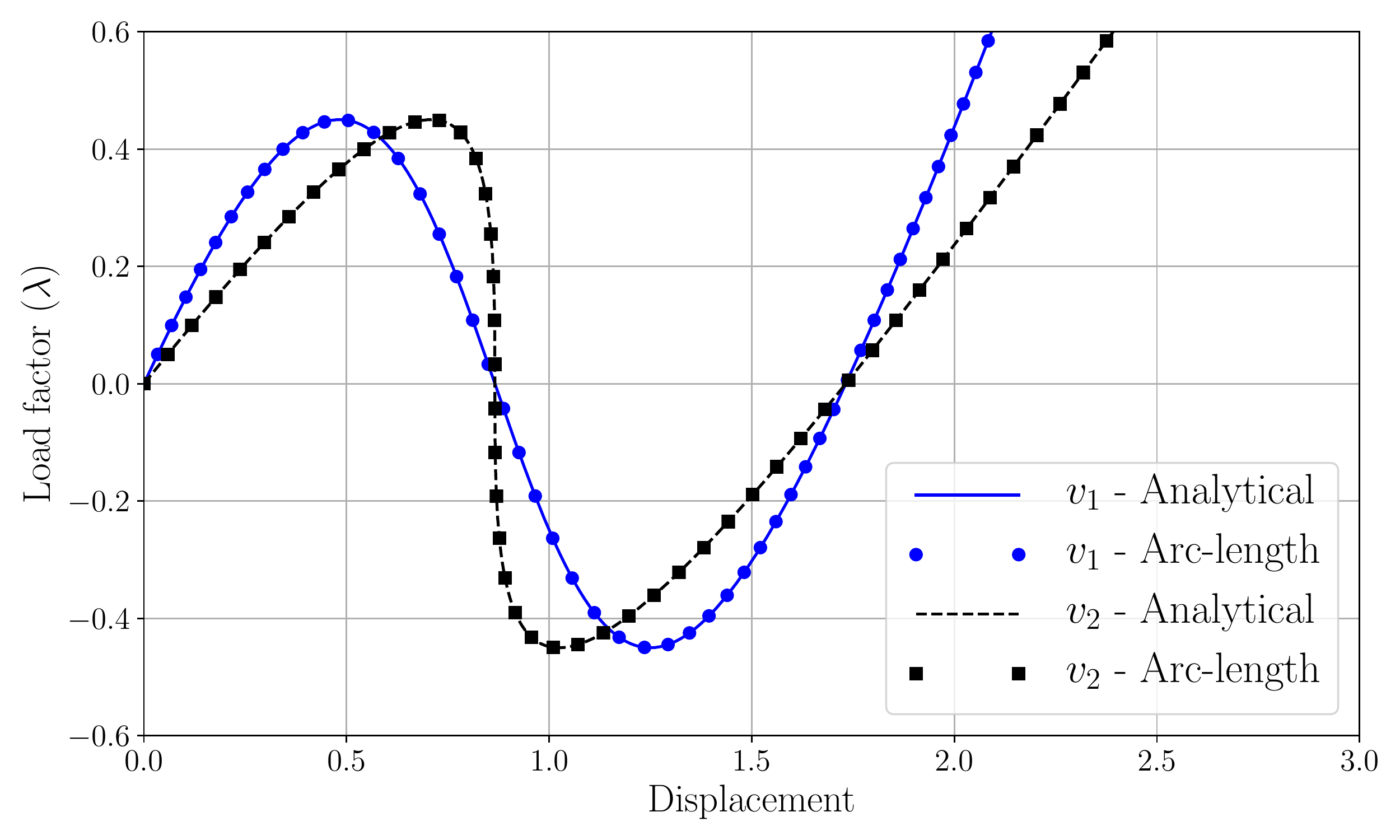}} \\
\subfloat[$E_1=0.75$]{\includegraphics[clip, scale=0.3]{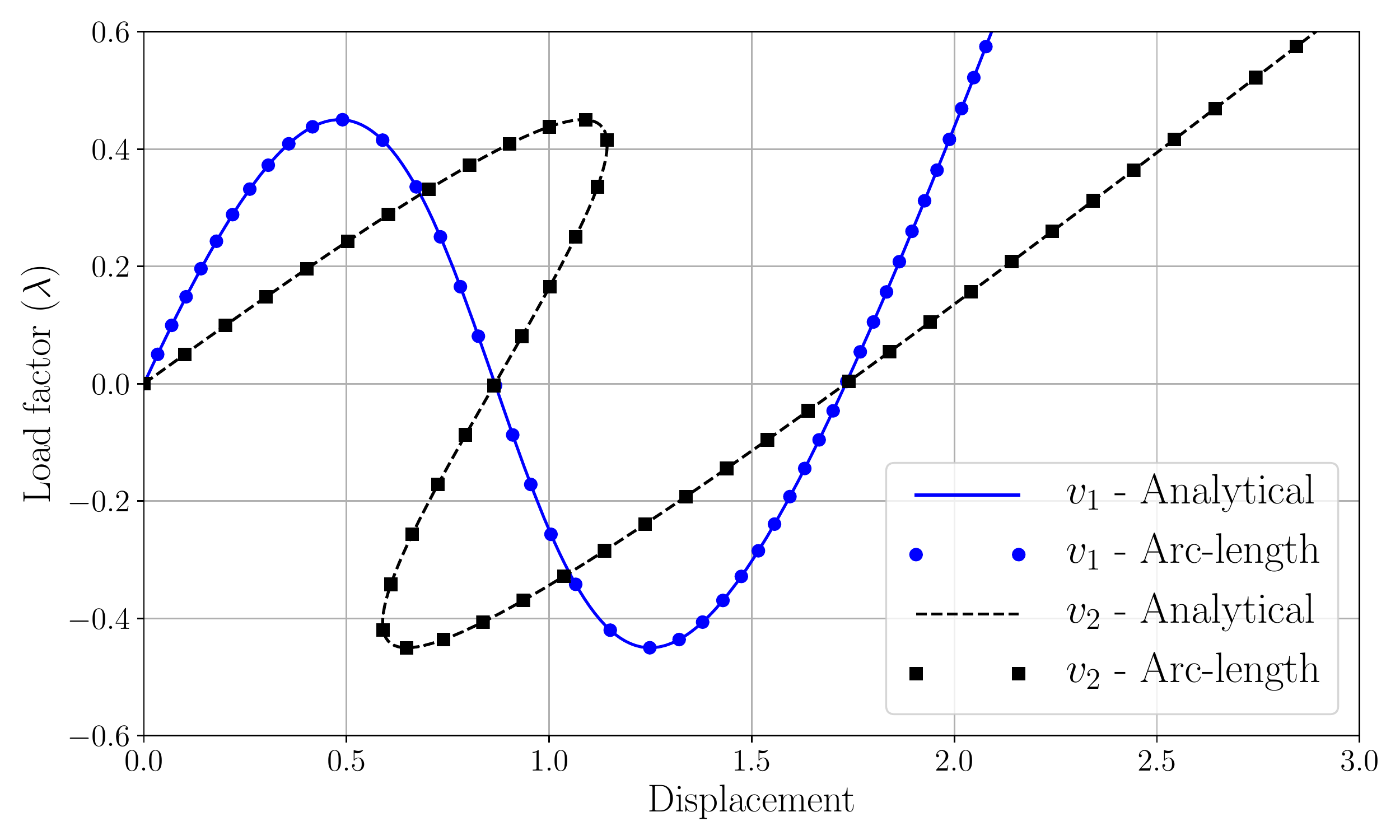}}
\subfloat[$E_1=0.5$]{\includegraphics[clip, scale=0.3]{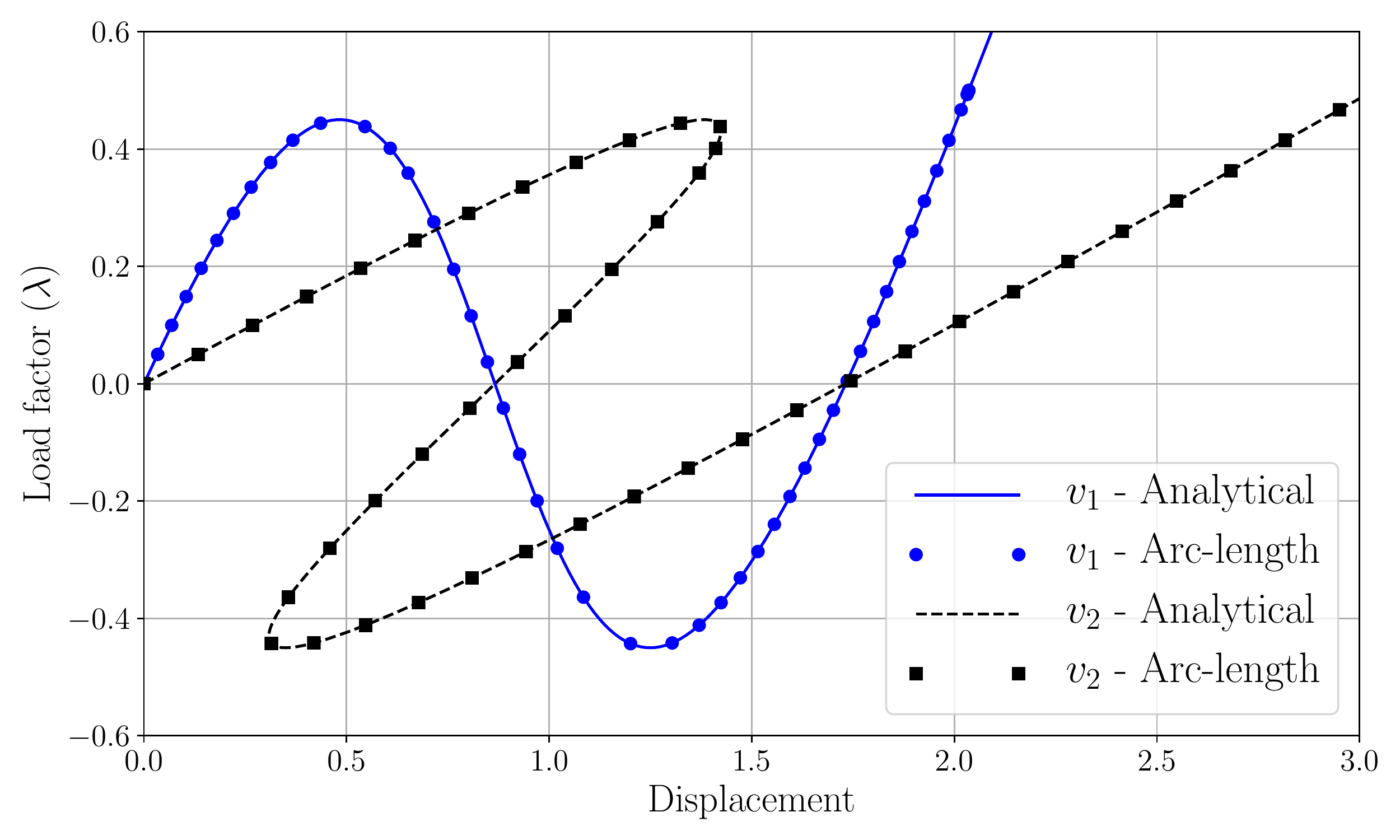}}
\caption{3-member planar truss: load-displacement curves for the free DOFs with $L_0=A=E_2=1$ and different values of $E_1$. The markers represent the converged load steps.}
\label{fig-planartruss-graph2}
\end{figure}

Numerical solutions are computed using the proposed arc-length implementation for four different values of $E_1$ and the load-displacement curves are presented in Fig. \ref{fig-planartruss-graph2} together with the analytical solution. As shown, the proposed technique captures equilibrium paths for all four cases quite accurately. It is also worth mentioning that the proposed implementation does not require prohibitively small increments for successful computation of numerical solutions.

\subsection{Space truss with 12 members}
\begin{figure}[H]
\centering
\subfloat[]{\includegraphics[trim=0mm 0mm 0mm 0mm, clip, scale=0.45]{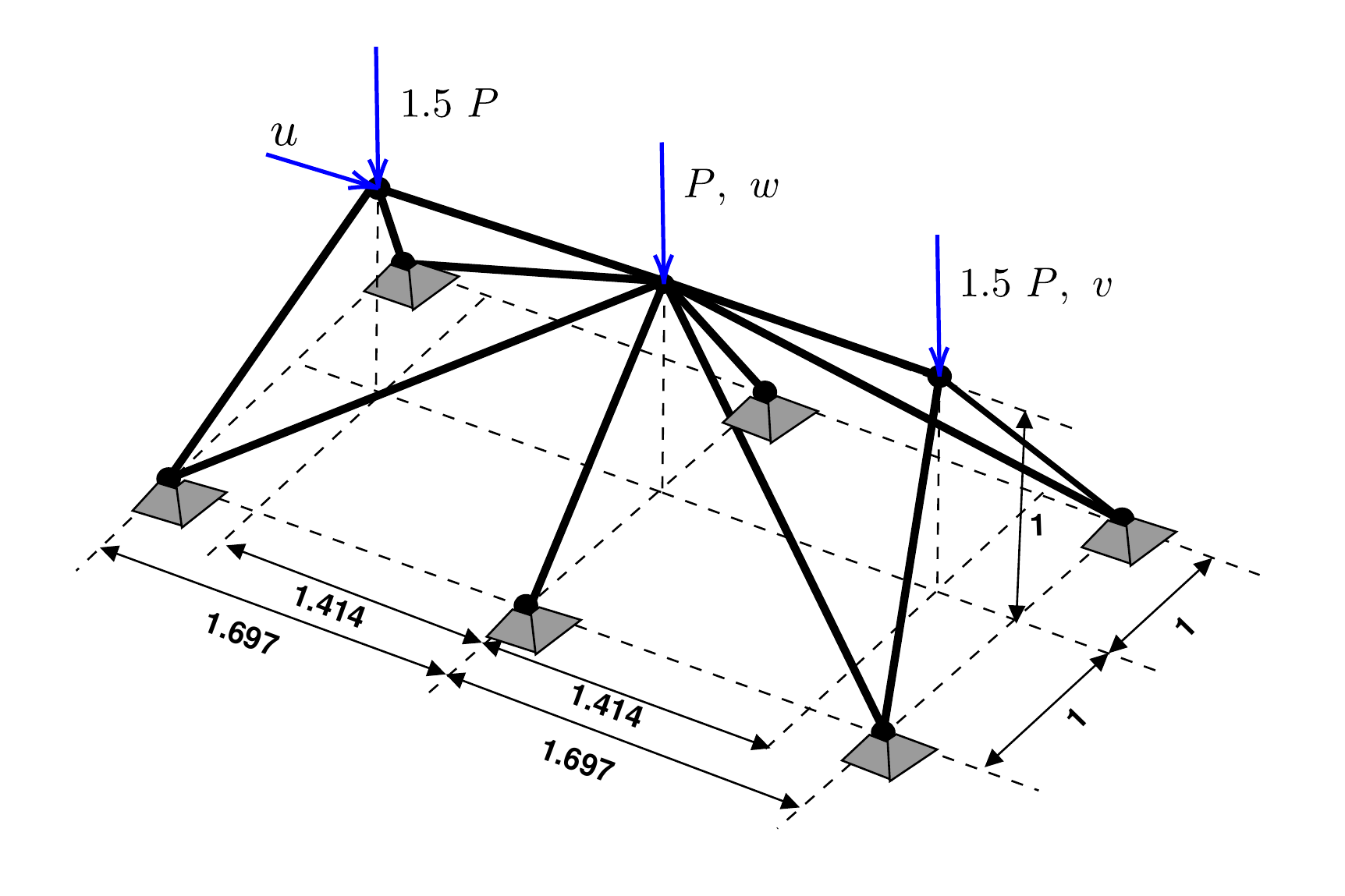} \label{fig-spacetruss12-geom} }
\subfloat[]{\includegraphics[clip, scale=0.3]{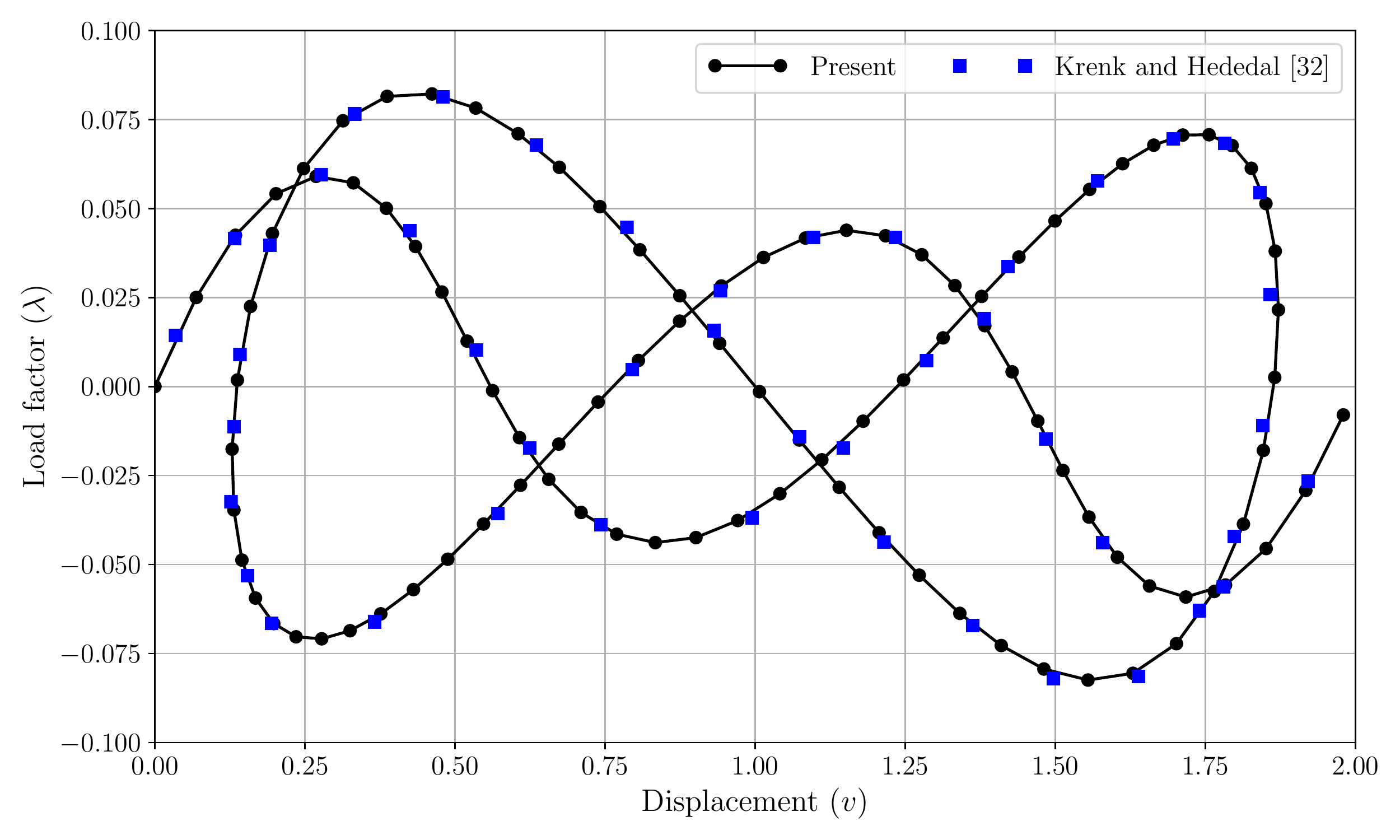} \label{fig-spacetruss12-graph1}} \\
\subfloat[]{\includegraphics[clip, scale=0.3]{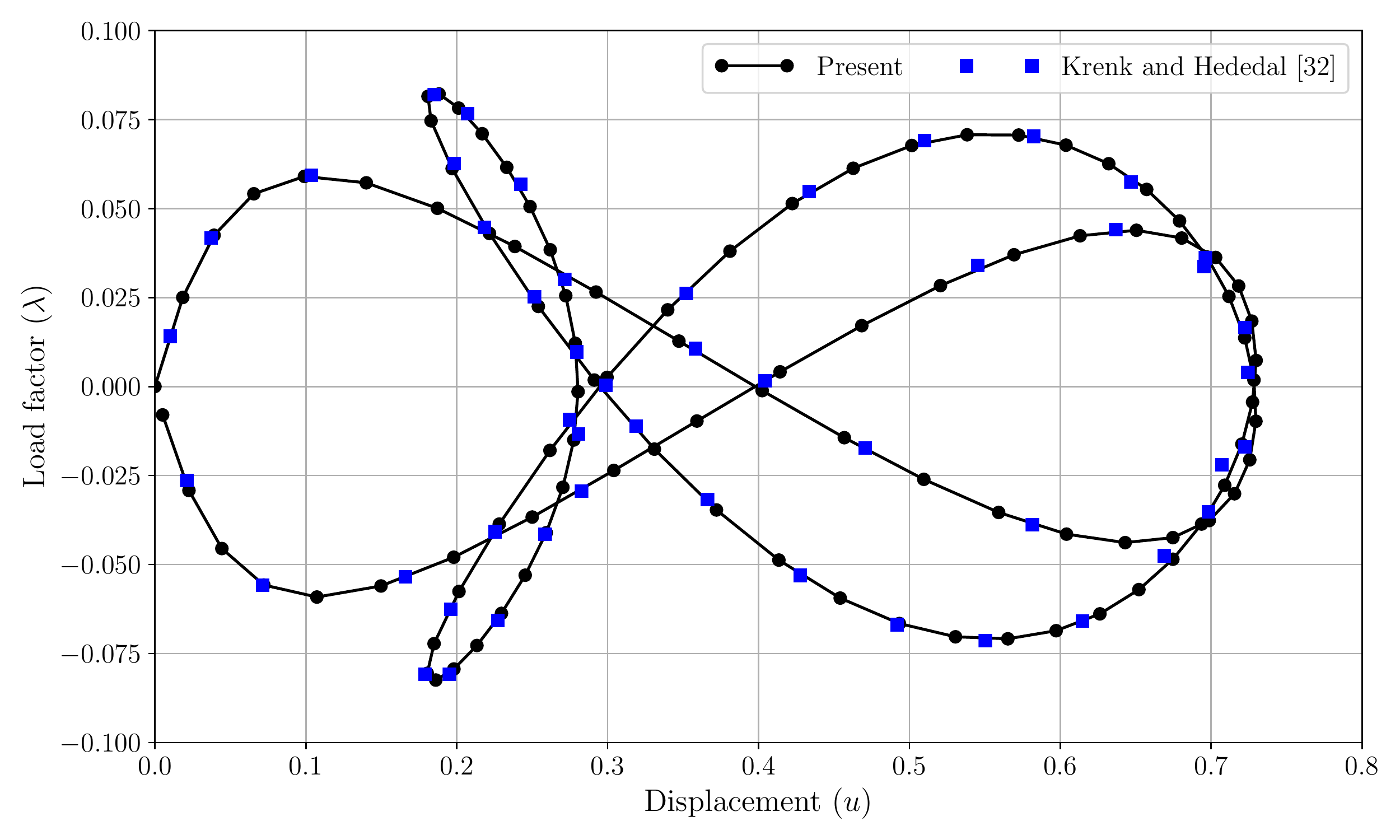} \label{fig-spacetruss12-graph2}}
\subfloat[]{\includegraphics[clip, scale=0.3]{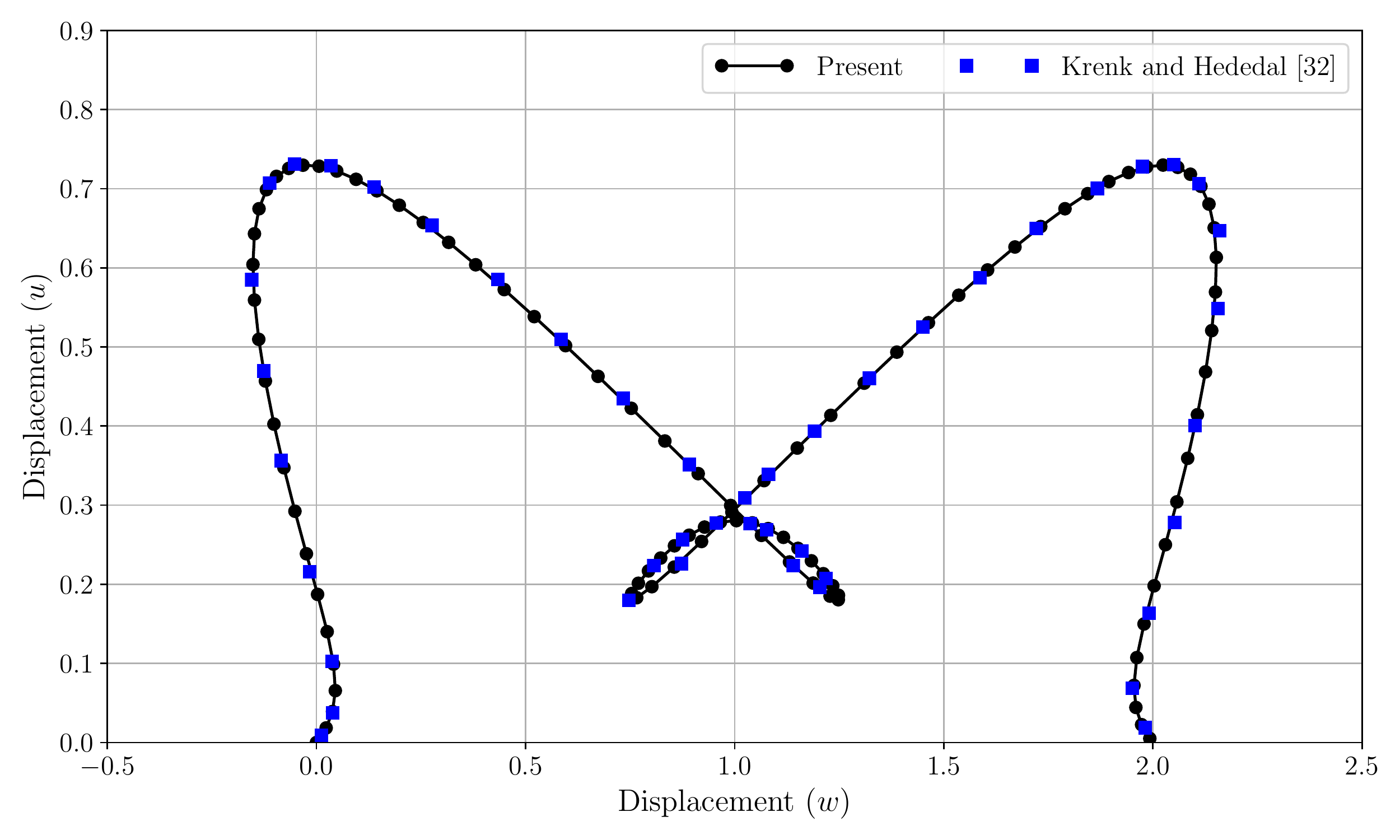} \label{fig-spacetruss12-graph3}}
\caption{12-member space truss: (a) geometry and boundary conditions, (b) $P$ - $u$ curve, (c) $P$ - $w$ curve, and (d) $u$ - $w$ curve.}
\label{fig-spacetruss12}
\end{figure}

This example consists of a 12-bar space truss structure whose geometry and boundary conditions are shown in Fig. \ref{fig-spacetruss12-geom}. This example has been previously studied by Yang and Leu \cite{YangCMAME1991}, Yang et al. \cite{YangMAMS2007}, Krenk and Hededal \cite{KrenkCMAME1995}, Leon et al. \cite{LeonAMR2011}, and Habibi and Bidmeshki \cite{HabibiIJSSD2019}. The parameters are: $A=1$ and $E=1$. Each bar is discretised with one nonlinear truss element based on the Green-Lagrange strain measure ($\varepsilon_G$), see \ref{sec-appendixA}. The analysis is started using a load increment of $\Delta \lambda = 0.025$, which corresponds to an arc-length of $\Delta s= 0.10636$. The response of the structure is presented in terms of load-displacement curves in Figs. \ref{fig-spacetruss12-graph1} and \ref{fig-spacetruss12-graph2} and displacement-displacement curve in Fig. \ref{fig-spacetruss12-graph3}. These graphs illustrate that the results obtained using the proposed technique match well with the solution obtained by Krenk and Hededal \cite{KrenkCMAME1995}. The results obtained for the present example show that the proposed arc-length implementation captures the complex nonlinear response of the structure quite well.

\subsection{Lee frame}
\begin{figure}[H]
\centering
\subfloat[]{\includegraphics[trim=0mm 0mm 0mm 0mm, clip, scale=0.5]{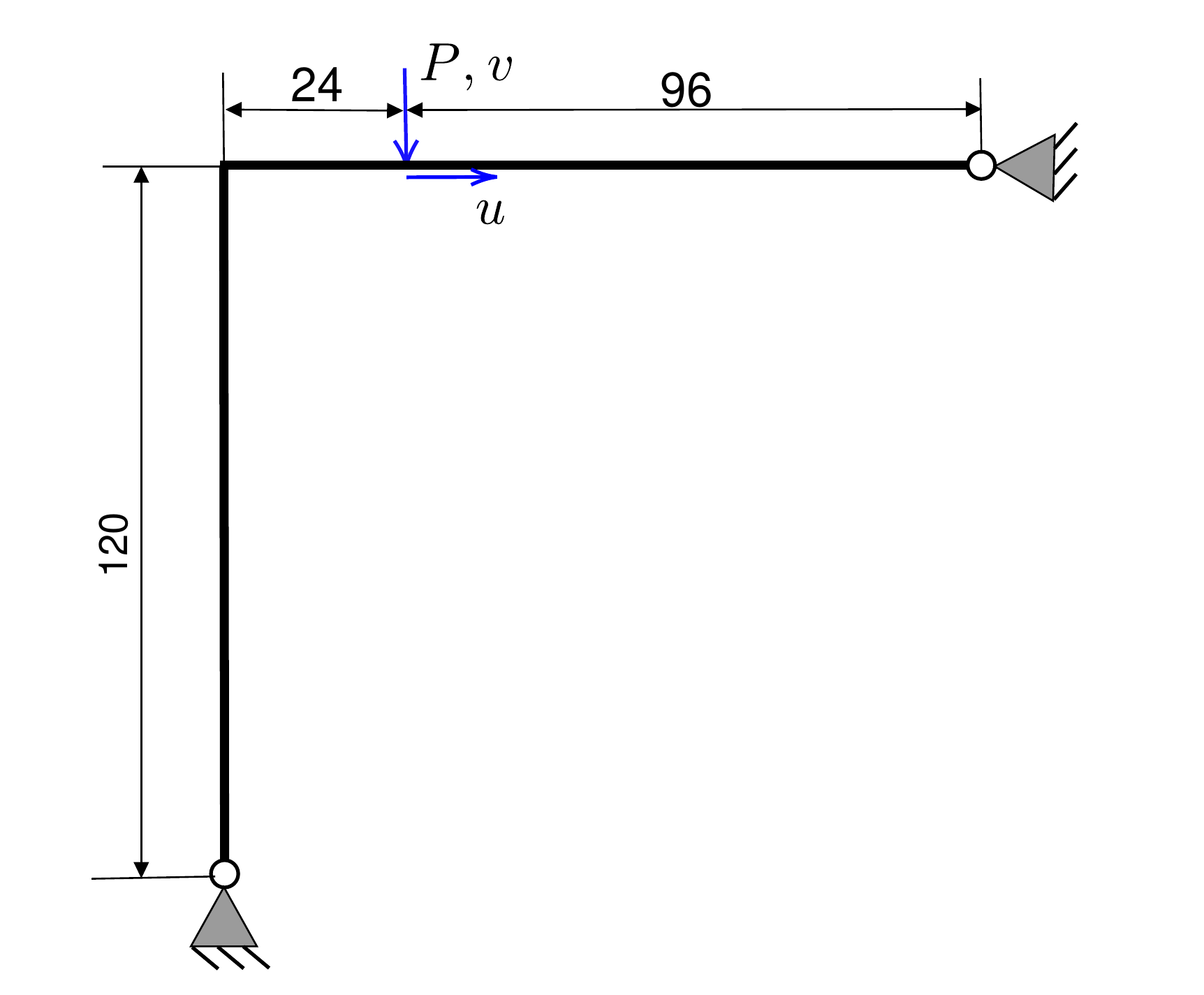} \label{fig-leeframe-geom} }
\subfloat[]{\includegraphics[clip, scale=0.45]{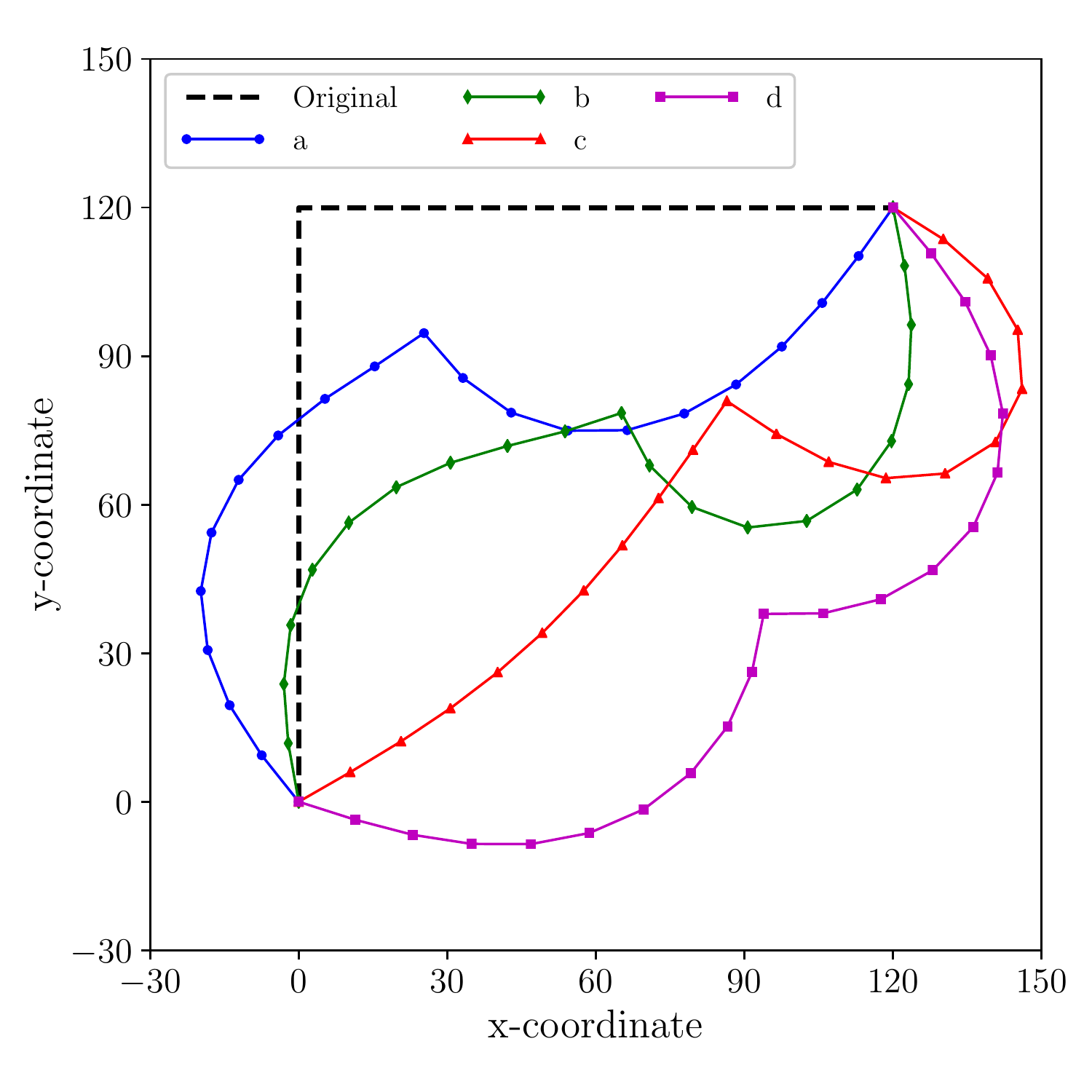} \label{fig-leeframe-defshapes} }
\caption{Lee frame: (a) geometry and boundary conditions and (b) deformed configurations of the frame at points a, b, c and d marked in Fig. \ref{fig-leeframe-graph}.}
\end{figure}

\begin{figure}[H]
\centering
\includegraphics[clip, scale=0.45]{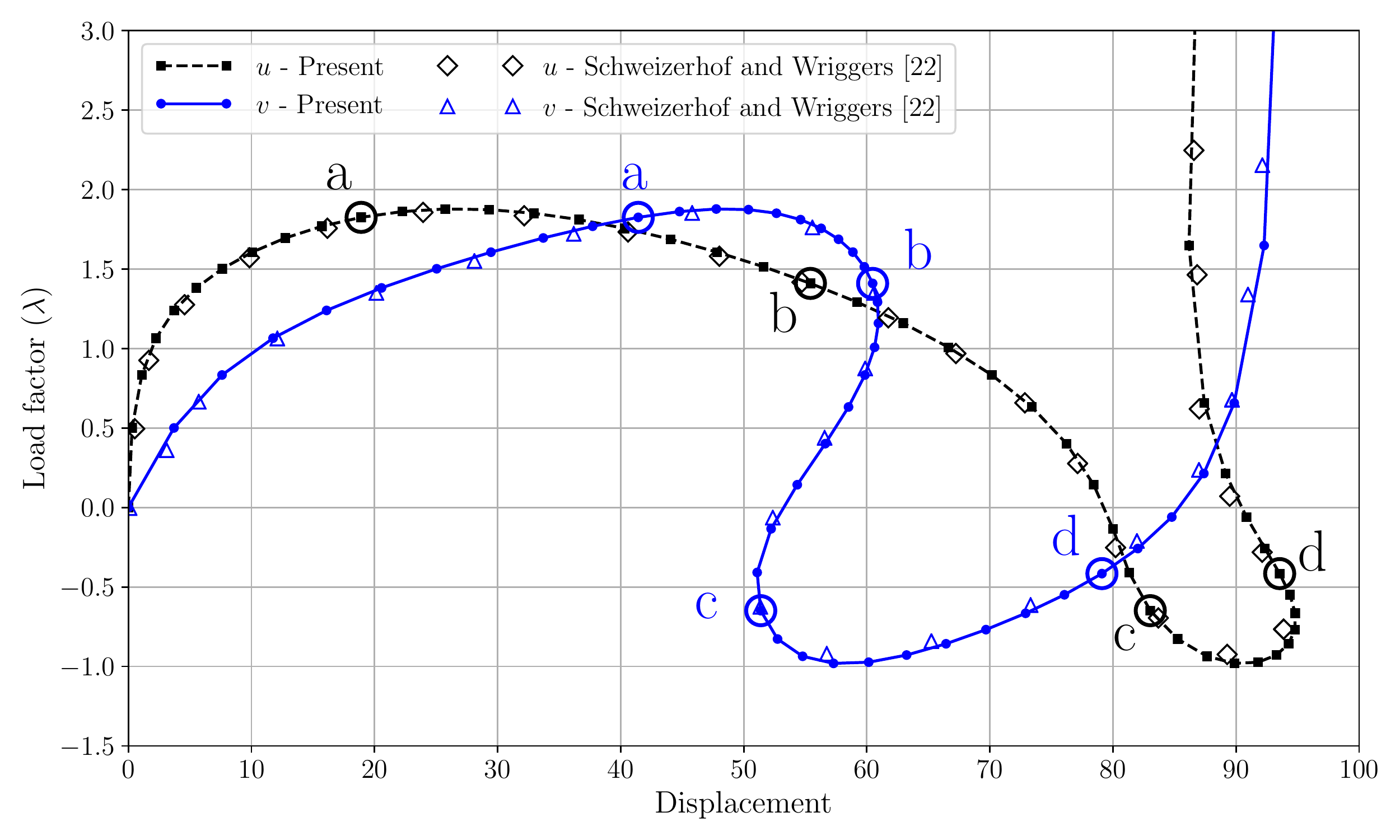}
\caption{Lee frame: load-displacement curve. The markers for the present work represent the converged load steps.}
\label{fig-leeframe-graph}
\end{figure}

This example consists of a planar frame whose geometry and boundary conditions are as shown in Fig. \ref{fig-leeframe-geom}. The geometric and material parameters are: $A=6$ cm$^2$, $I=2$ cm$^4$, $E=720$ kN/cm$^2$, $\nu=0.3$ and $\kappa=1$. The point load $P$ is taken as 1 kN. In this example, the snap-back behaviour occurs once the frame undergoes a significantly large deformation. The analysis is performed using 20 nonlinear beam-column elements and with an initial load increment $\Delta \lambda = 0.5$, which corresponds to an arc-length increment of $\Delta s = 14.24$. The load-displacement curve for the node at which the point load is applied is shown in Fig. \ref{fig-leeframe-graph} along with a reference solution from Schweizerhof and Wriggers \cite{SchweizerhofCMAME1986}. Deformed shapes of the frame at four different points a, b, c and d in Fig. \ref{fig-leeframe-graph} are shown in \ref{fig-leeframe-defshapes}. The load-displacement graphs presented for the example illustrate that the proposed technique successfully captures the load limit point as well as the displacement limit point.

\subsection{Hinged-clamped 215-degree arch}
\begin{figure}[H]
\centering
\subfloat[]{\includegraphics[clip, scale=0.4]{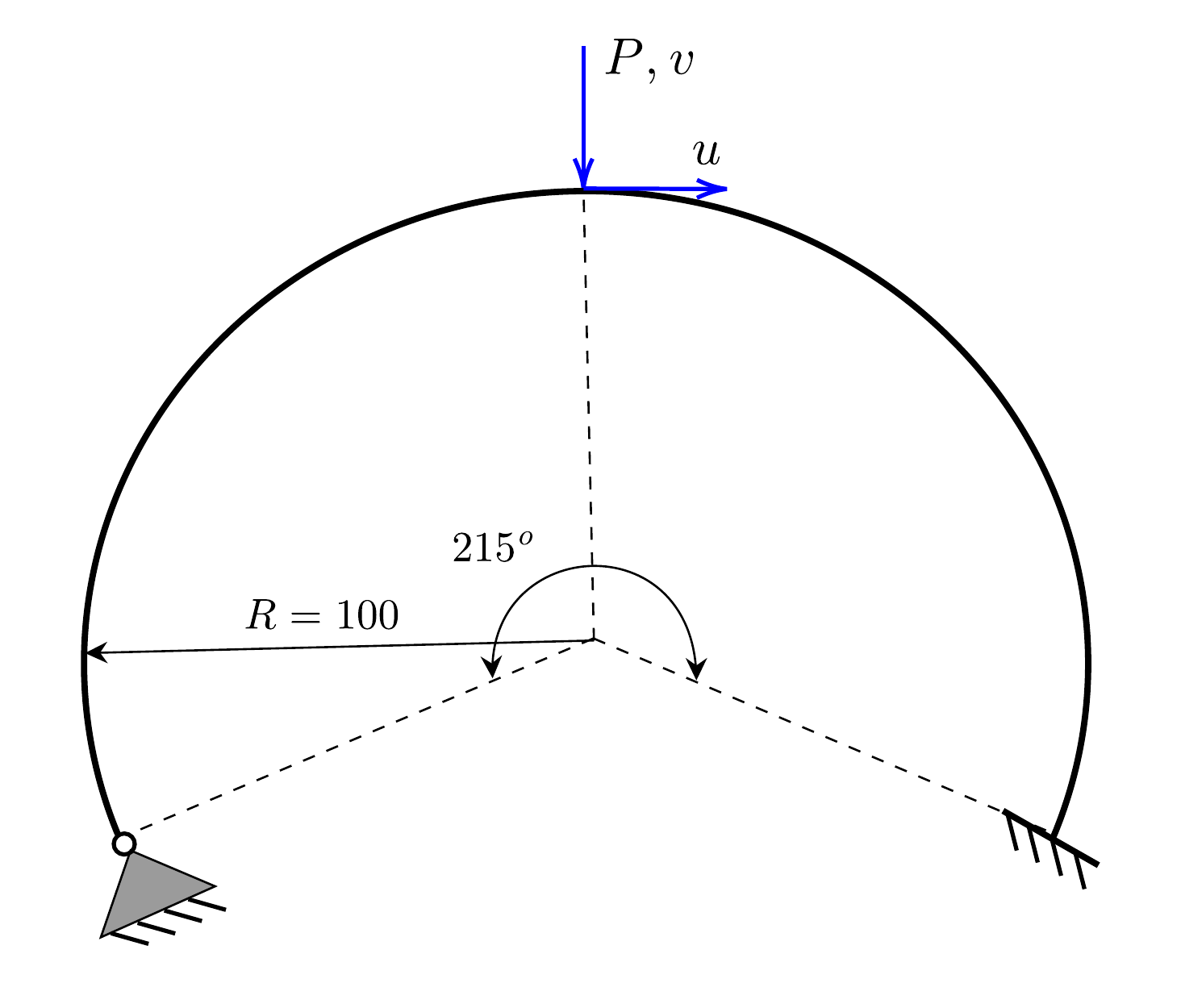} \label{fig-arch215deg-geom}}
\subfloat[]{\includegraphics[clip, scale=0.4]{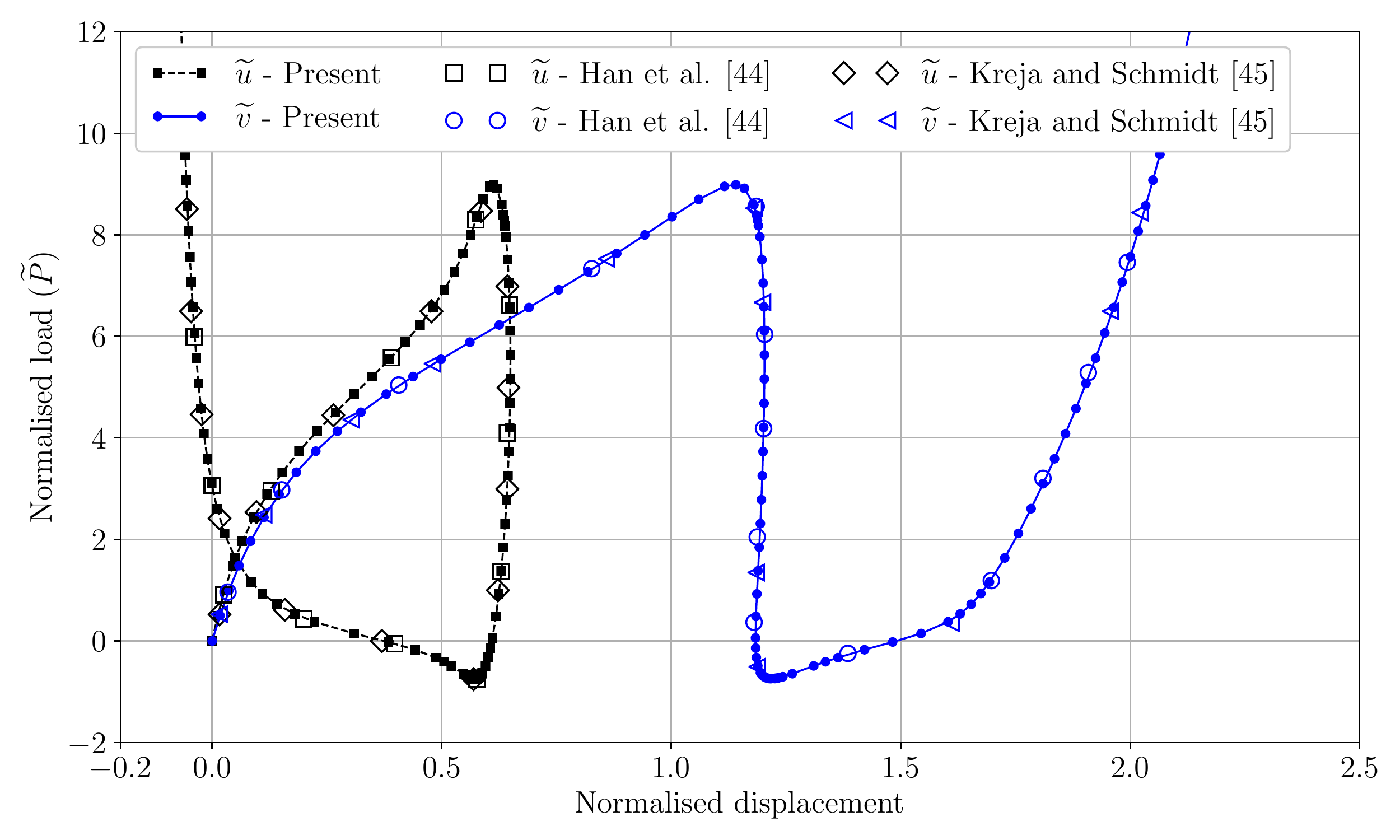} \label{fig-arch215deg-graph}}
\caption{Hinged-clamped 215$^o$ arch: (a) geometry and boundary conditions and (b) the load-displacement curve. The markers for the present work represent the converged load steps.}
\label{fig-arch215deg-geom-graph}
\end{figure}

\begin{figure}[H]
\centering
\subfloat[$\widetilde{P} =  4.508$]{\includegraphics[clip, scale=0.25]{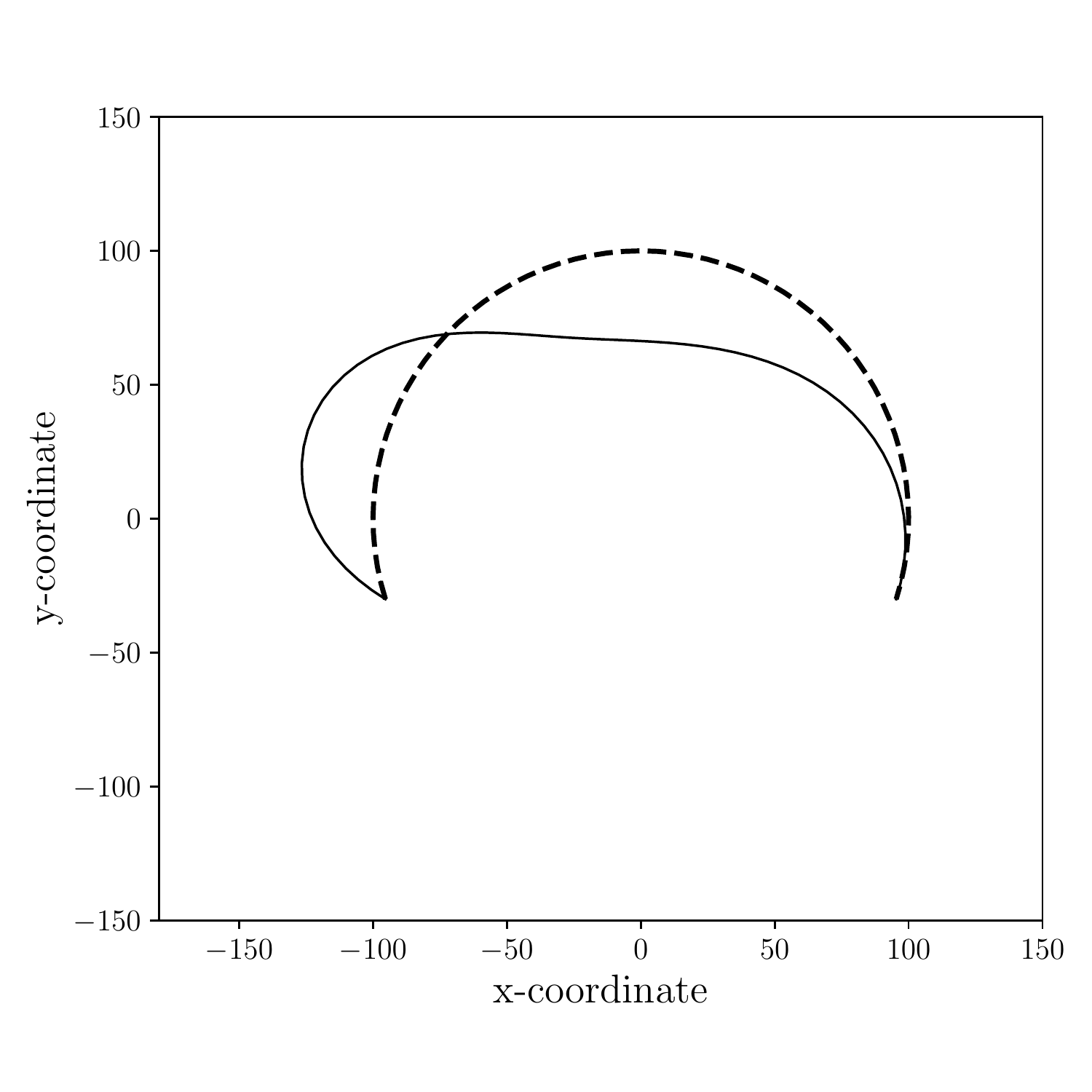}}
\subfloat[$\widetilde{P} =  8.003$]{\includegraphics[clip, scale=0.25]{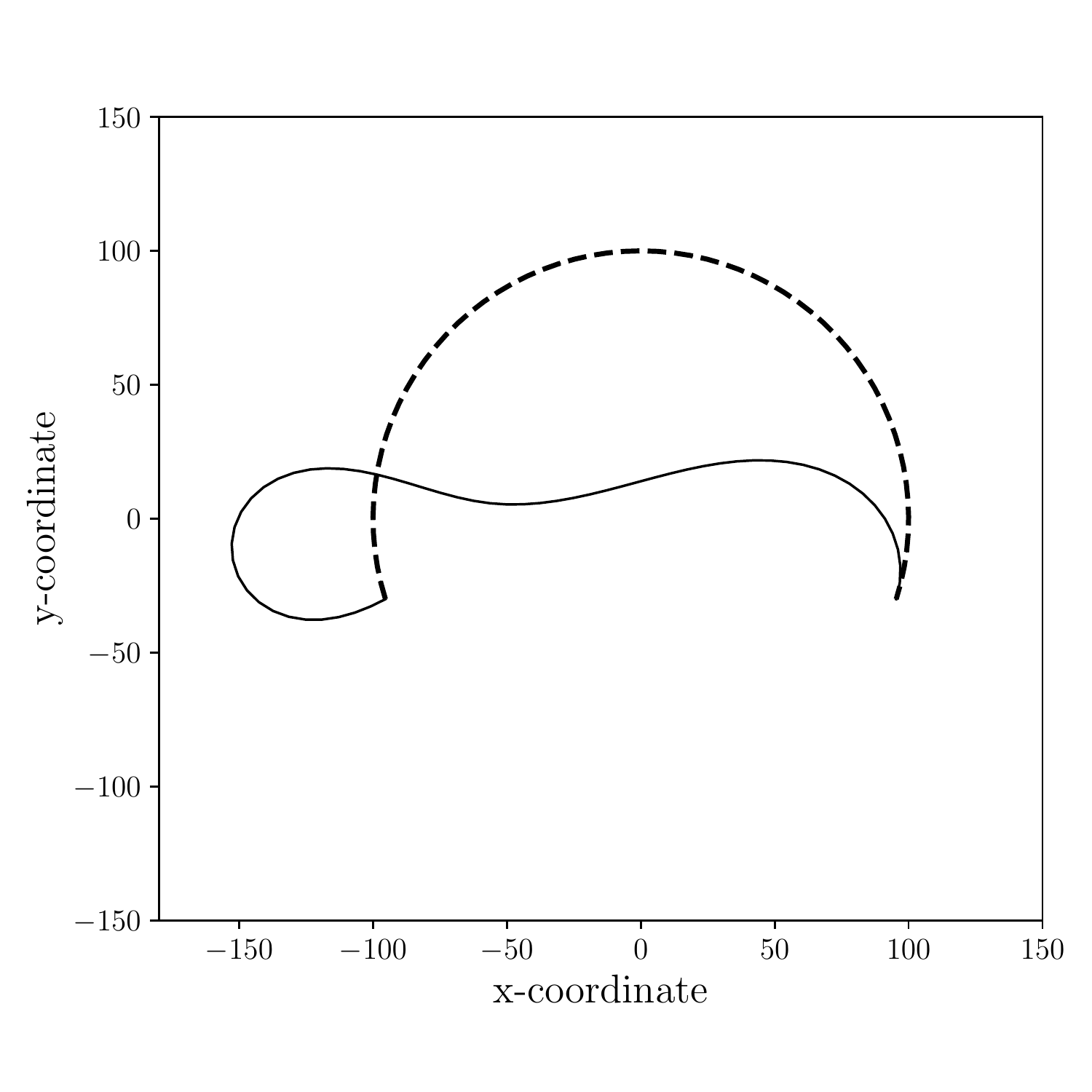}}
\subfloat[$\widetilde{P} =  8.993$]{\includegraphics[clip, scale=0.25]{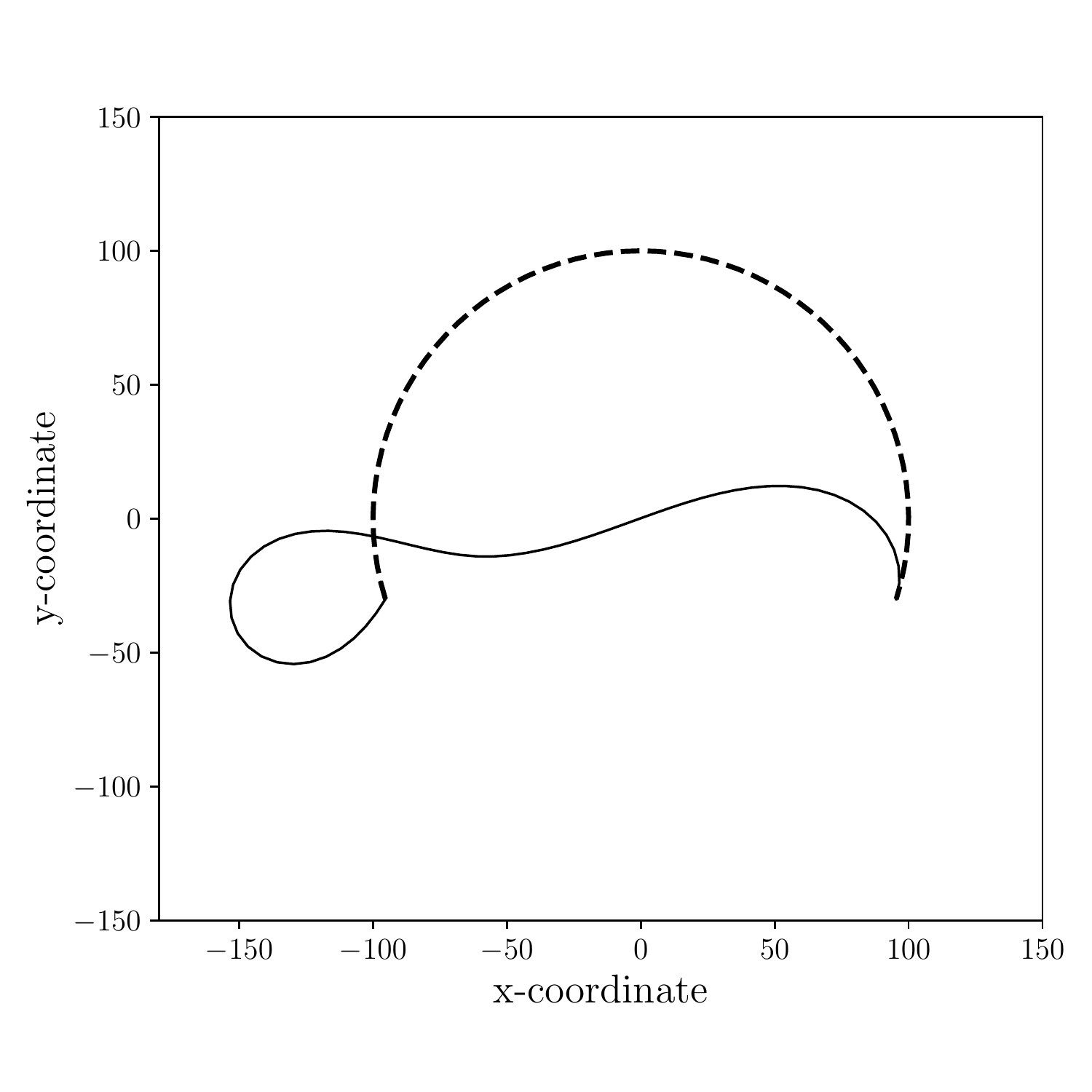}}
\subfloat[$\widetilde{P} =  7.967$]{\includegraphics[clip, scale=0.25]{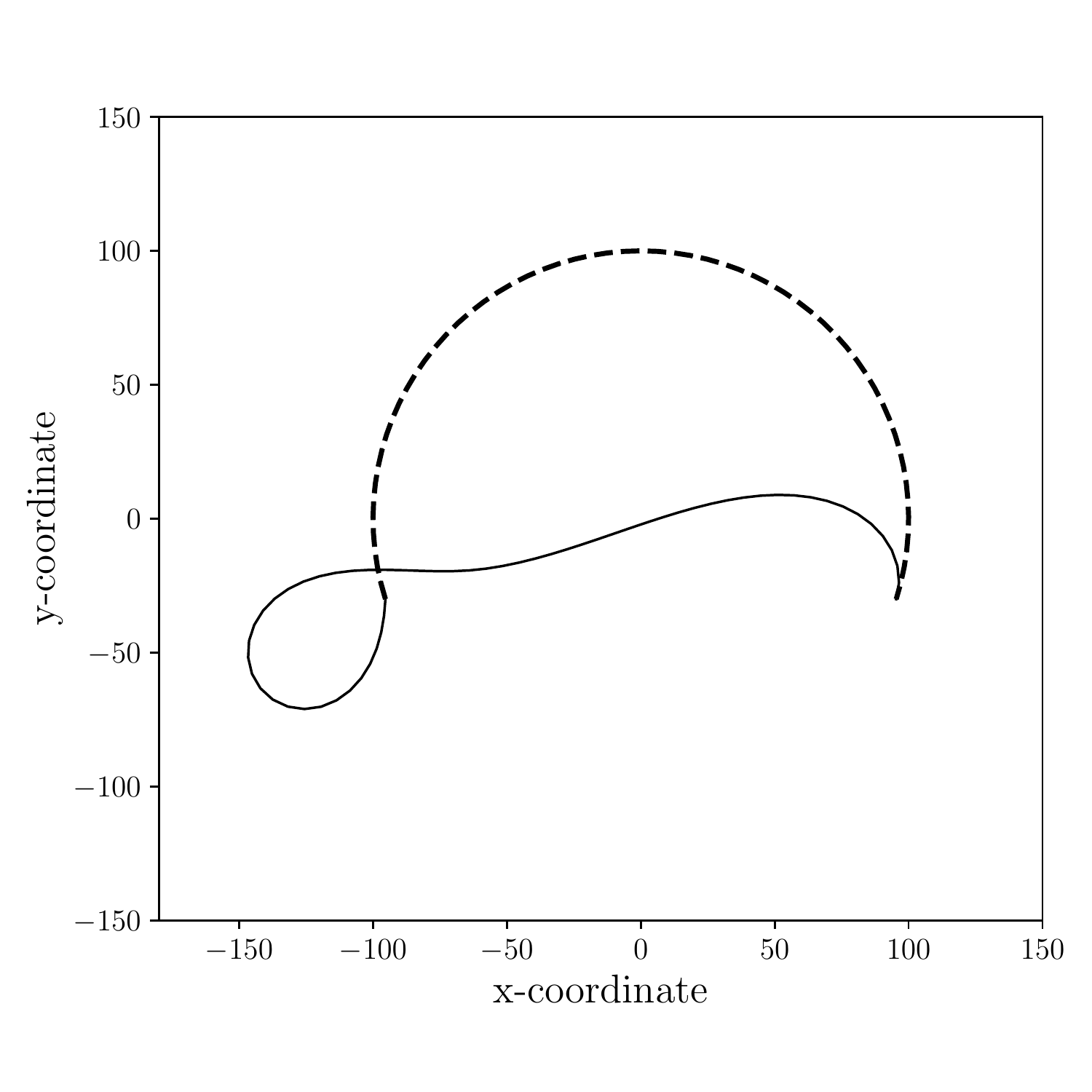}} \\
\subfloat[$\widetilde{P} =  3.258$]{\includegraphics[clip, scale=0.25]{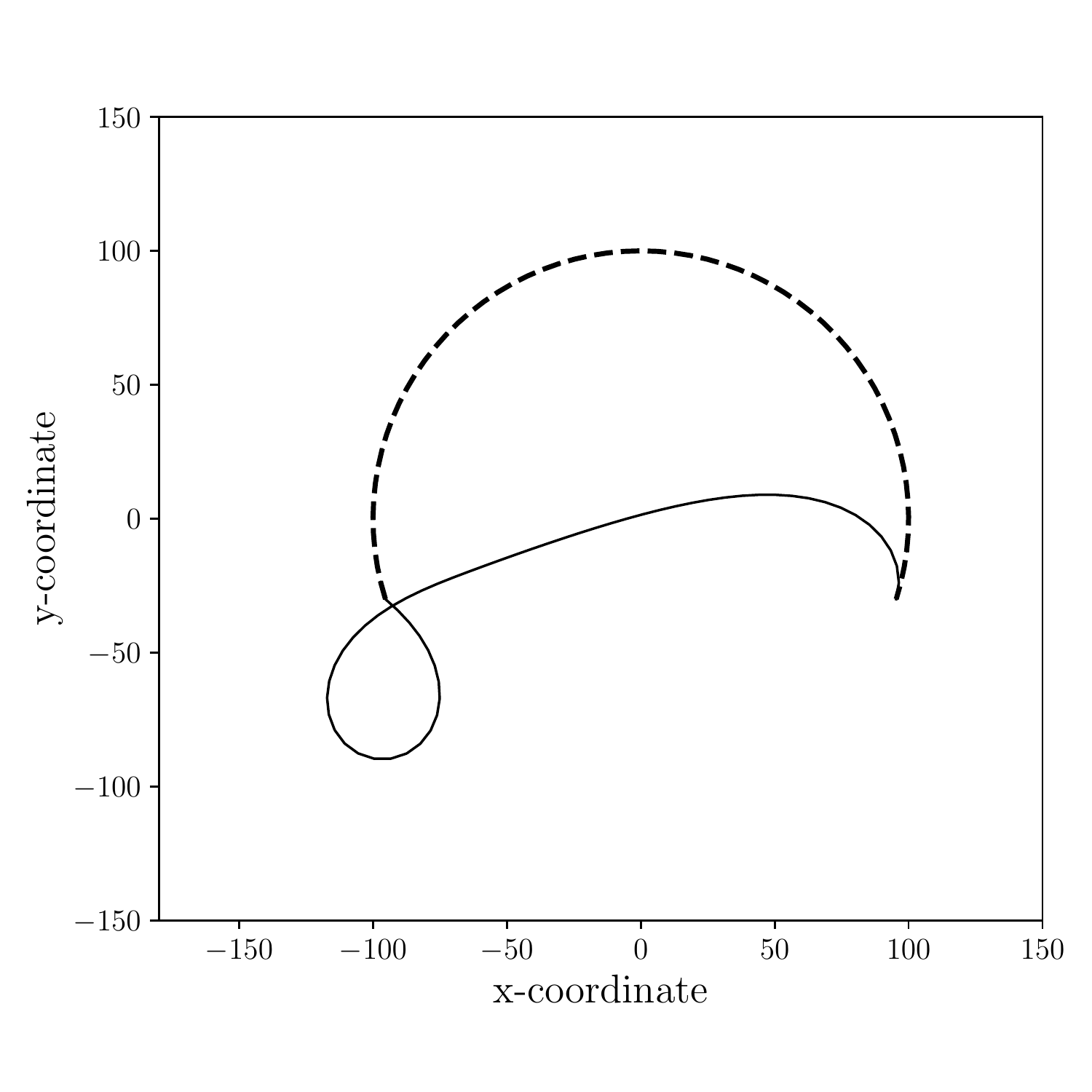}}
\subfloat[$\widetilde{P} = -0.624$]{\includegraphics[clip, scale=0.25]{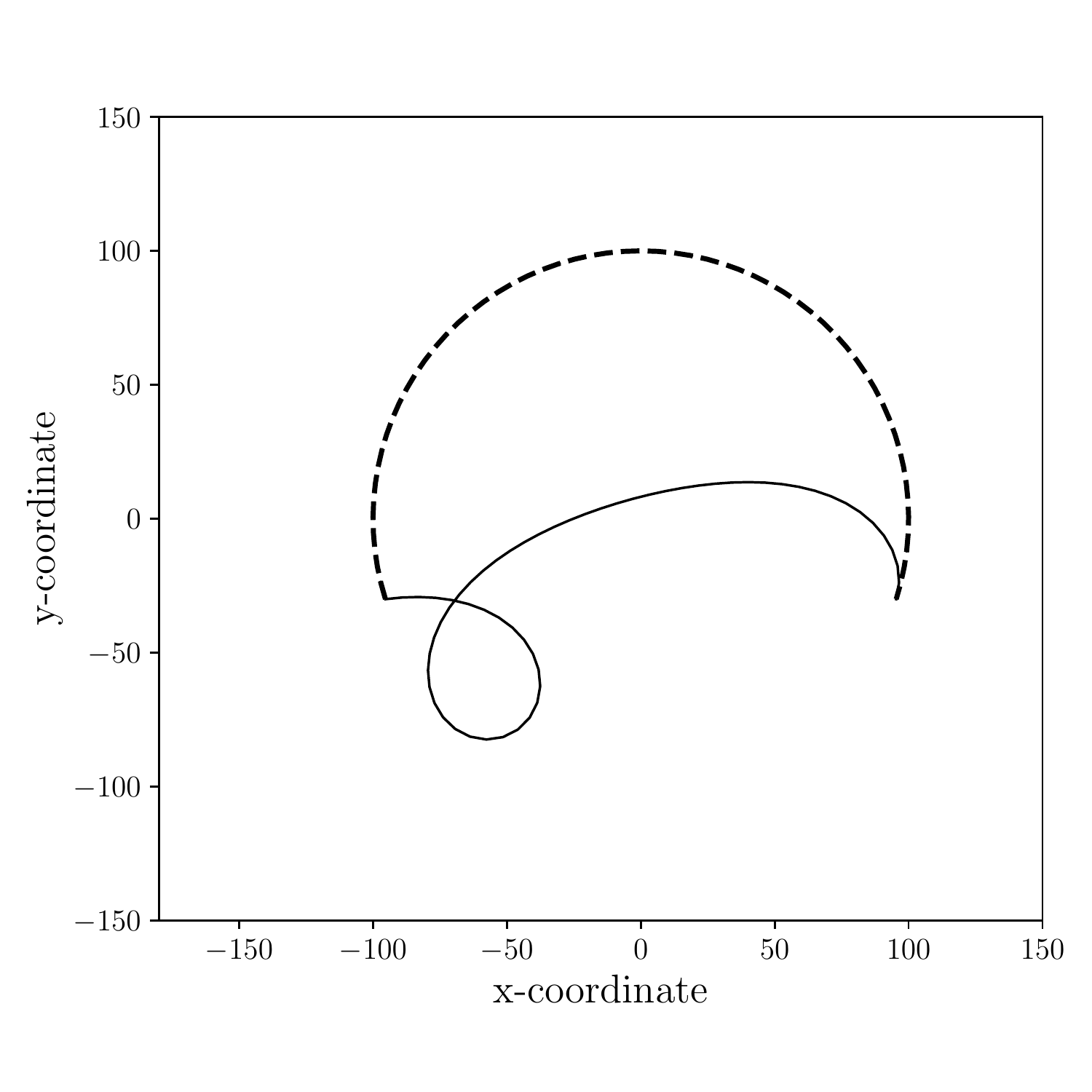}}
\subfloat[$\widetilde{P} =  2.120$]{\includegraphics[clip, scale=0.25]{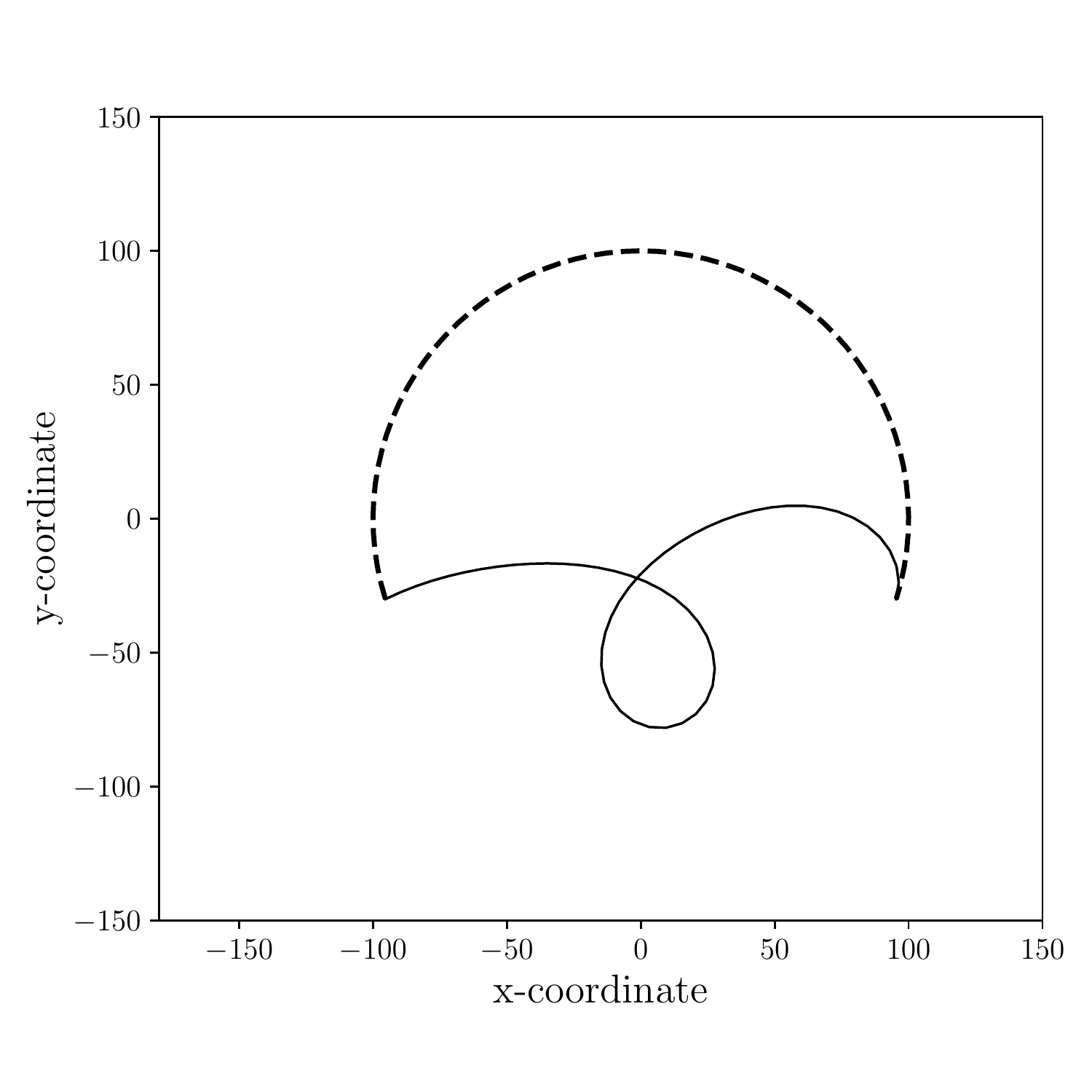}}
\subfloat[$\widetilde{P} = 11.611$]{\includegraphics[clip, scale=0.25]{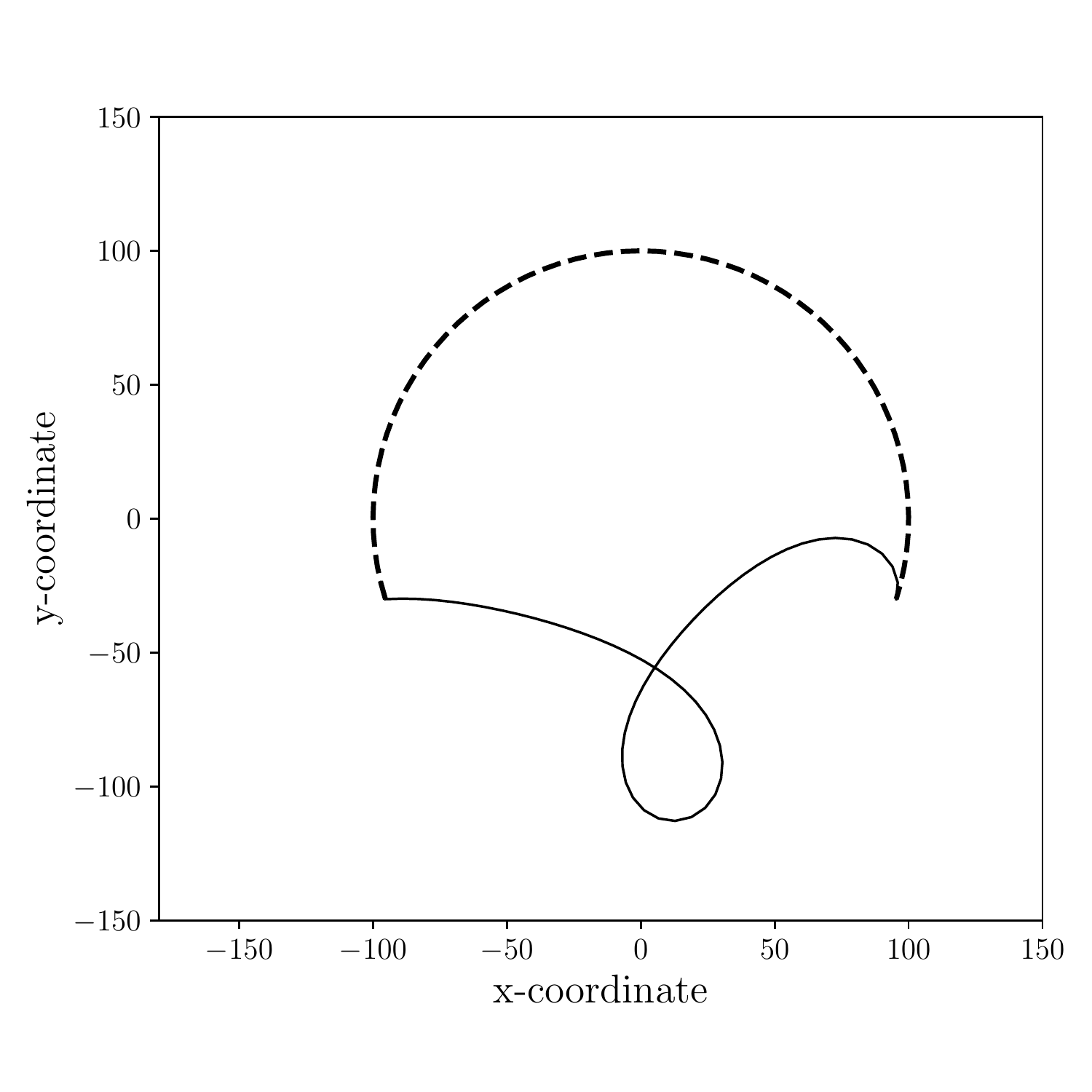}}
\caption{Hinged-clamped 215$^o$ arch: deformed shapes at eight different load steps. \mydashedline \, : original configuration and \mysolidline \, : deformed configuration.}
\label{fig-arch215deg-defshapes}
\end{figure}

This is another widely-used benchmark example for demonstrating the instabilities in structural mechanics problems. This example consists of a 215$^o$ degree circular arch of radius $R=100$ cm
which is hinged on its one end and clamped on the other end, as depicted in Fig. \ref{fig-arch215deg-geom}. A point load $P$ is applied at the crown. The parameters are : $A=2.29$, $I=1.0$, $E=1.0 \times 10^6$, $\nu=0$, and $\kappa=1.0$, in consistent units. The problem is discretised with 60 nonlinear beam-column elements. The initial load increment used for this problem is $\Delta \lambda=50$. The load-deflection curve in terms of normalised load $\left( \widetilde{P} = \frac{P \, R^2}{E \, I} \right)$ and normalised displacement $\left( \widetilde{u}=\frac{u}{R}, \widetilde{v}=\frac{v}{R} \right)$ is presented in Fig. \ref{fig-arch215deg-graph}. The load-displacement curve for this problem consists of an almost vertical jump in $v$, and this very difficult equilibrium path is captured quite successfully using the proposed technique. The result obtained with the proposed arc-length implementation is in good agreement with the reference solutions taken from Han et al. \cite{HanIJNLM2008} and Kreja and Schmidt. \cite{KrejaIJNLM2006}. The buckling load $8.993 \, EI/R^2$ obtained in the present work has less than 1\% error relative to the analytical buckling value of $8.97 \, EI/R^2$. Deformed shapes of the arch at eight different load instants are shown in Fig. \ref{fig-arch215deg-defshapes}.

\newpage
\subsection{Semi-circular hinged arch}
\begin{figure}[H]
\centering
\includegraphics[clip, scale=0.6]{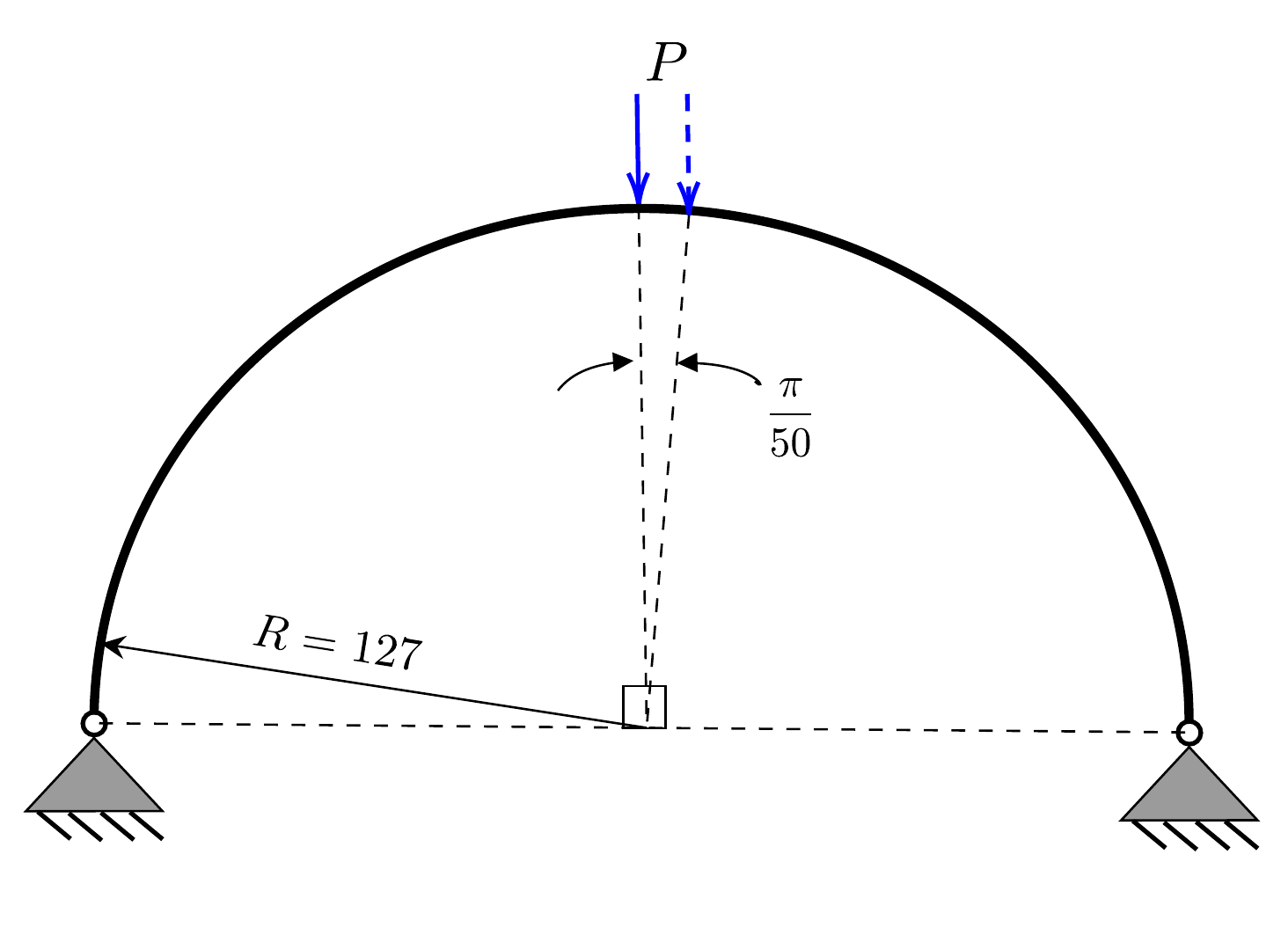}
\caption{Semi-circular arch: geometry and boundary conditions. Length units are centimeters.}
\label{fig-archsemicircle-geom}
\end{figure}

This example consists of a semi-circular arch of radius $R=127$ cm that is hinged on its two ends, as shown in Fig. \ref{fig-archsemicircle-geom}. The parameters are : $A=64.52$ cm$^2$, $I=41.62$ cm$^4$, $E=0.1378$ N/cm$^2$, $\nu=0.5$, and $\kappa=1.0$. The problem is discretised with 50 nonlinear beam-column elements. Similar to the studies conducted in Yang and Shieh \cite{YangAIAA1990}, the analysis is performed for two different loading conditions, as shown in Fig. \ref{fig-archsemicircle-geom}: (i) a point load at the crown of the arch which yields symmetrical deformation of the arch and (ii) a point load at an off-set angle of $\pi/50$ which yields an asymmetrical deformation of the arch.

\begin{figure}[H]
\centering
\subfloat[Symmetric loading]{\includegraphics[clip, scale=0.4]{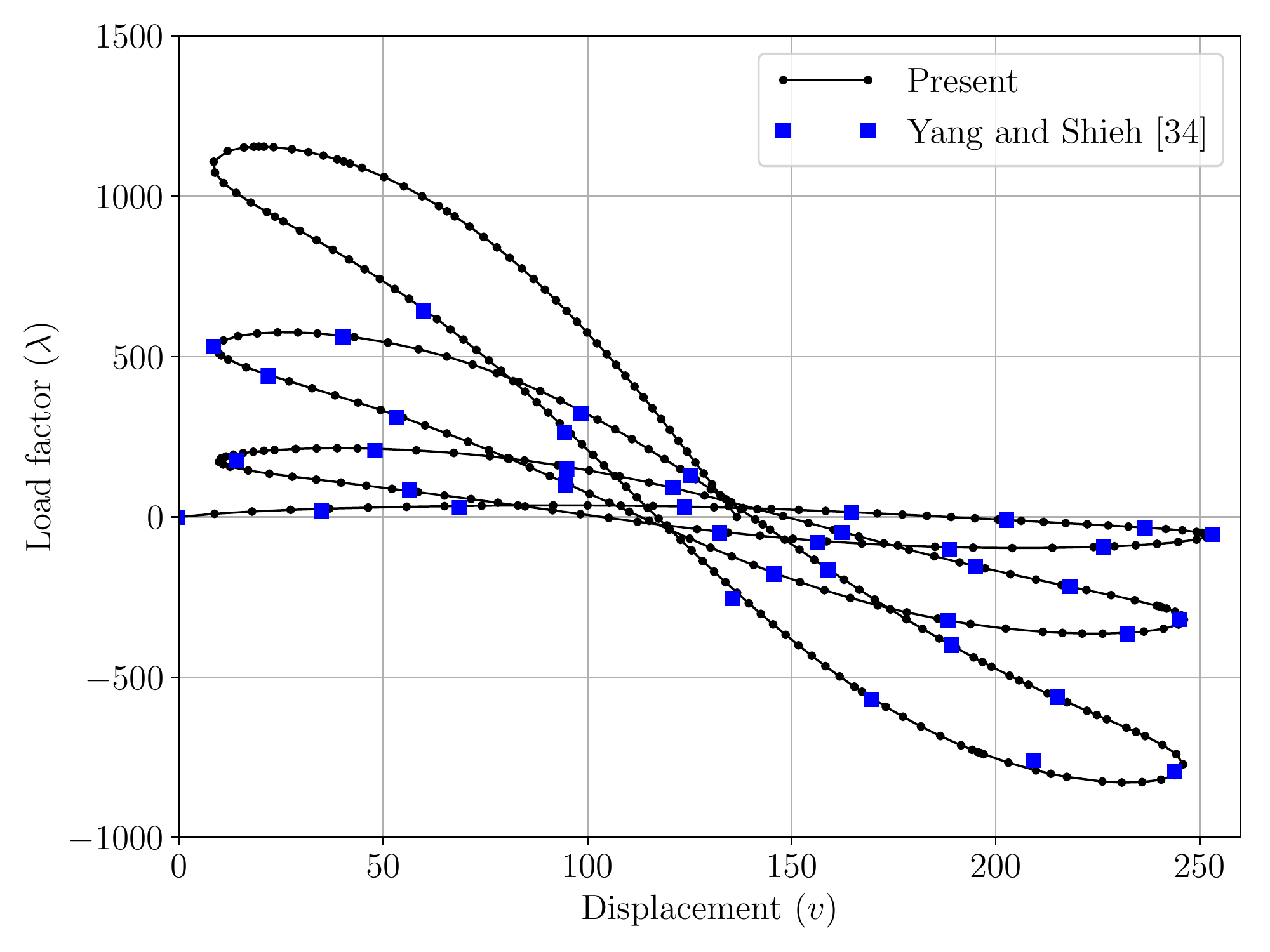} \label{fig-archsemicircle-graph-sym}}
\subfloat[Asymmetric loading]{\includegraphics[clip, scale=0.4]{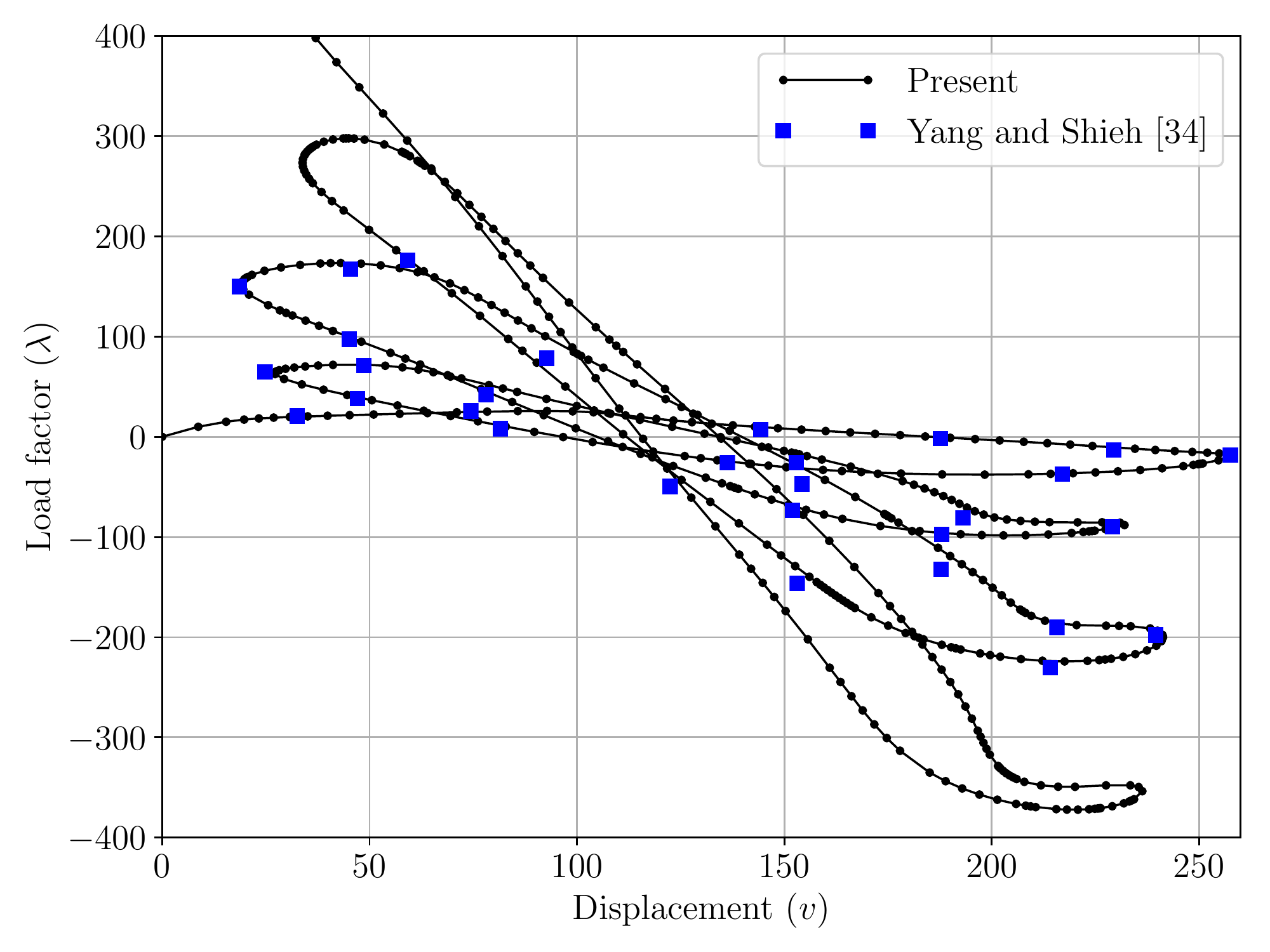} \label{fig-archsemicircle-graph-asym}}
\caption{Semi-circular arch: load-displacement curves for symmetric and asymmetric loading.}
\label{fig-archsemicircle-graphs}
\end{figure}

\begin{figure}[H]
\centering
\subfloat[]{\includegraphics[clip, scale=0.18]{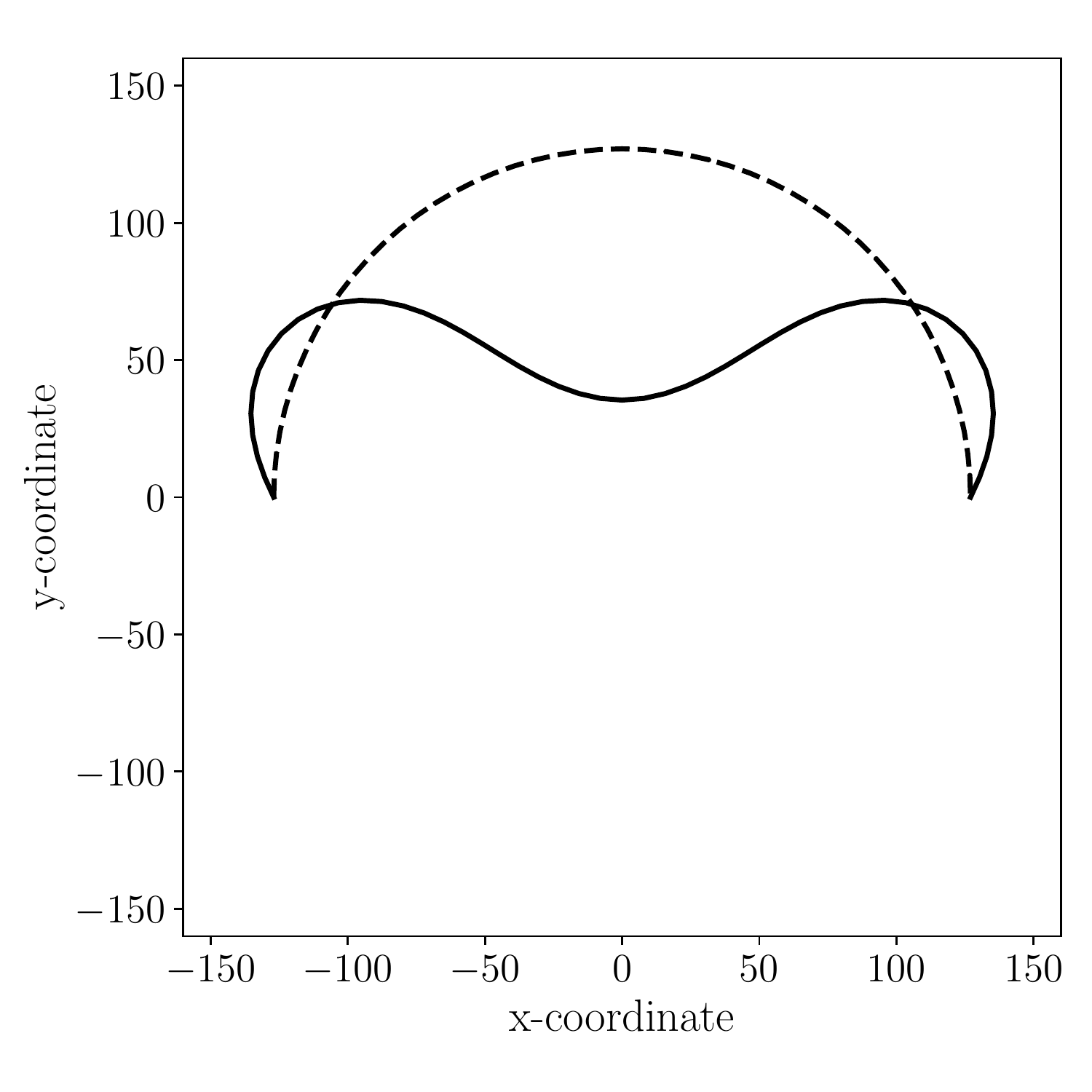}}
\subfloat[]{\includegraphics[clip, scale=0.18]{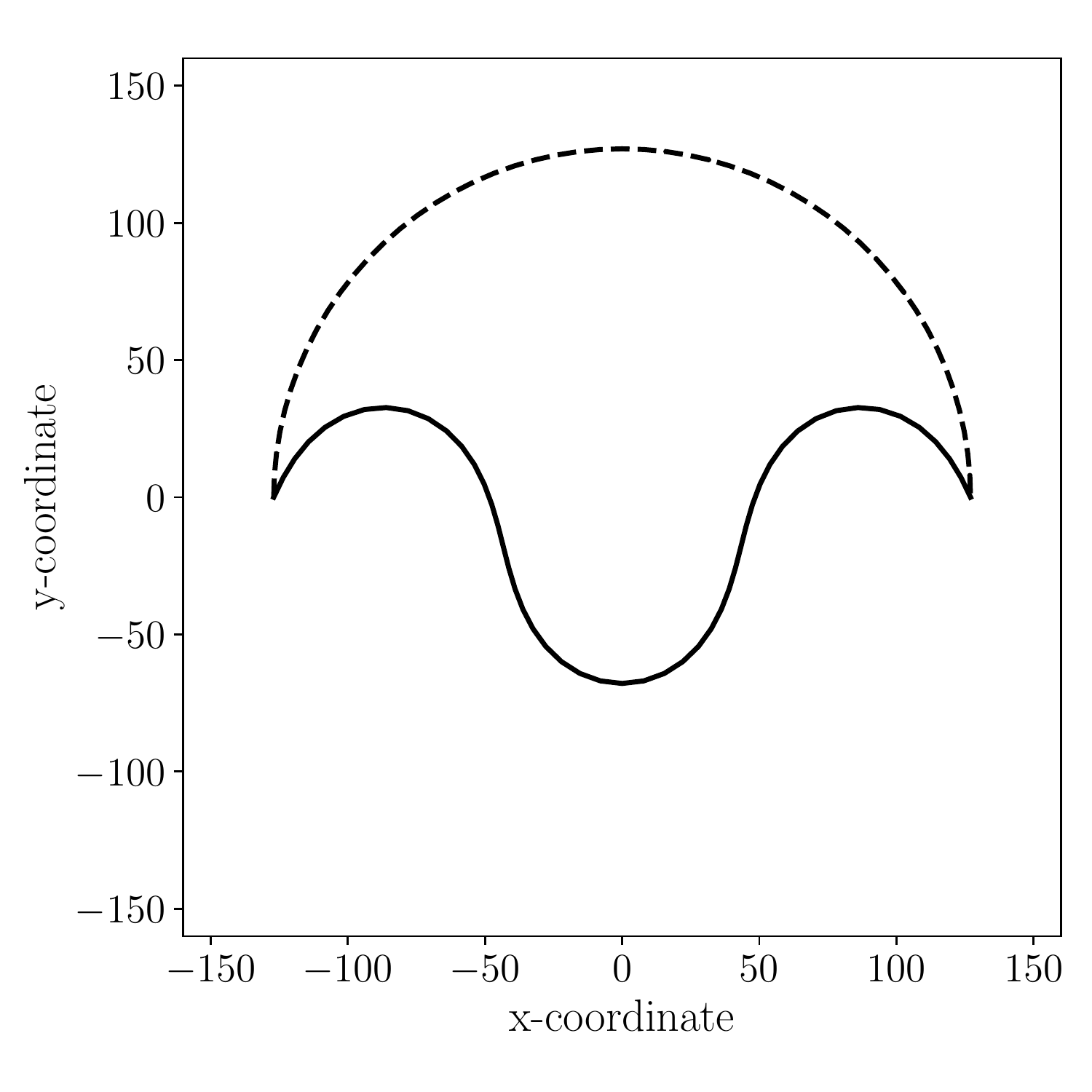}}
\subfloat[]{\includegraphics[clip, scale=0.18]{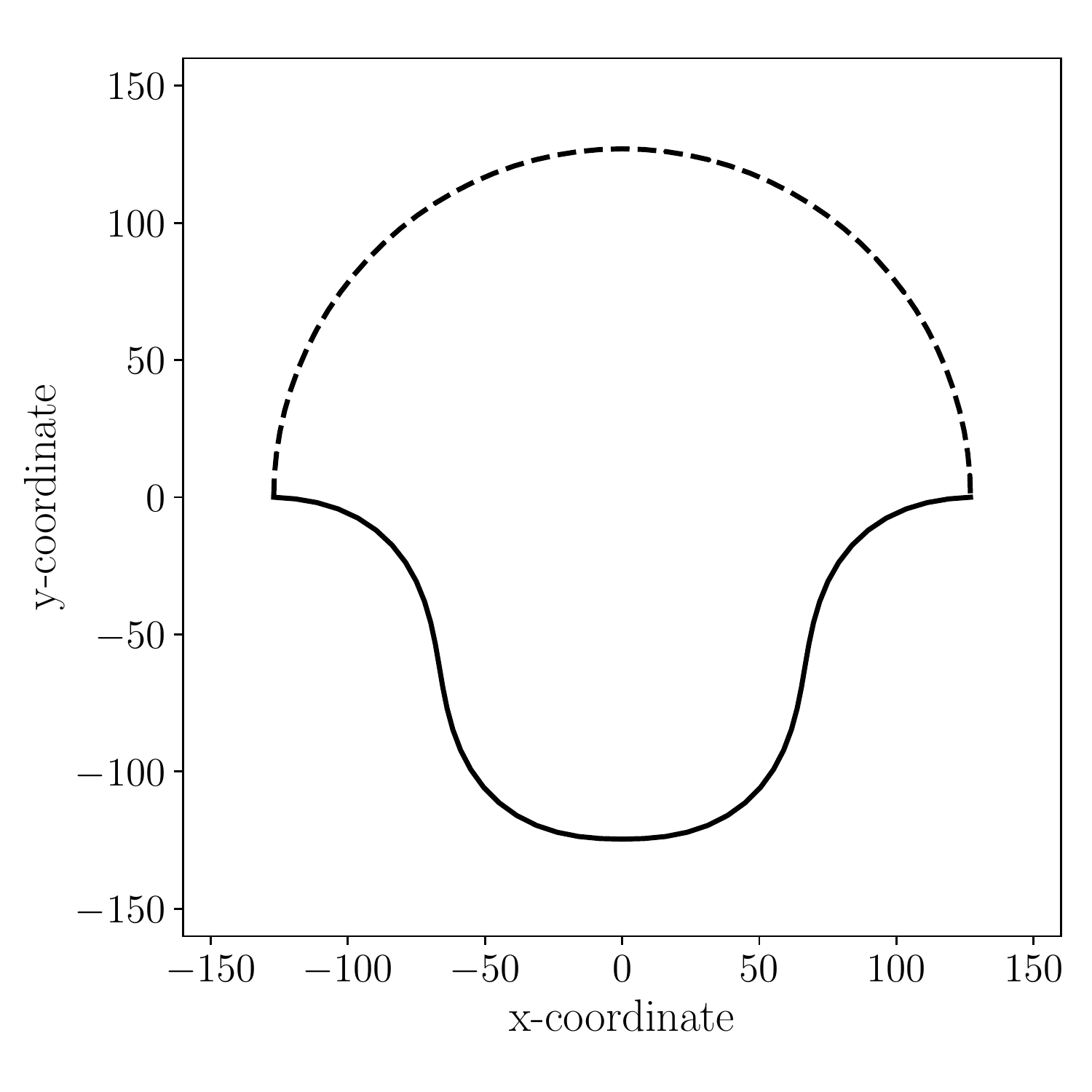}}
\subfloat[]{\includegraphics[clip, scale=0.18]{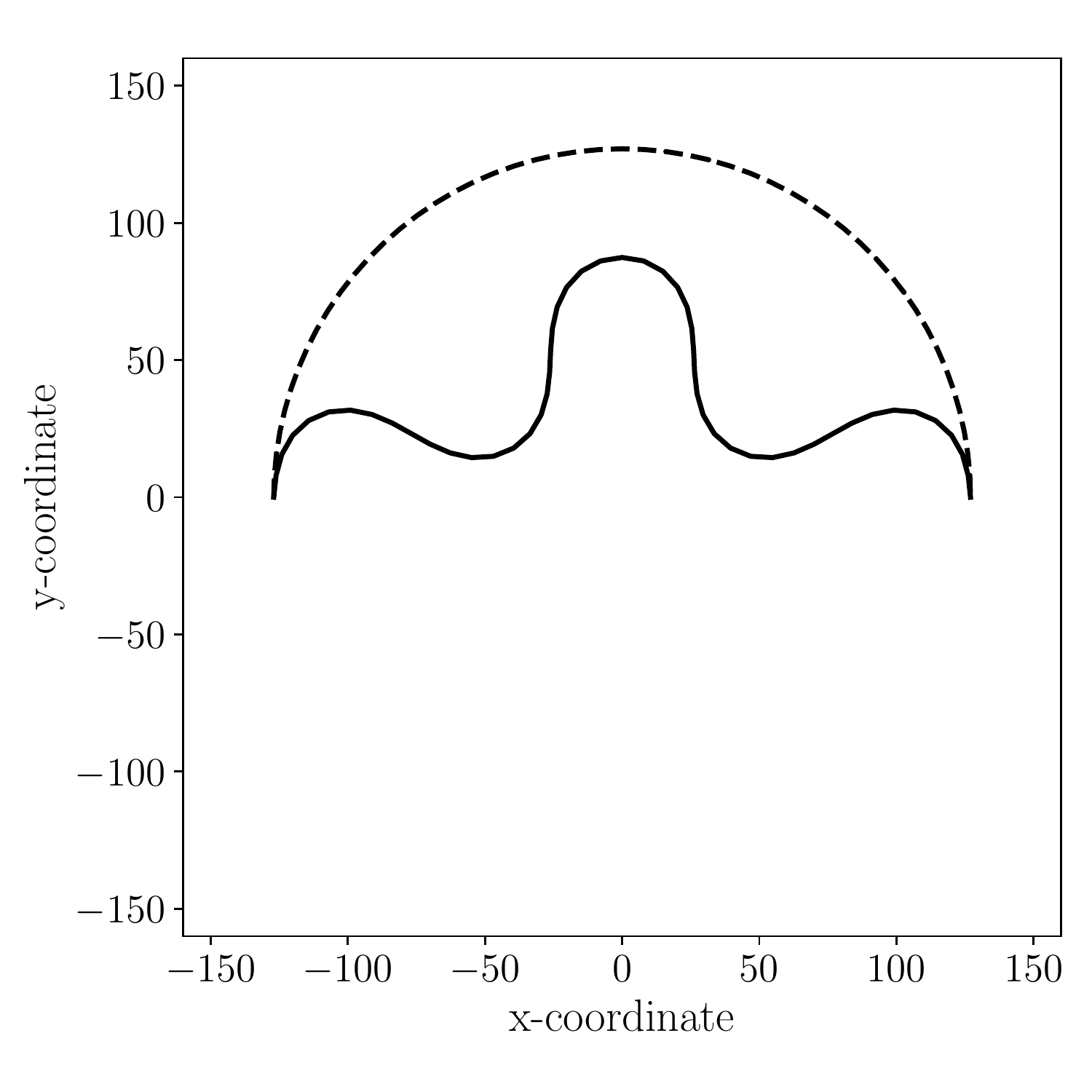}}
\subfloat[]{\includegraphics[clip, scale=0.18]{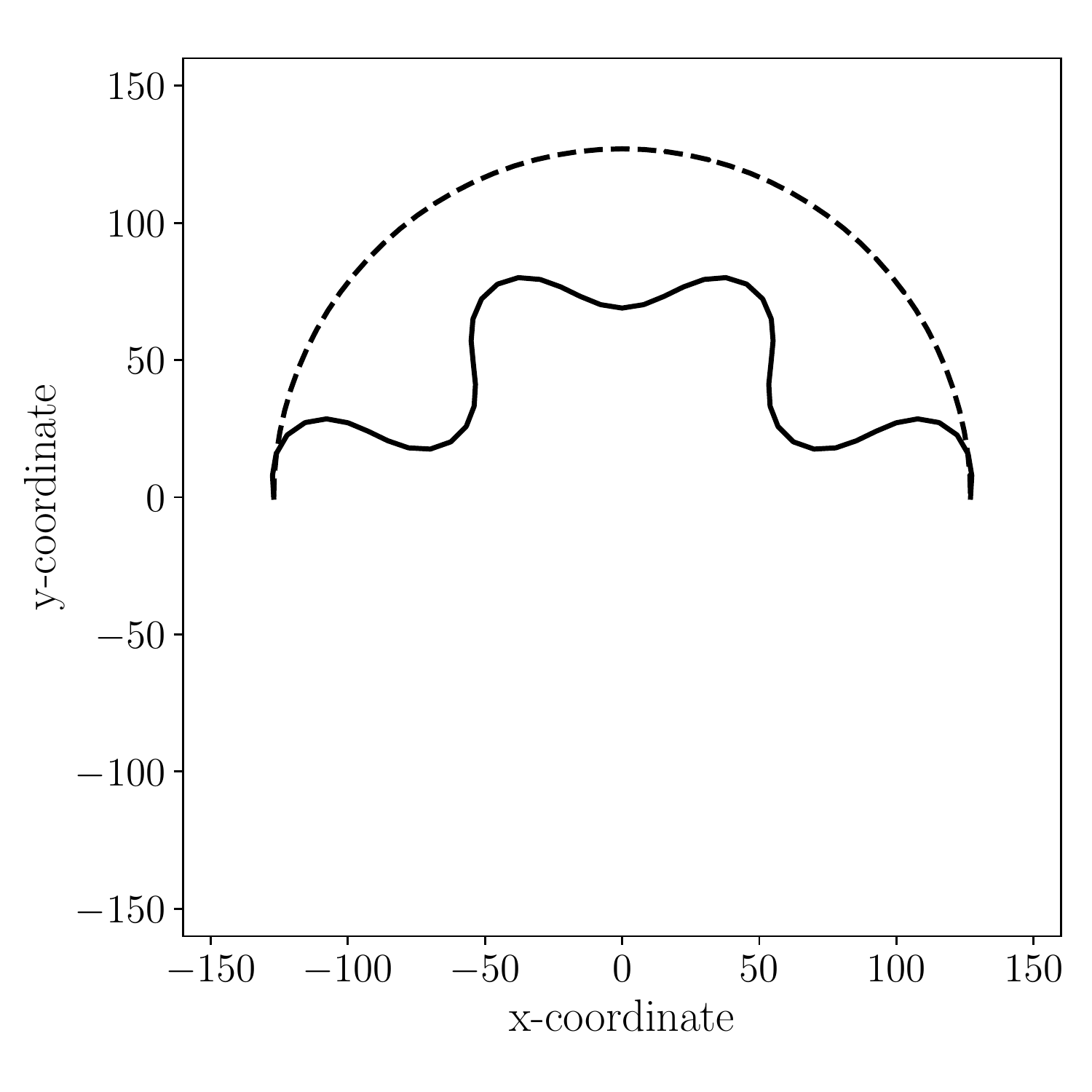}} \\
\subfloat[]{\includegraphics[clip, scale=0.18]{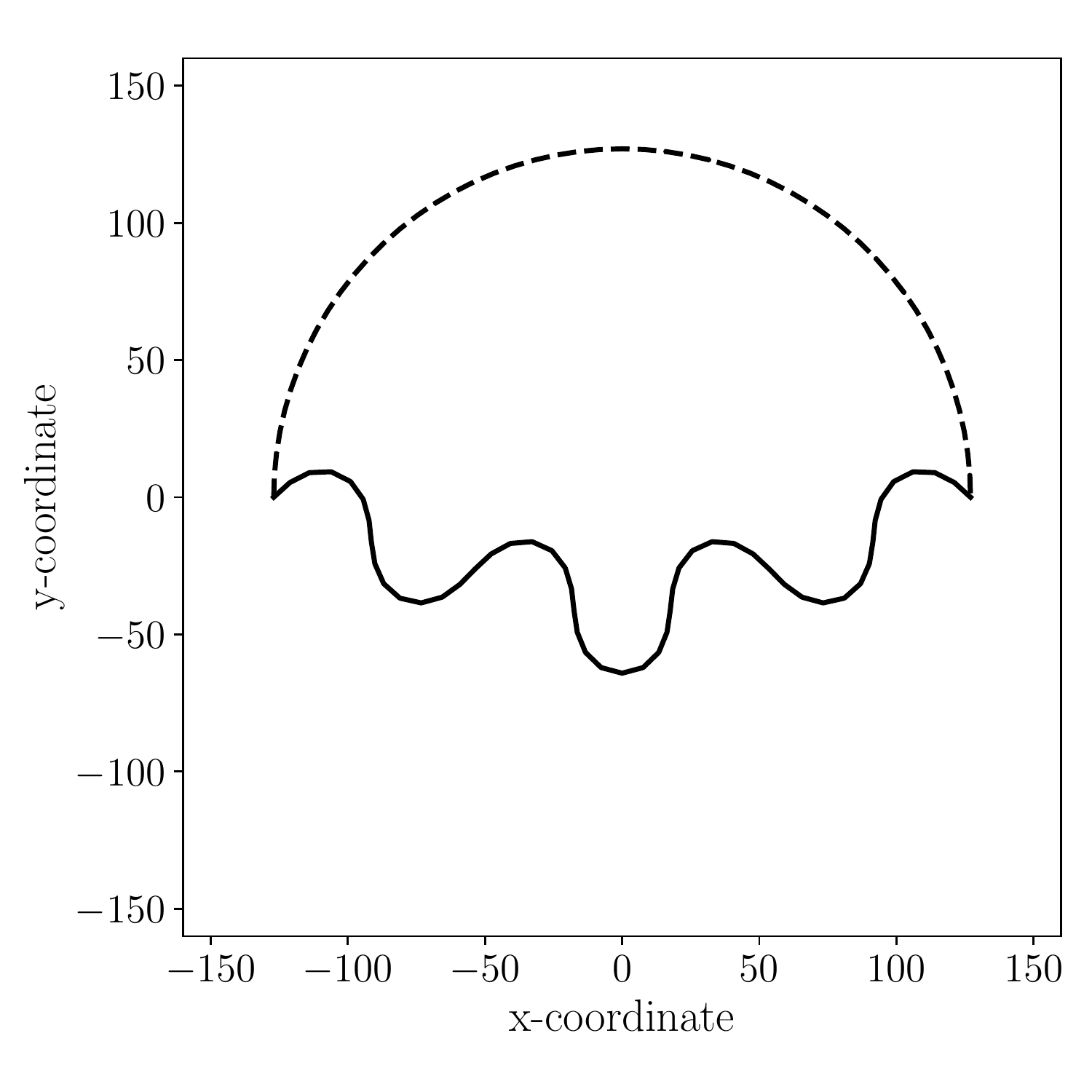}}
\subfloat[]{\includegraphics[clip, scale=0.18]{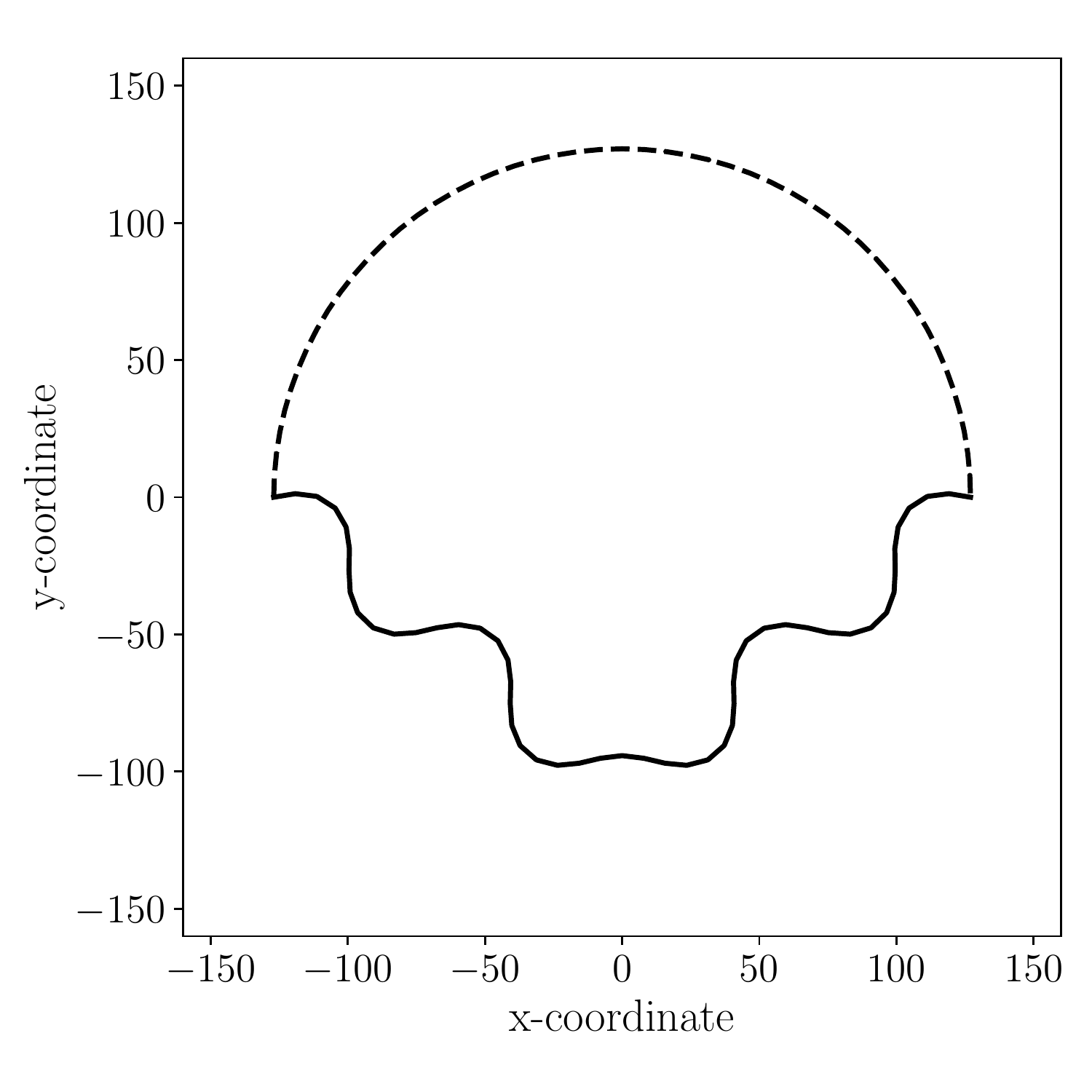}}
\subfloat[]{\includegraphics[clip, scale=0.18]{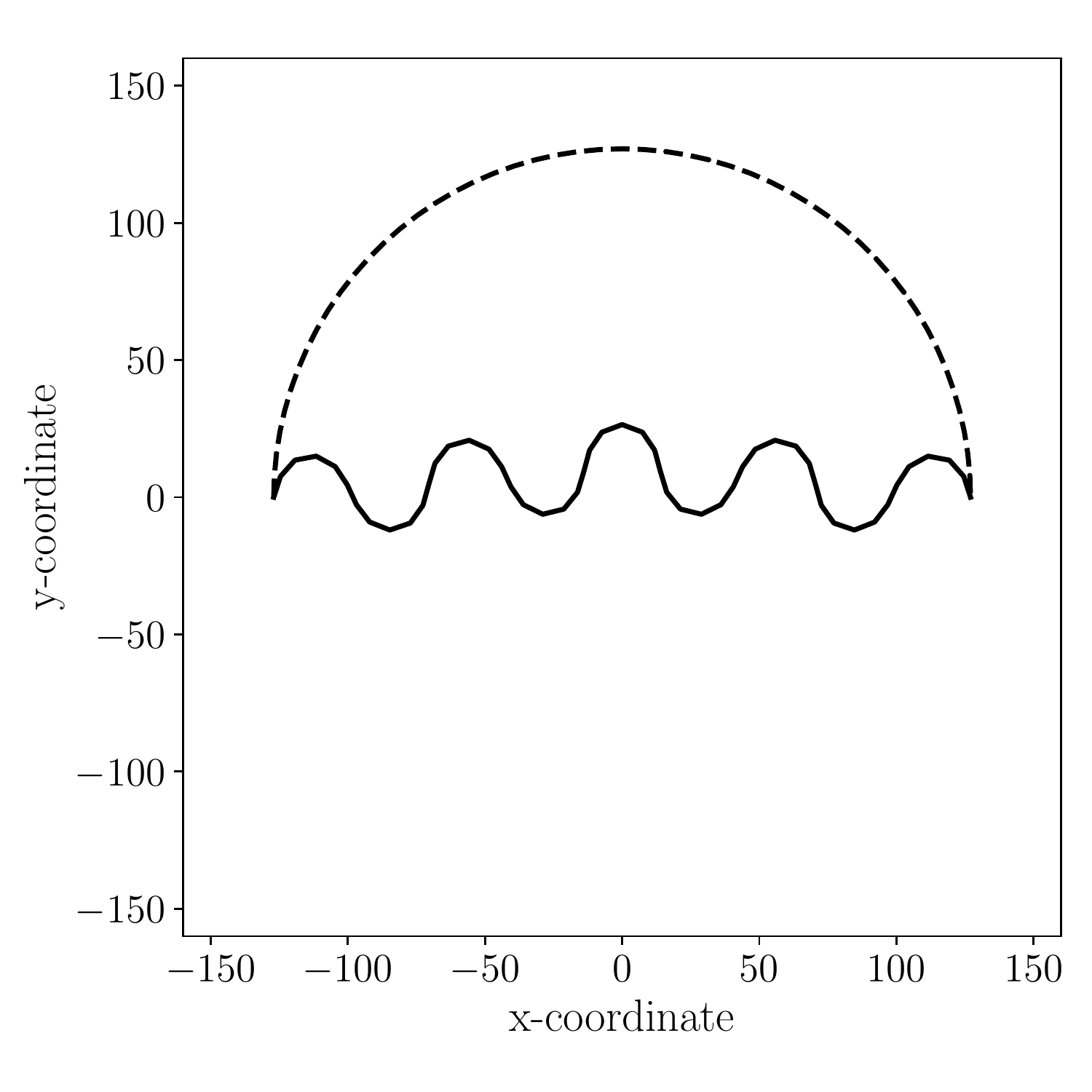}}
\subfloat[]{\includegraphics[clip, scale=0.18]{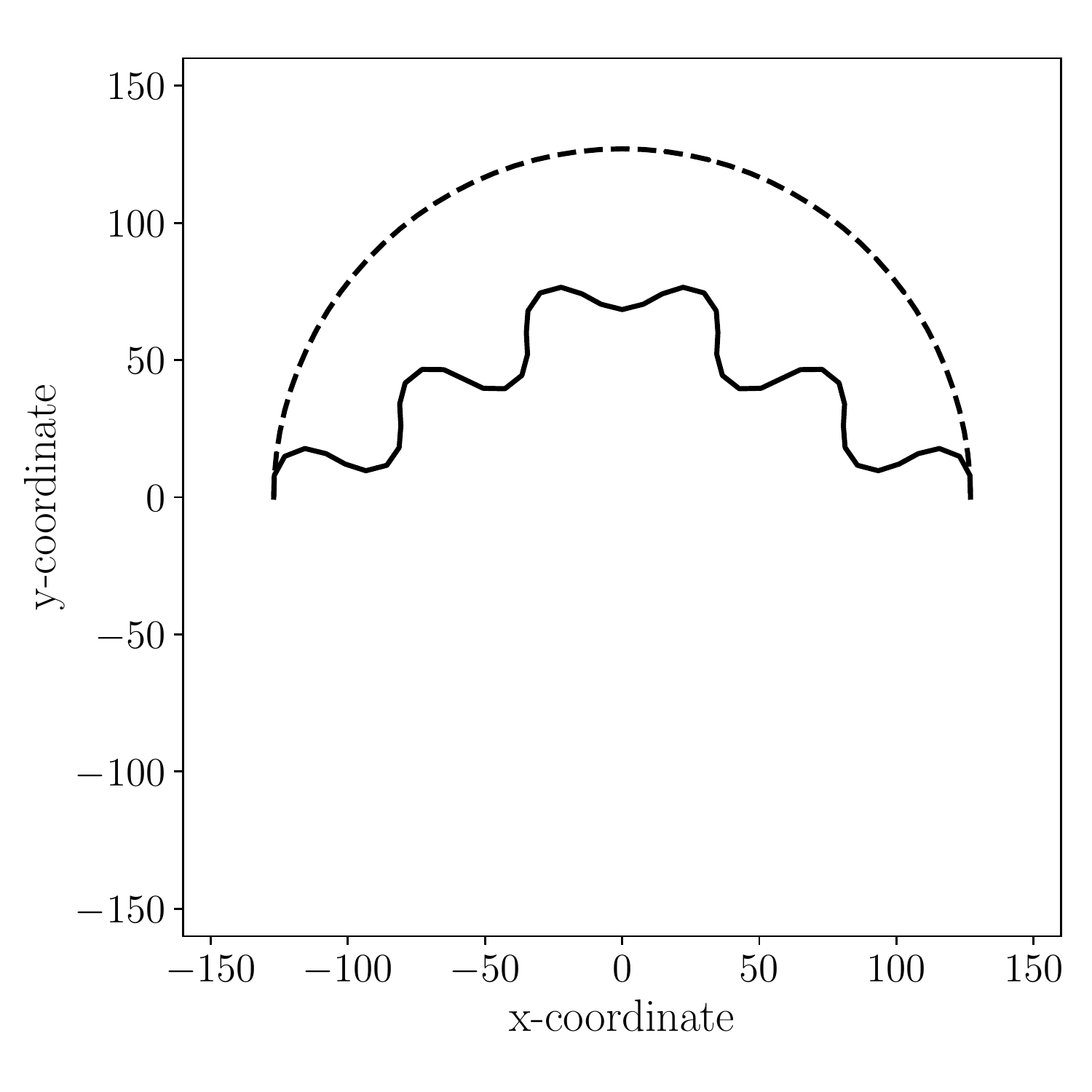}}
\subfloat[]{\includegraphics[clip, scale=0.18]{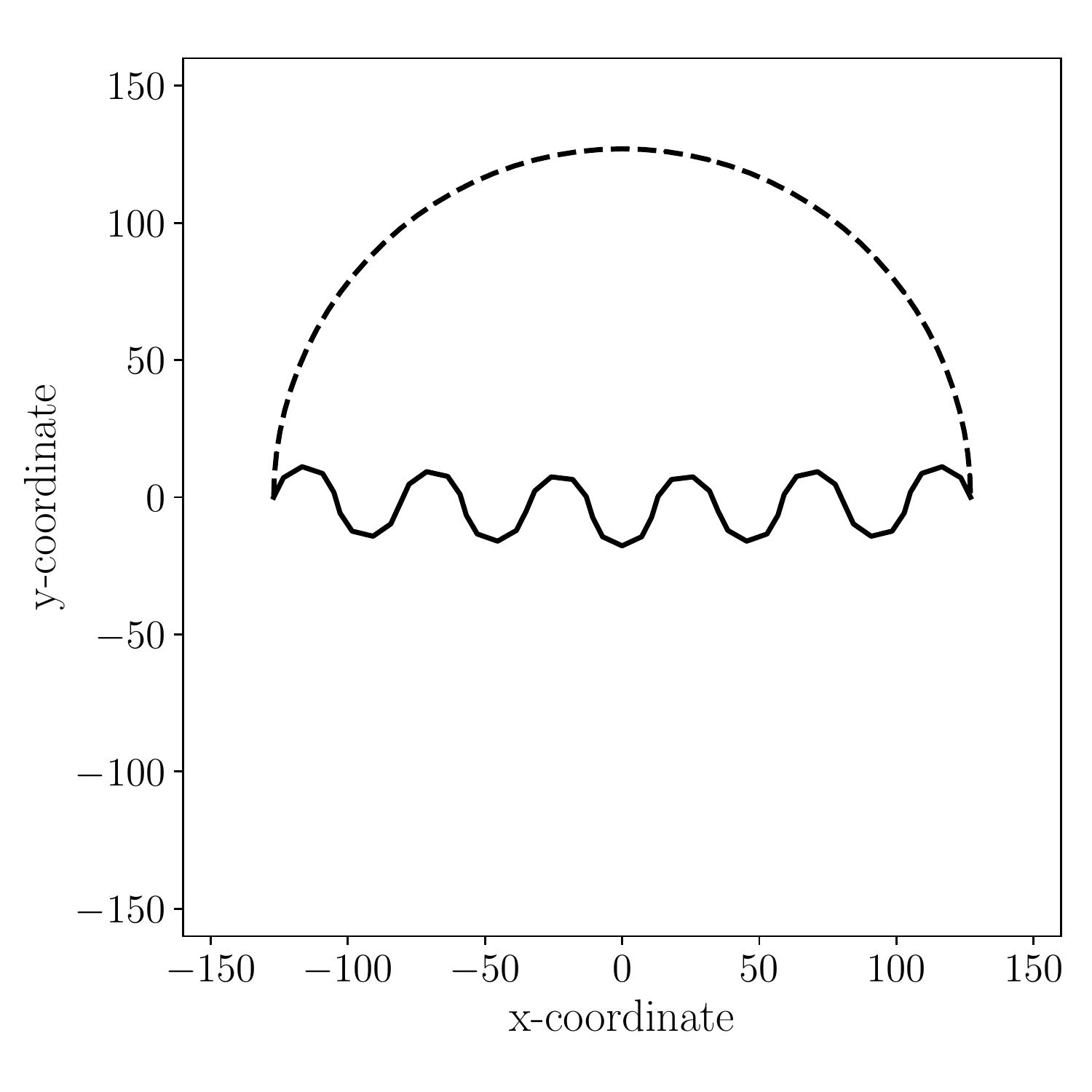}} \\
\subfloat[]{\includegraphics[clip, scale=0.18]{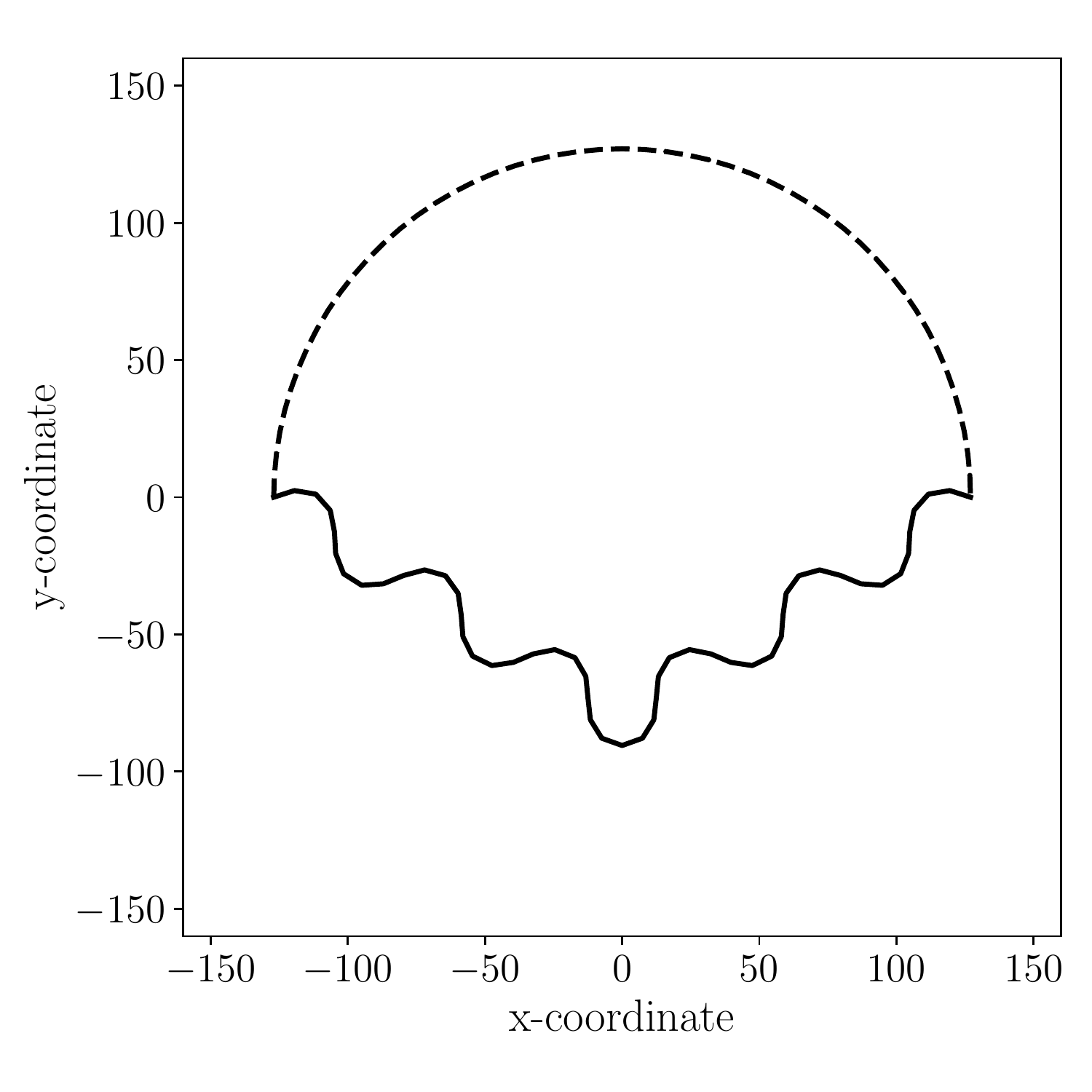}}
\subfloat[]{\includegraphics[clip, scale=0.18]{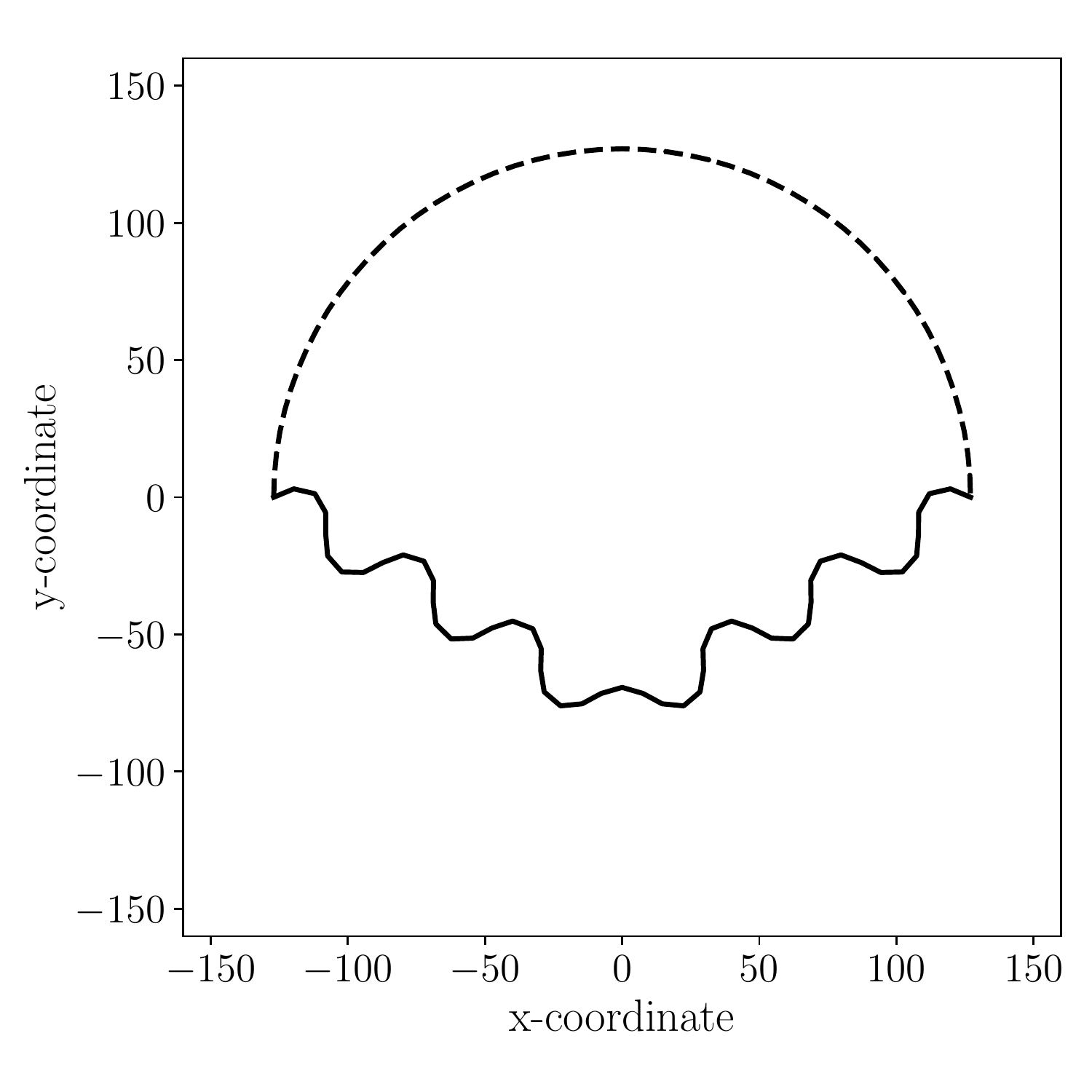}}
\subfloat[]{\includegraphics[clip, scale=0.18]{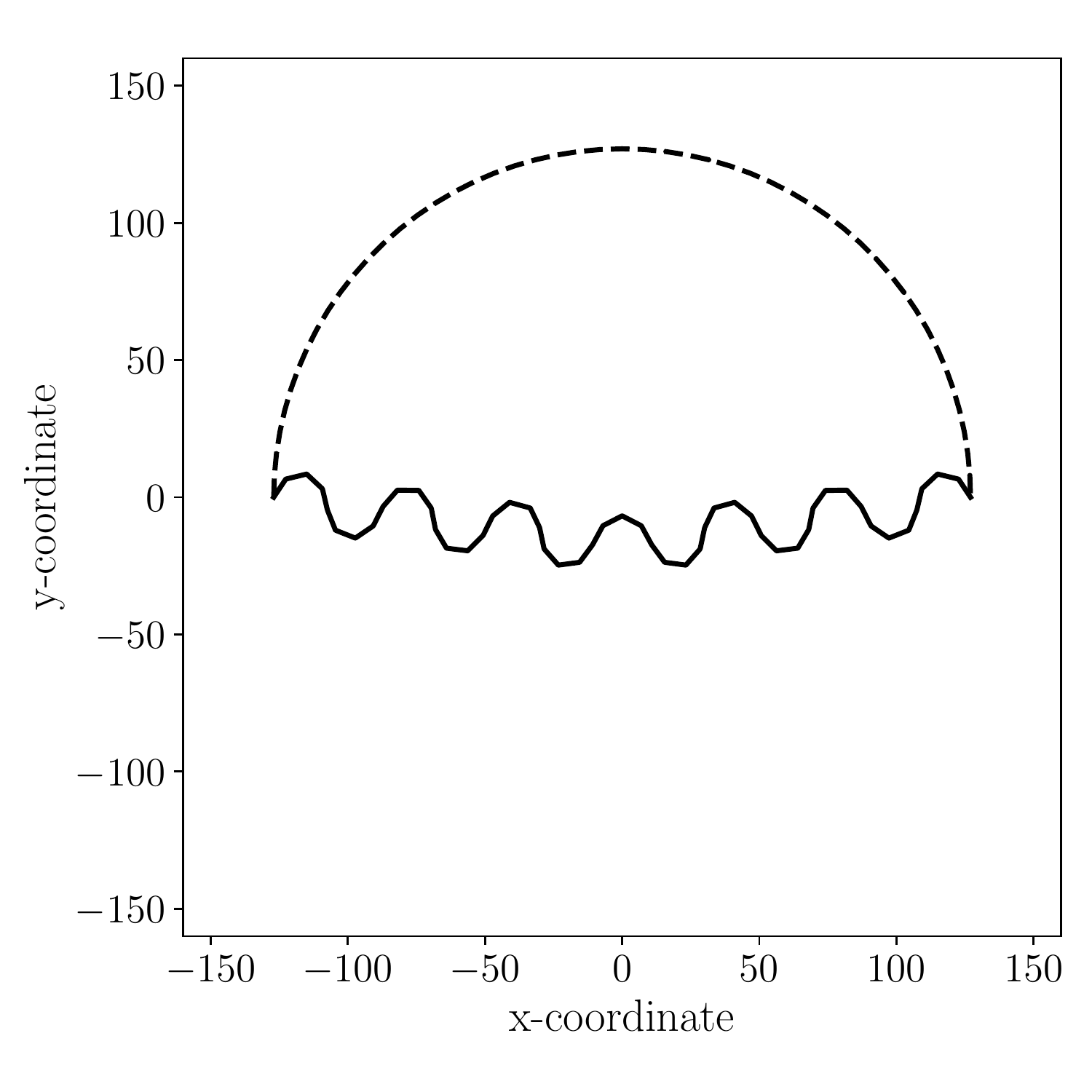}}
\subfloat[]{\includegraphics[clip, scale=0.18]{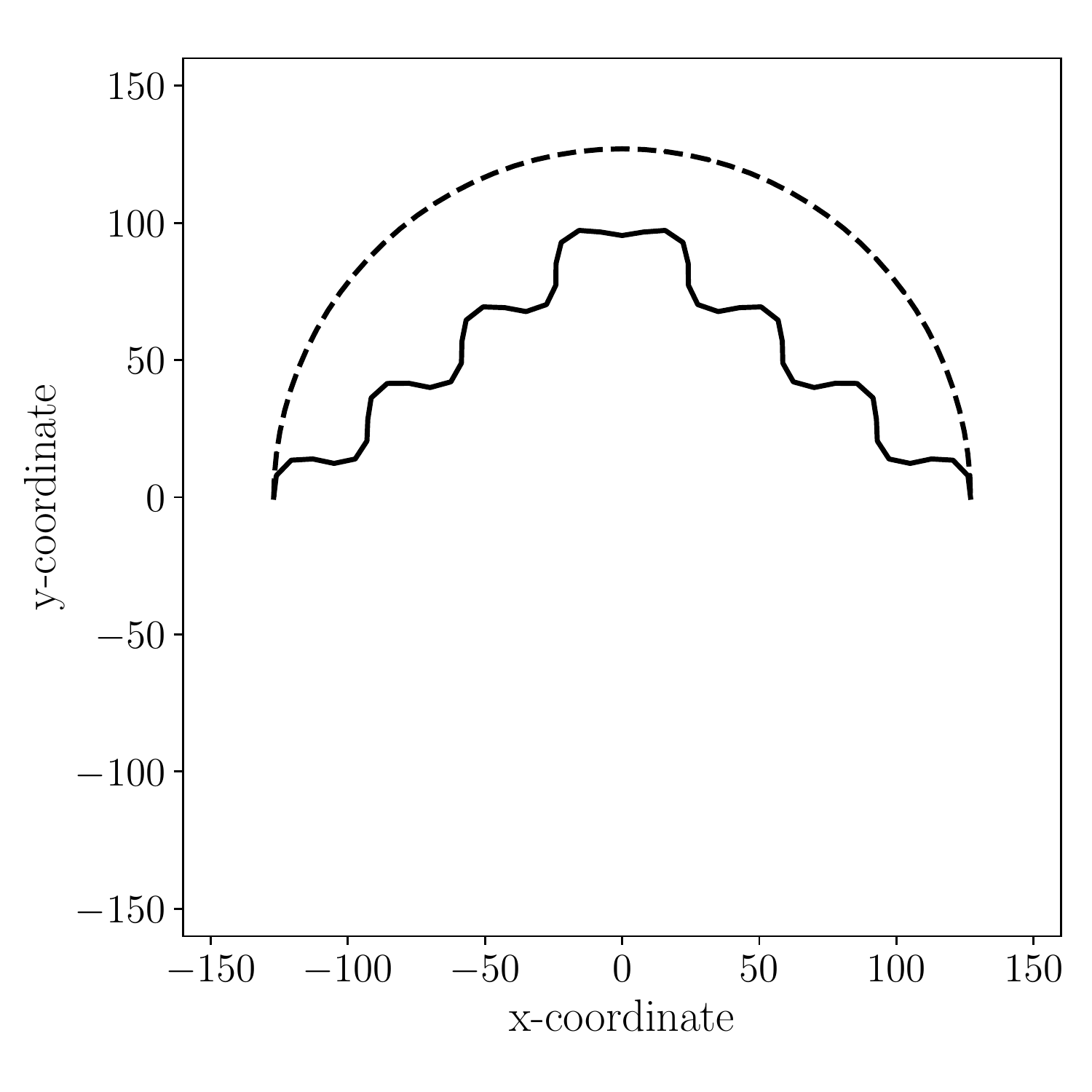}}
\subfloat[]{\includegraphics[clip, scale=0.18]{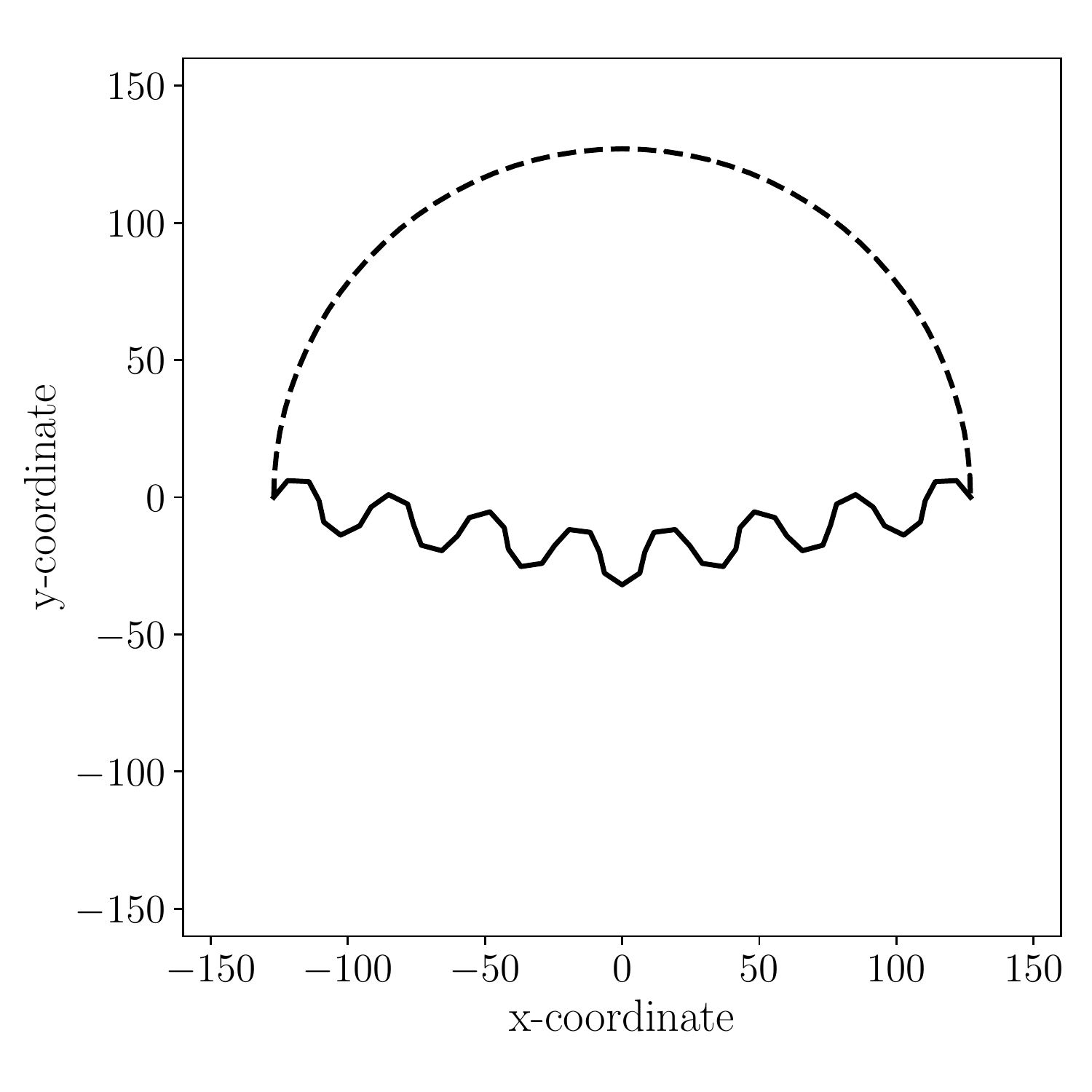}}
\caption{Semi-circular arch: deformed shapes at different load steps for symmetric loading. \mydashedline \, : original configuration and \mysolidline \, : deformed configuration.}
\label{fig-archsemicircle-defshapes-sym}
\end{figure}

\begin{figure}[H]
\centering
\subfloat[]{\includegraphics[clip, scale=0.18]{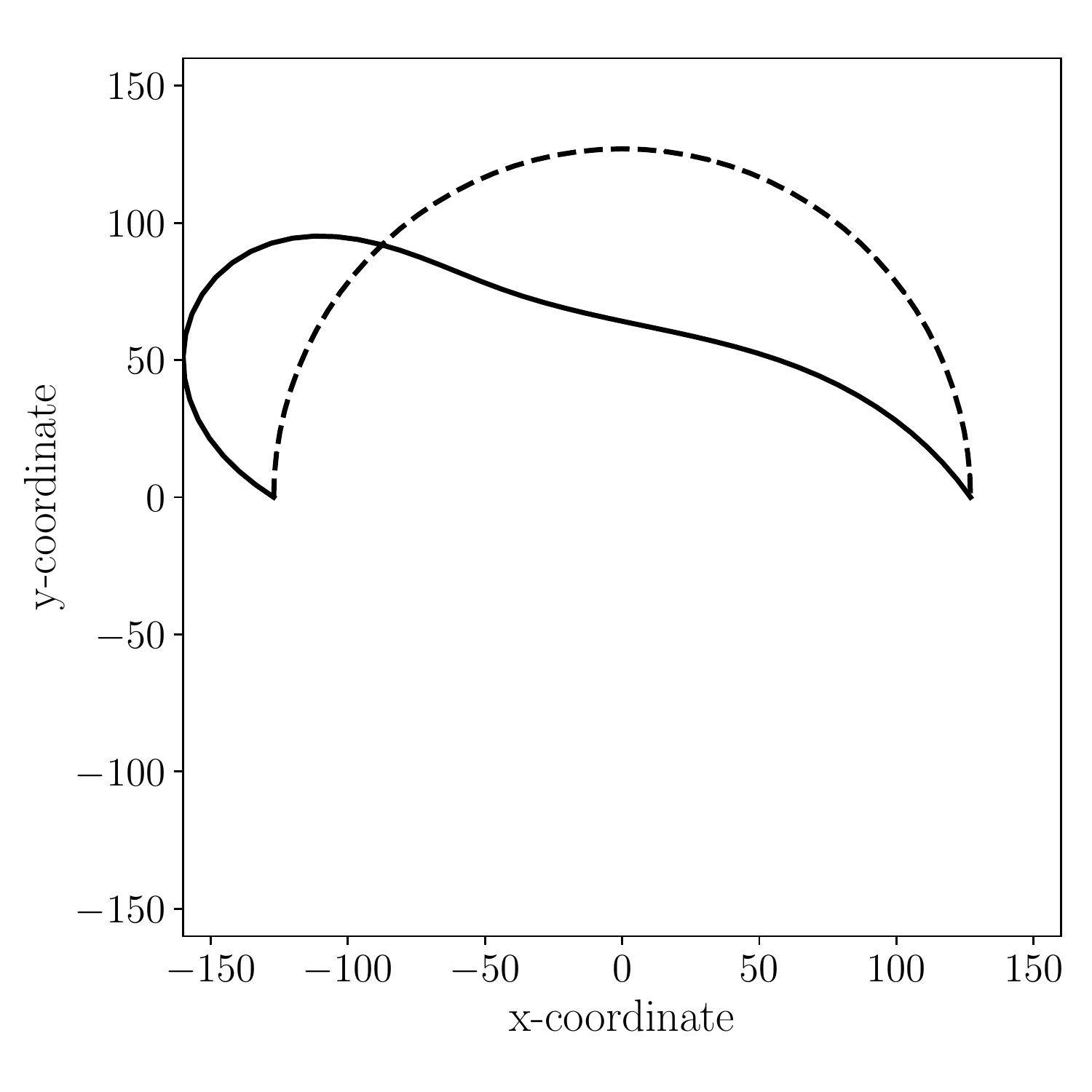}}
\subfloat[]{\includegraphics[clip, scale=0.18]{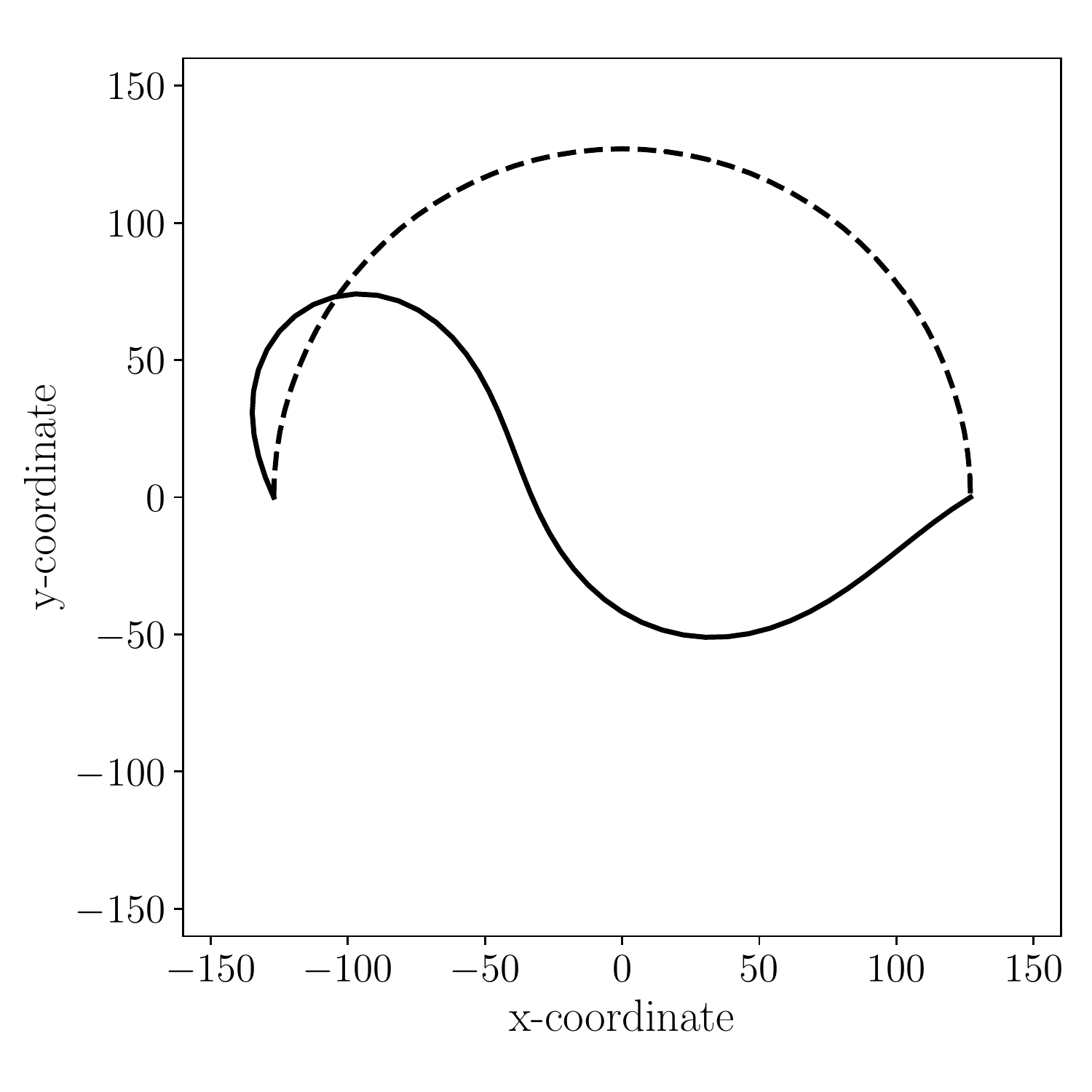}}
\subfloat[]{\includegraphics[clip, scale=0.18]{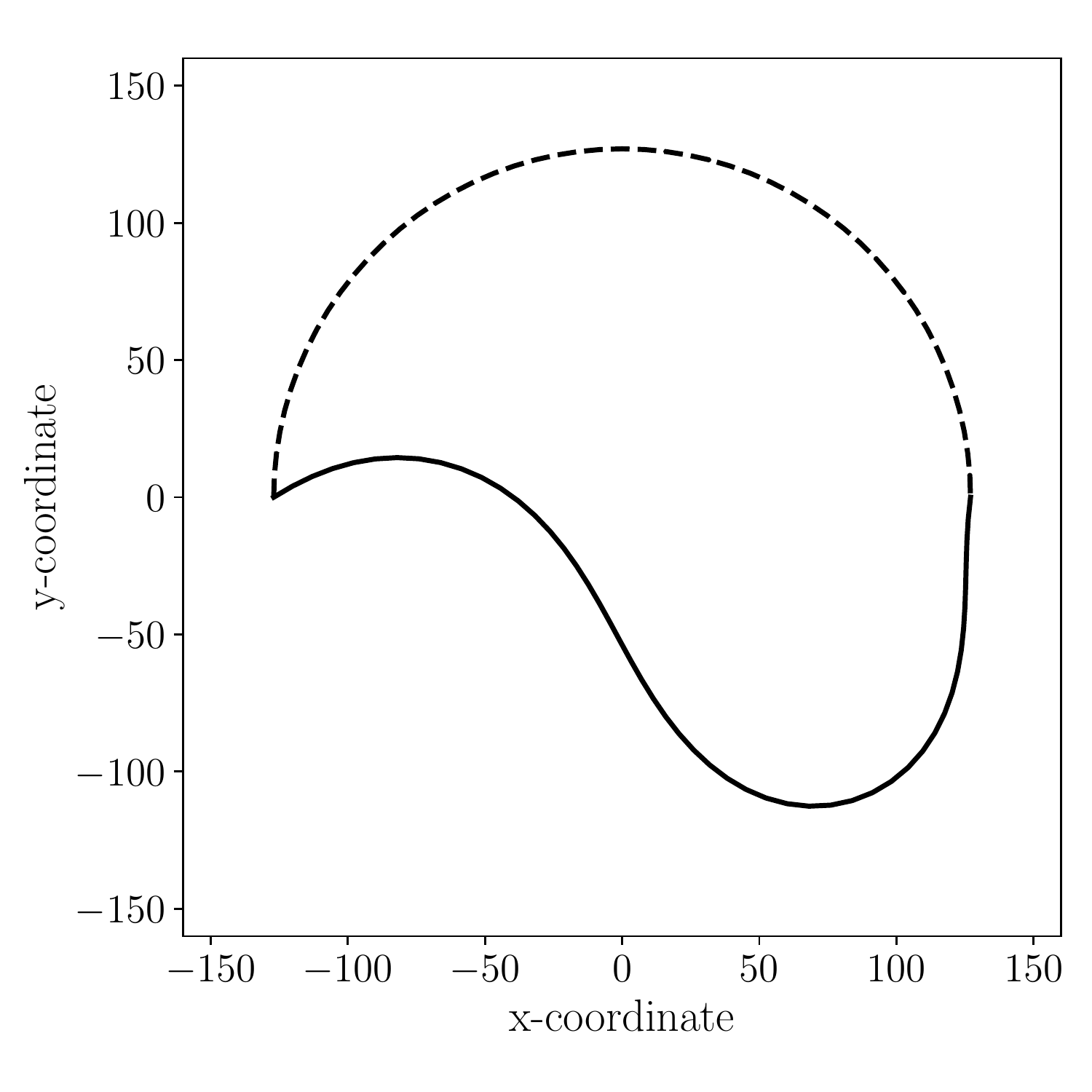}}
\subfloat[]{\includegraphics[clip, scale=0.18]{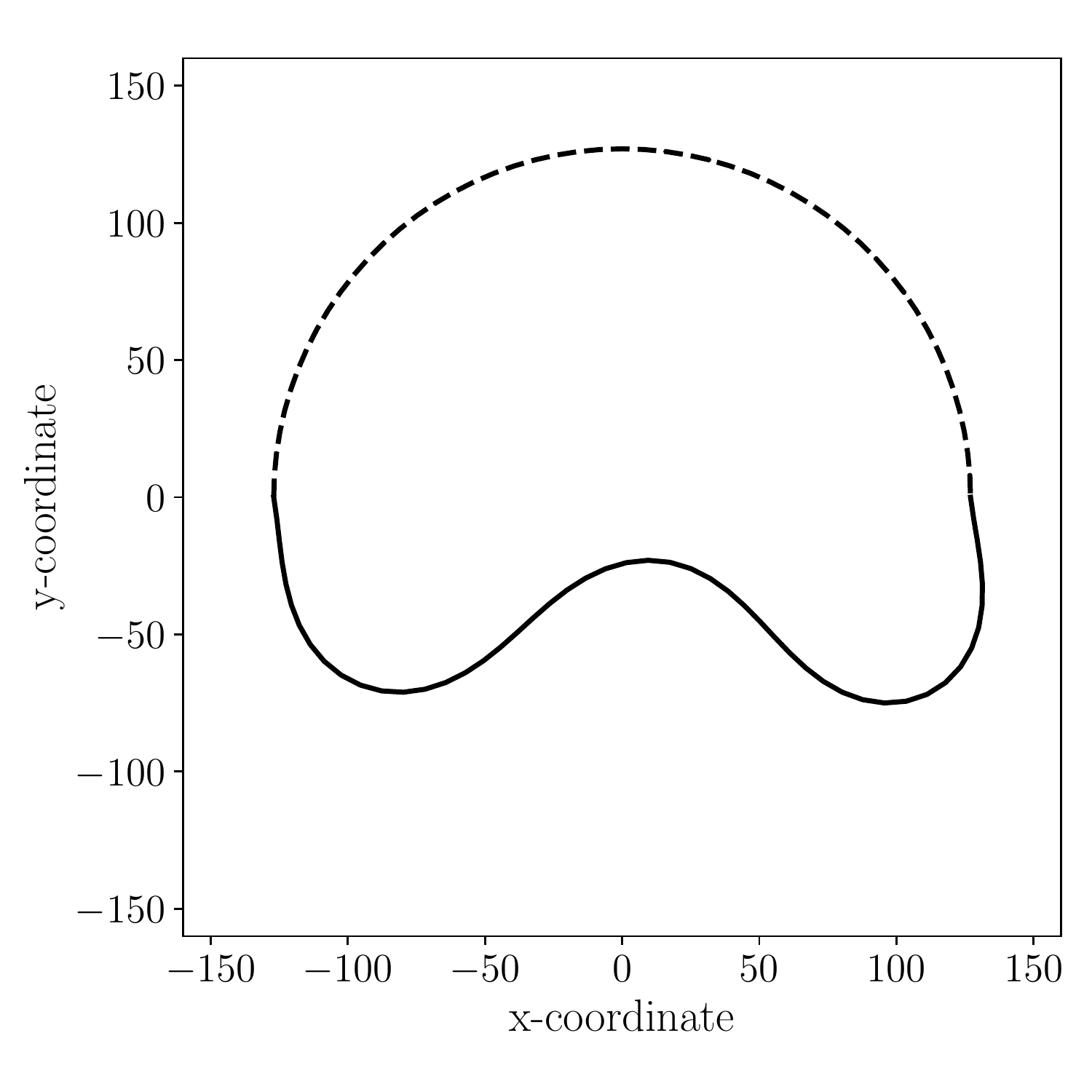}}
\subfloat[]{\includegraphics[clip, scale=0.18]{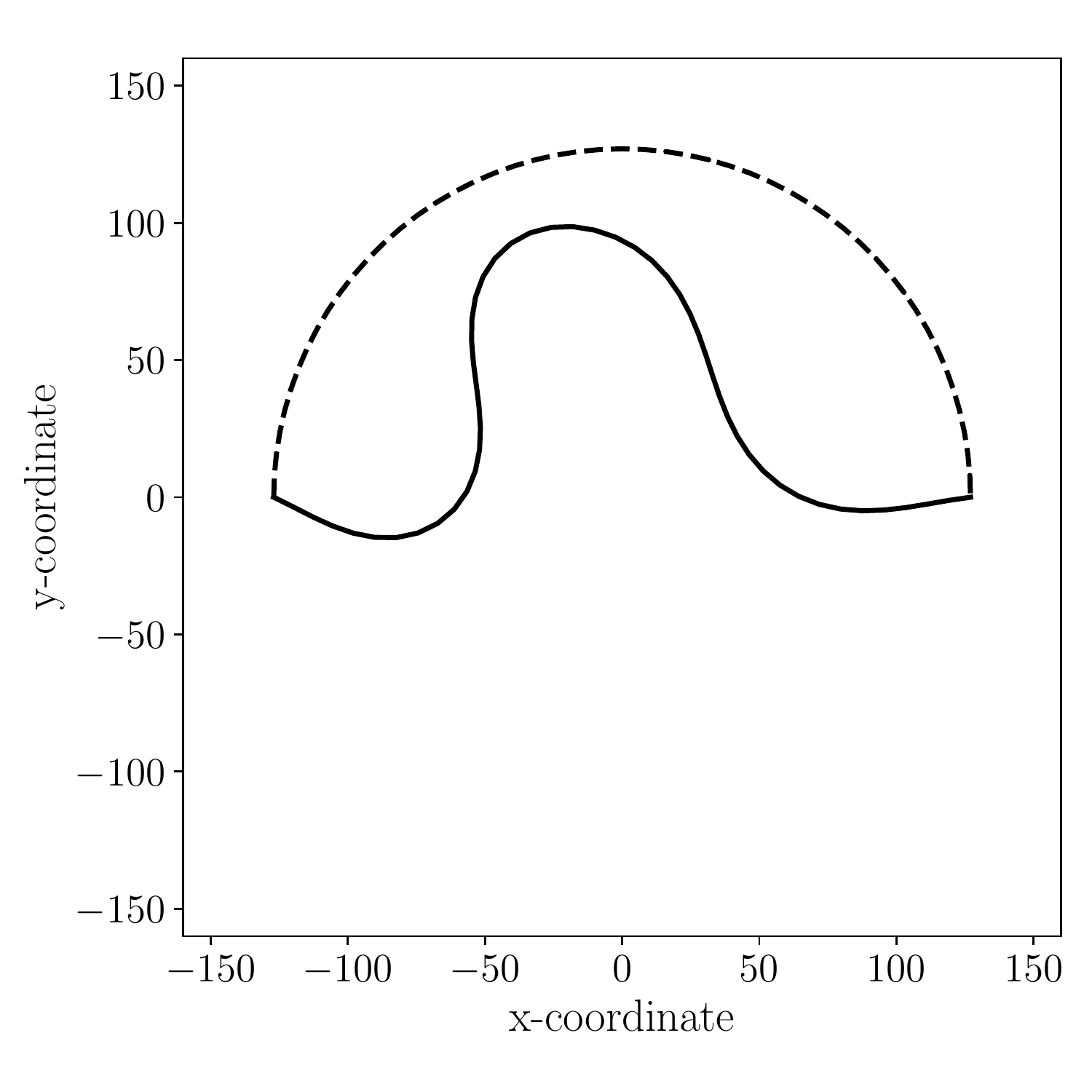}} \\
\subfloat[]{\includegraphics[clip, scale=0.18]{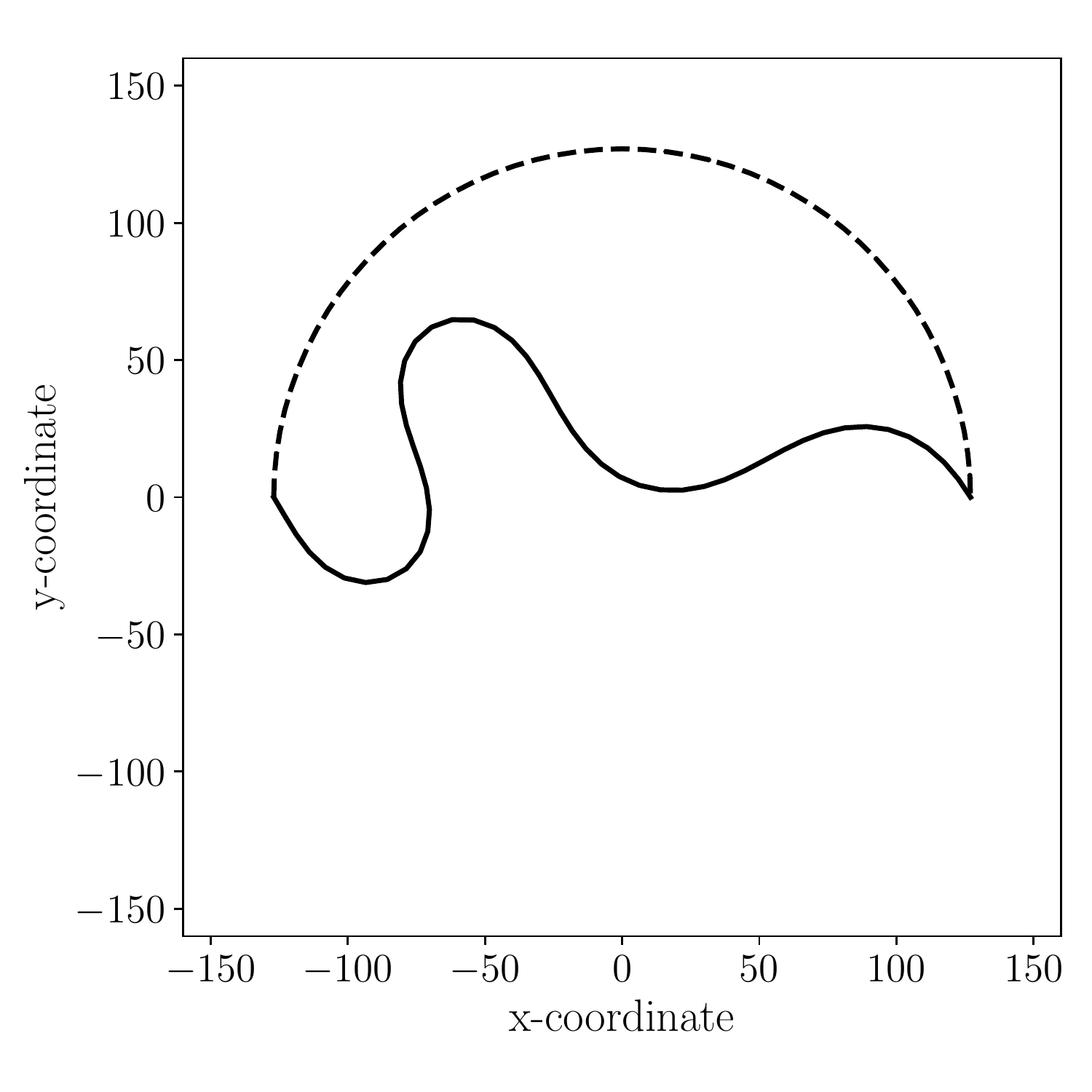}}
\subfloat[]{\includegraphics[clip, scale=0.18]{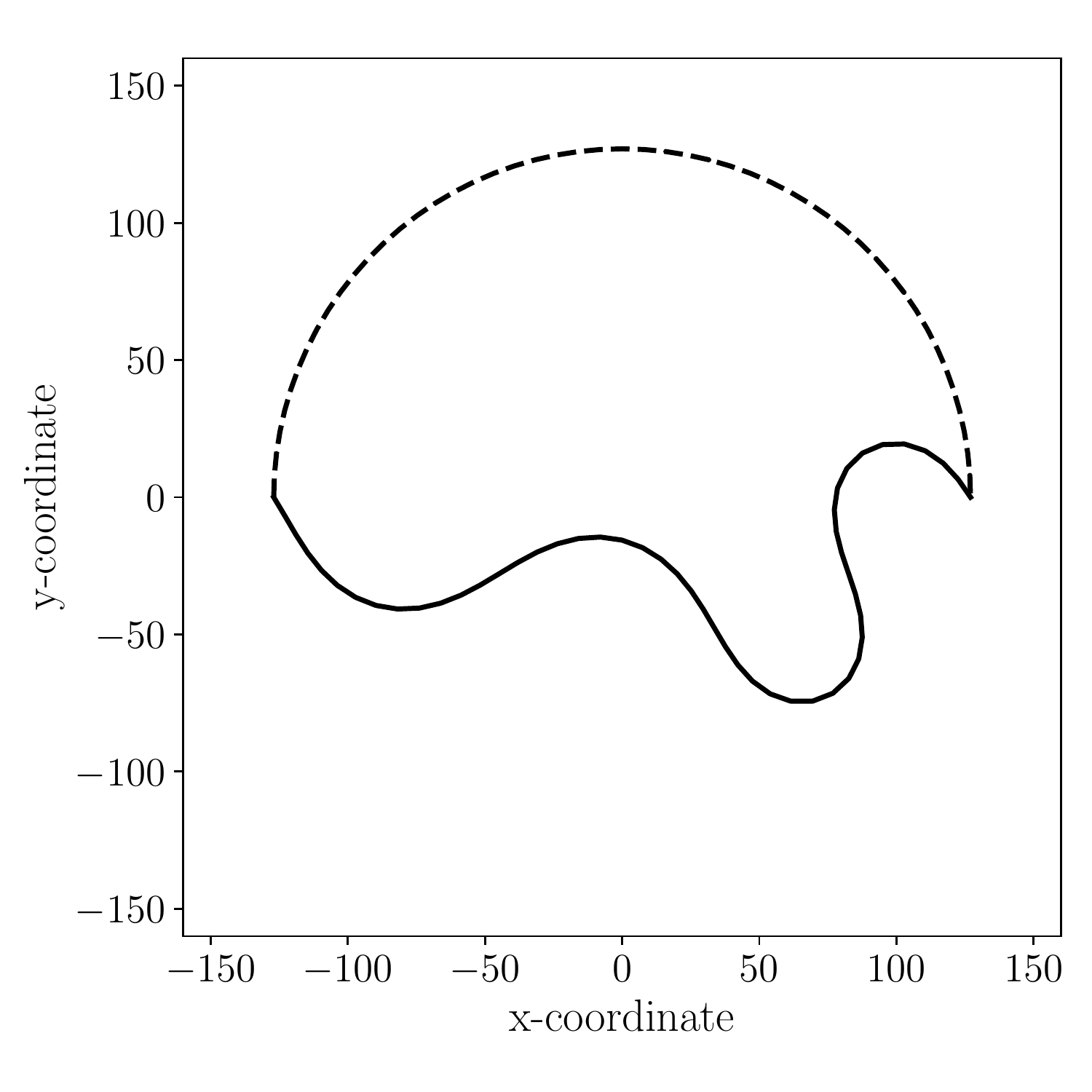}}
\subfloat[]{\includegraphics[clip, scale=0.18]{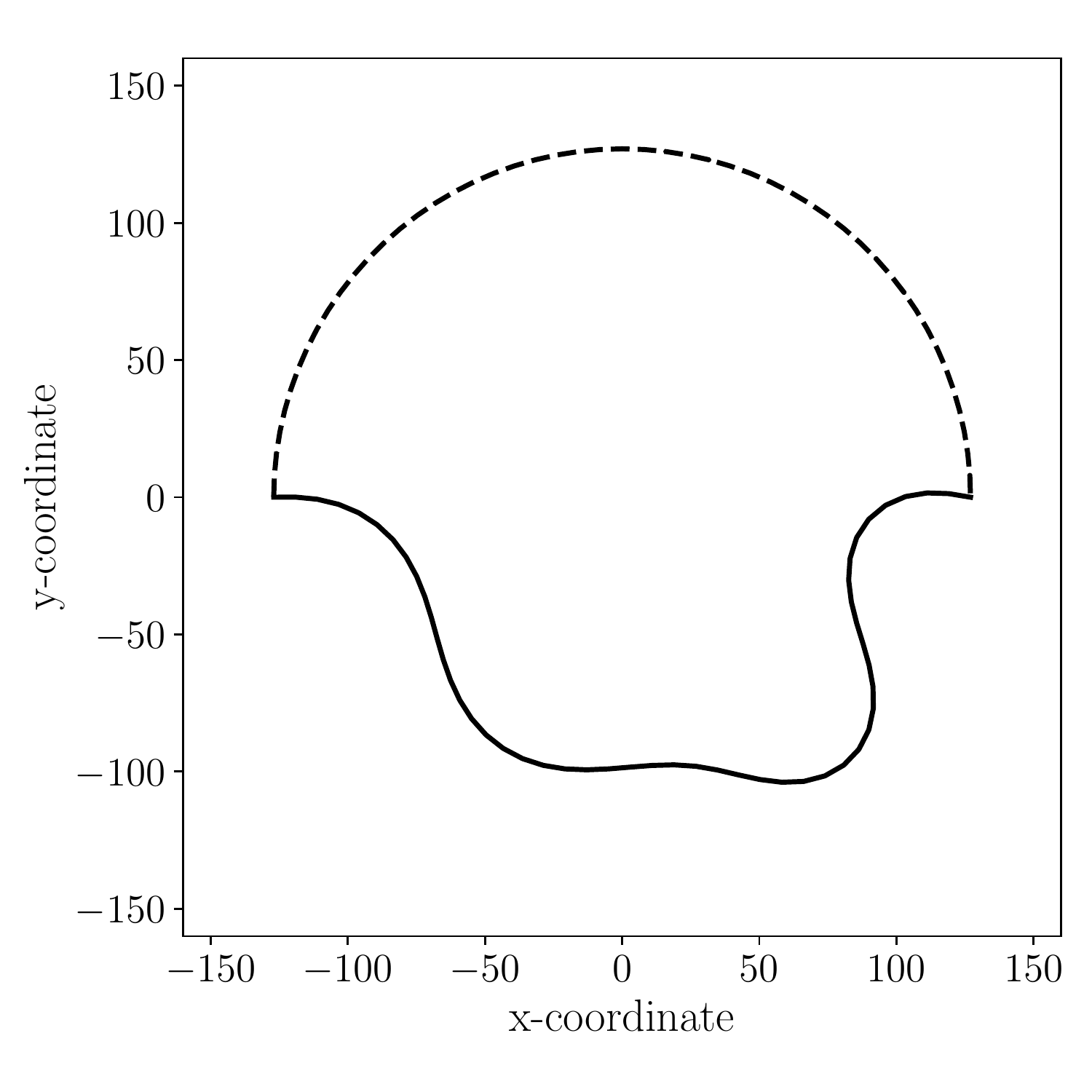}}
\subfloat[]{\includegraphics[clip, scale=0.18]{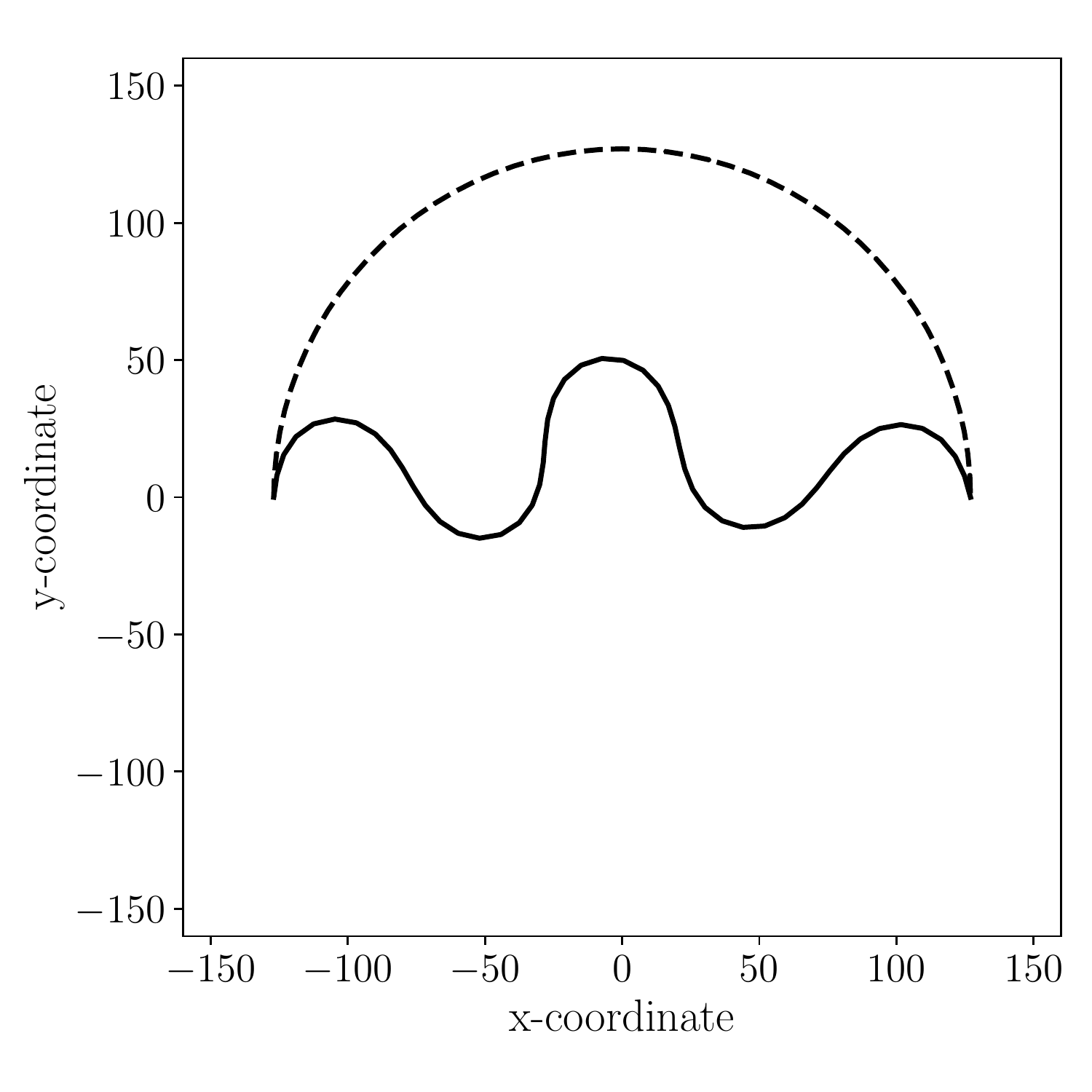}}
\subfloat[]{\includegraphics[clip, scale=0.18]{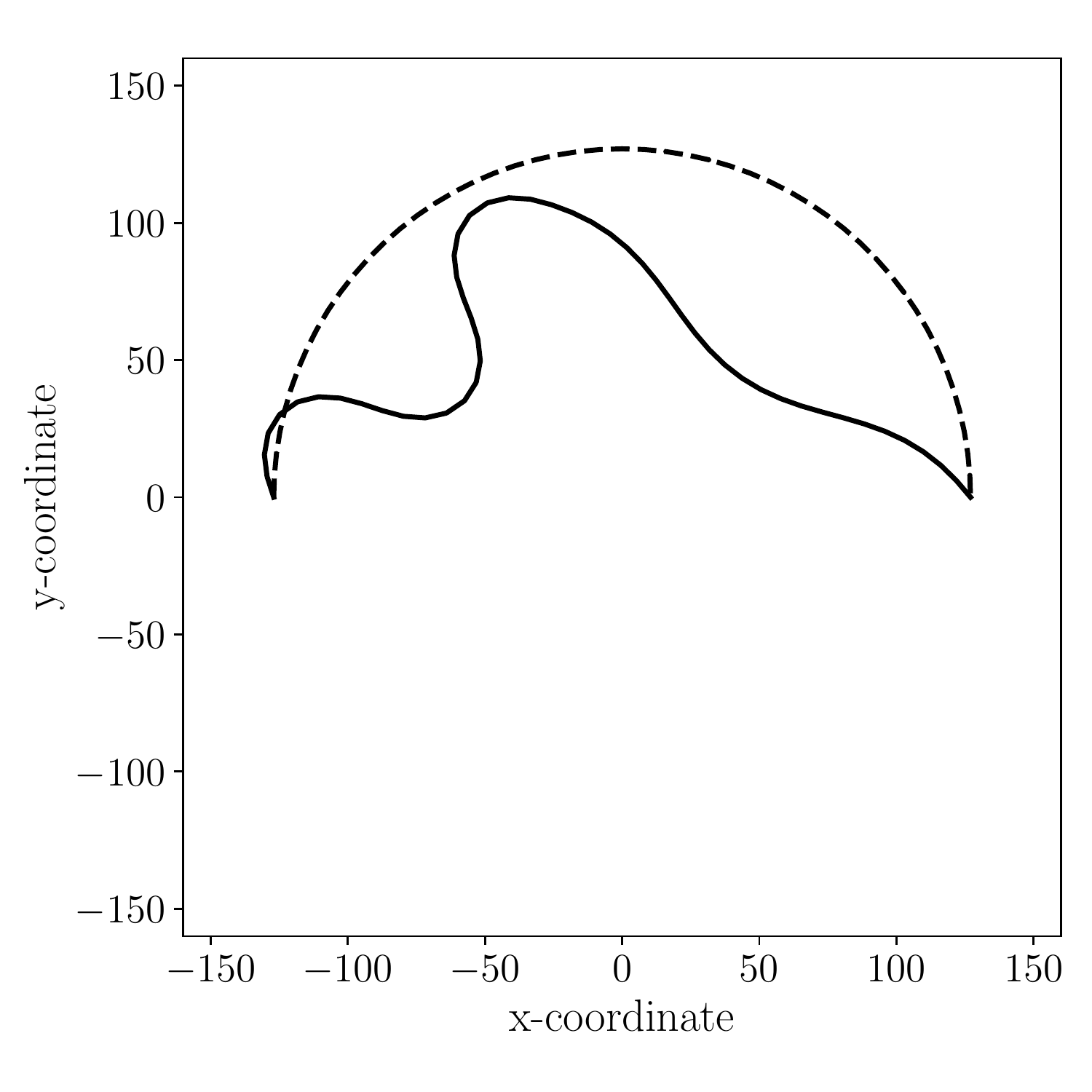}} \\
\subfloat[]{\includegraphics[clip, scale=0.18]{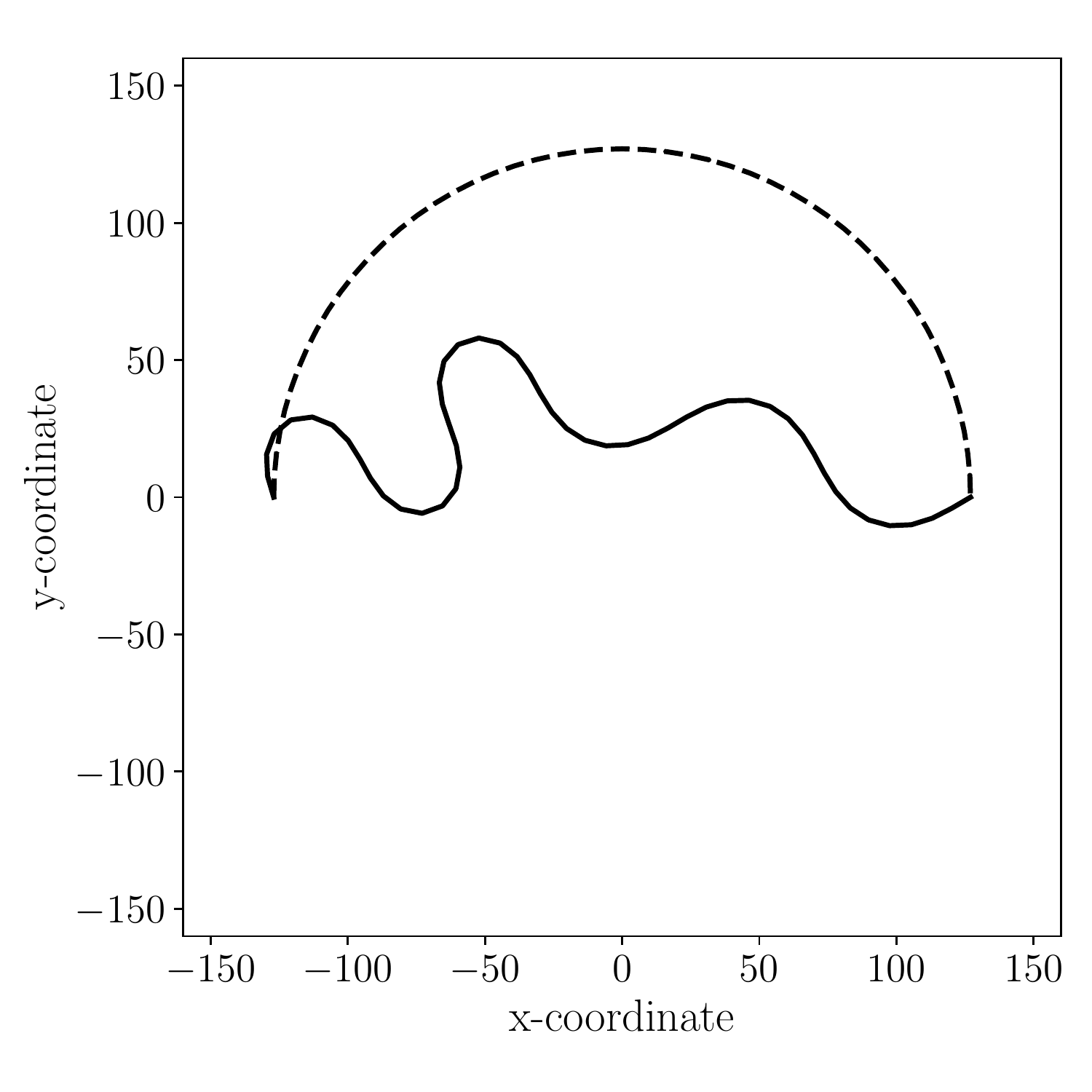}}
\subfloat[]{\includegraphics[clip, scale=0.18]{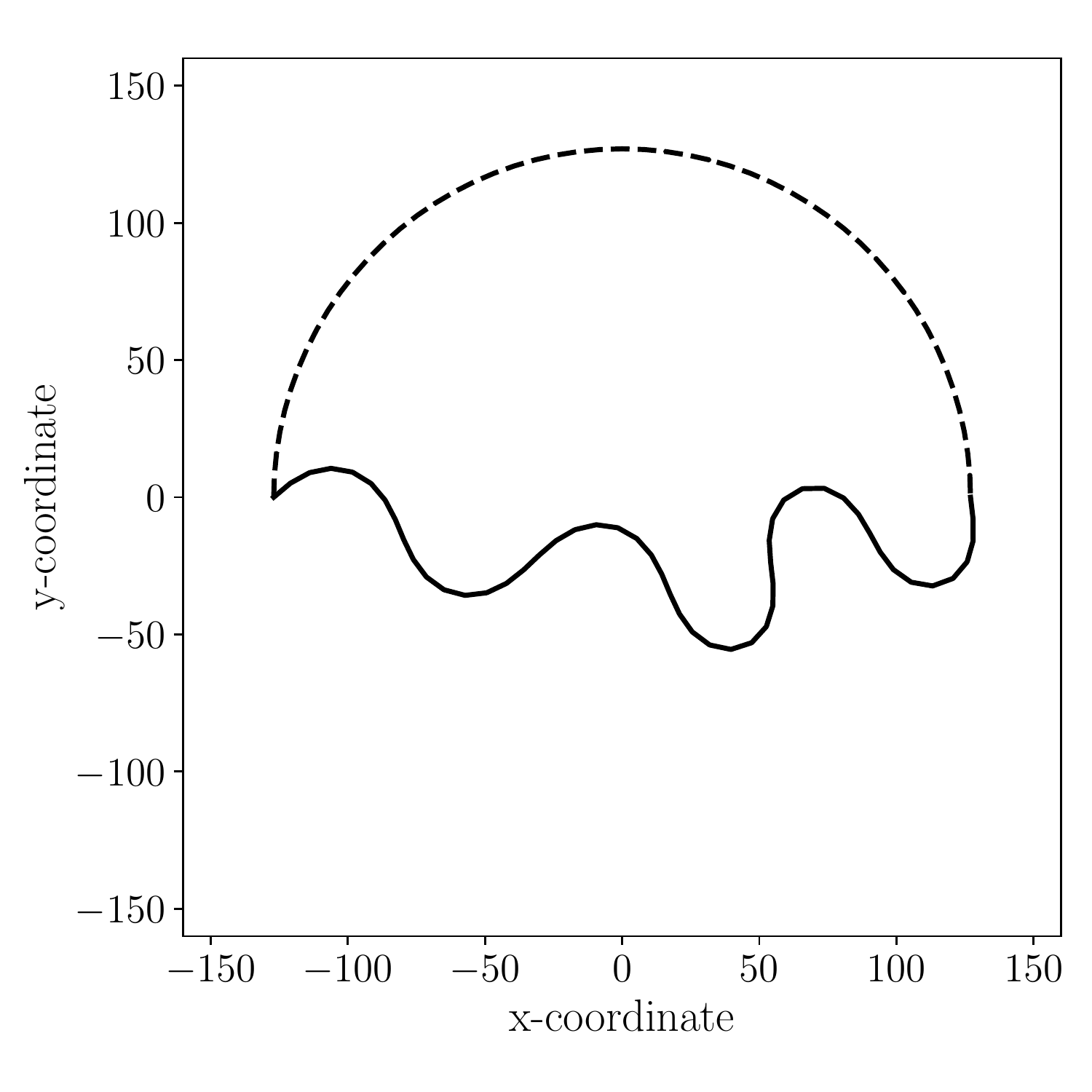}}
\subfloat[]{\includegraphics[clip, scale=0.18]{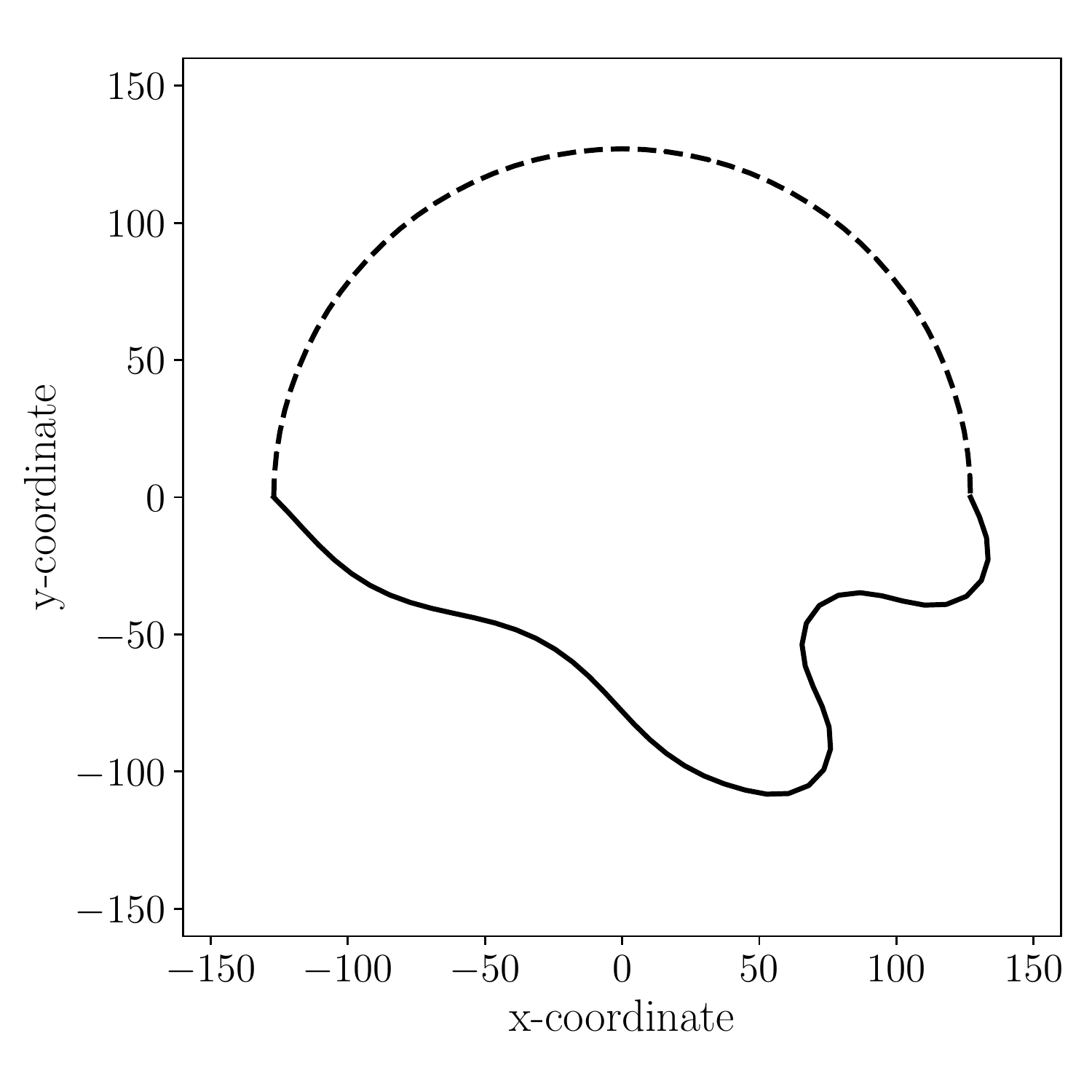}}
\subfloat[]{\includegraphics[clip, scale=0.18]{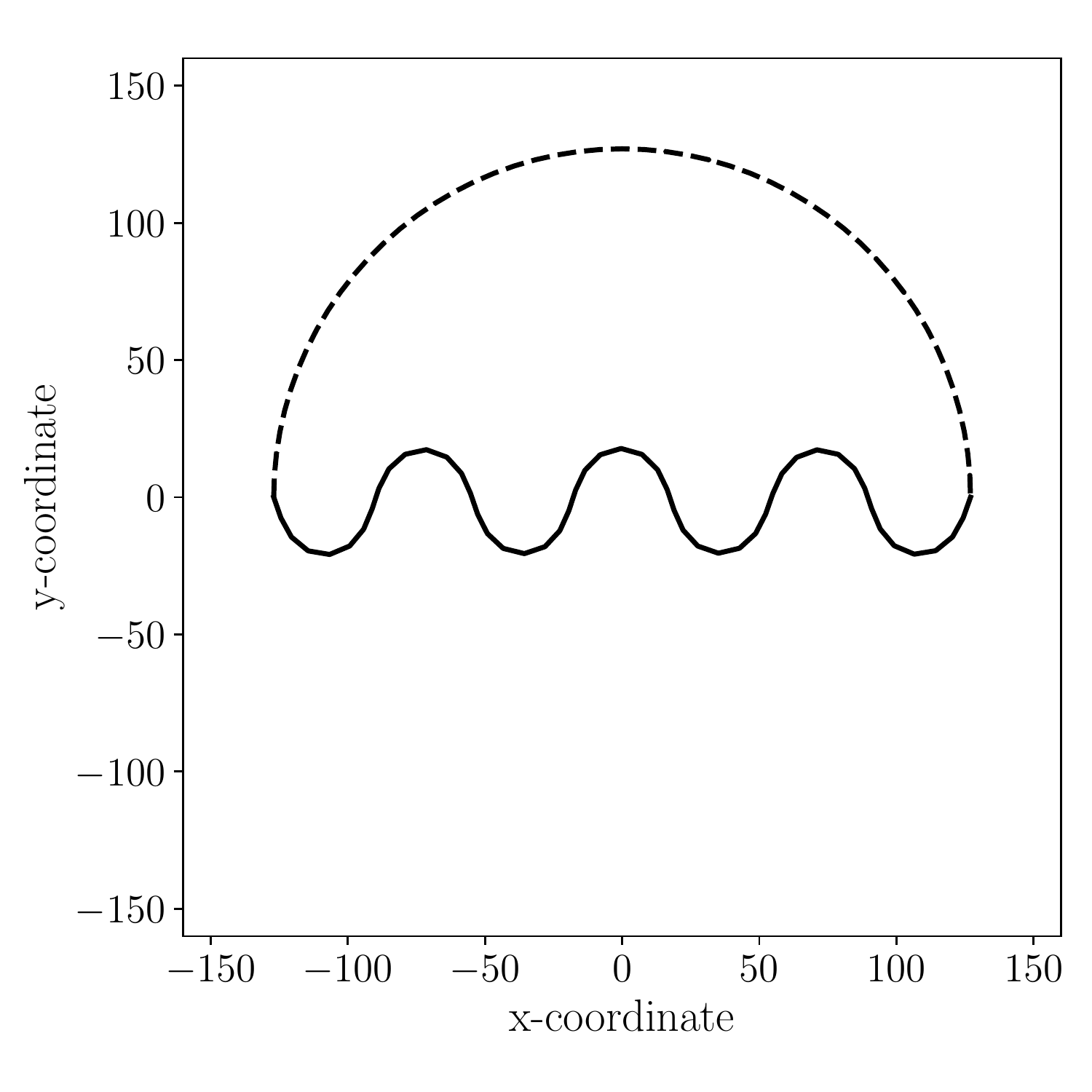}}
\subfloat[]{\includegraphics[clip, scale=0.18]{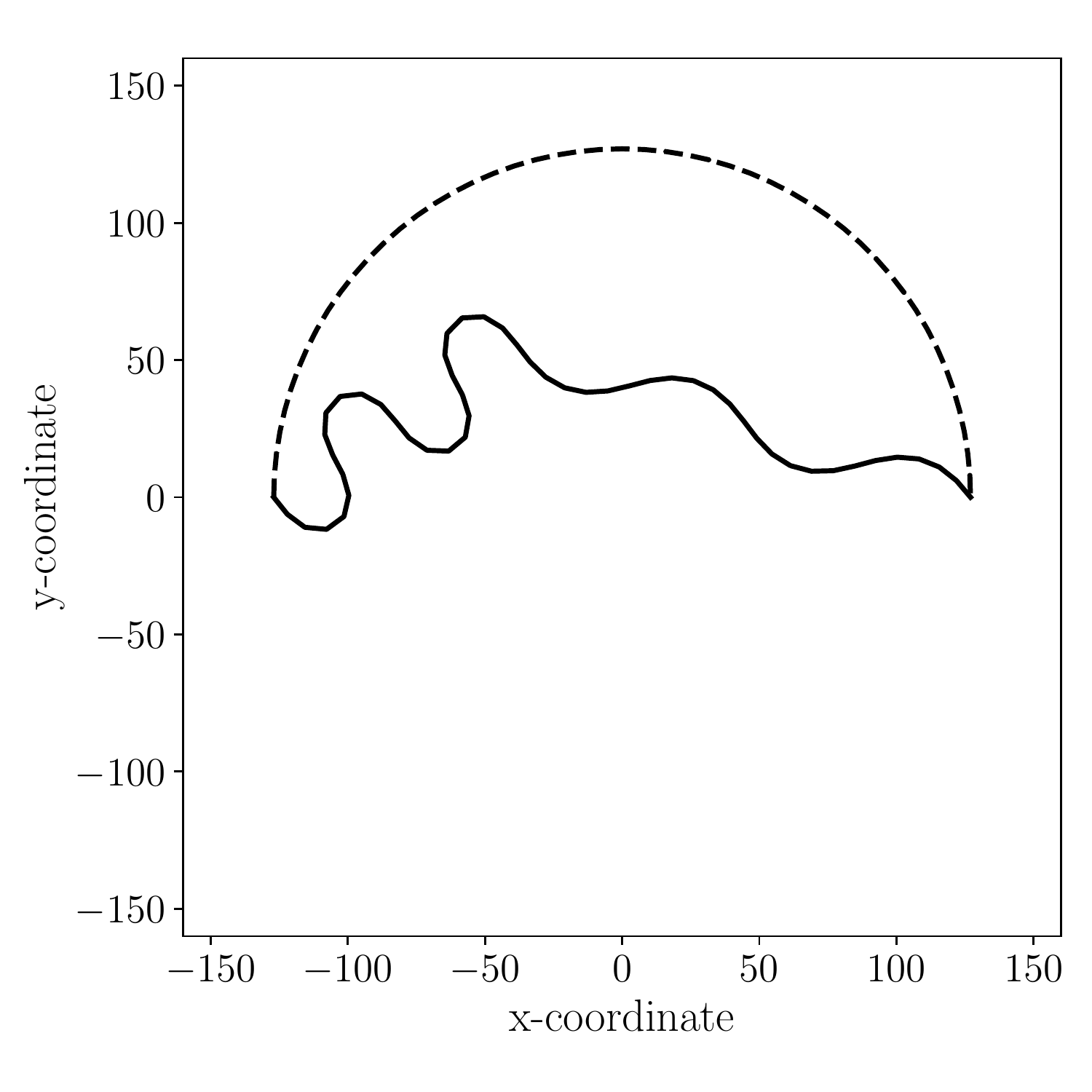}}
\caption{Semi-circular arch: deformed shapes at different load steps for asymmetric loading.  \mydashedline \, : original configuration and \mysolidline \, : deformed configuration.}
\label{fig-archsemicircle-defshapes-asym}
\end{figure}

The equilibrium paths for the symmetric and asymmetric loading are presented, respectively, in Figs. \ref{fig-archsemicircle-graph-sym} and \ref{fig-archsemicircle-graph-asym}. The deformed shapes at 15 different instants are shown in Figs. \ref{fig-archsemicircle-defshapes-sym} and \ref{fig-archsemicircle-defshapes-asym}, respectively, for the symmetric and asymmetric loading. The load-displacement curves in Fig. \ref{fig-archsemicircle-graphs} show that the proposed technique captures the looping paths of the load-displacement curve quite well, and that the solution obtained with the proposed technique matches well with those presented in Yang and Shieh \cite{YangAIAA1990}. The ability of the proposed technique in computing such complex equilibrium paths is quite remarkable considering especially that it is completely devoid of any sophisticated techniques used for explicitly tracking the forward movement along the equilibrium path in the classical implementations of the arc-length methods.

\subsection{Hinged cylindrical panel with point load}
\begin{figure}[H]
\centering
\subfloat[]{\includegraphics[trim=0mm 0mm 20mm 0mm, clip, scale=0.45]{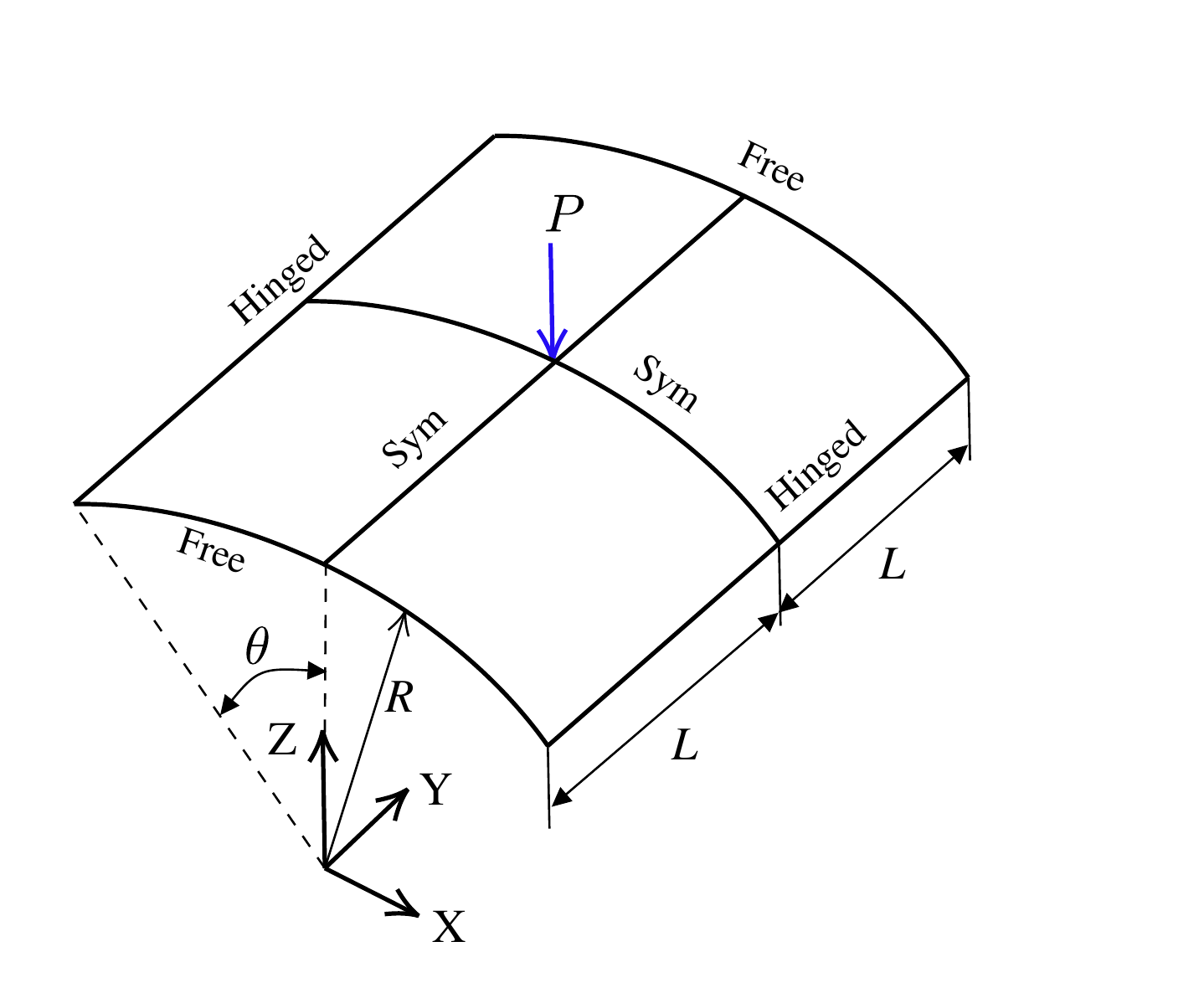} \label{fig-hingedpanel-geom}}
\subfloat[]{\includegraphics[trim=0mm 0mm 0mm 0mm, clip, scale=0.50]{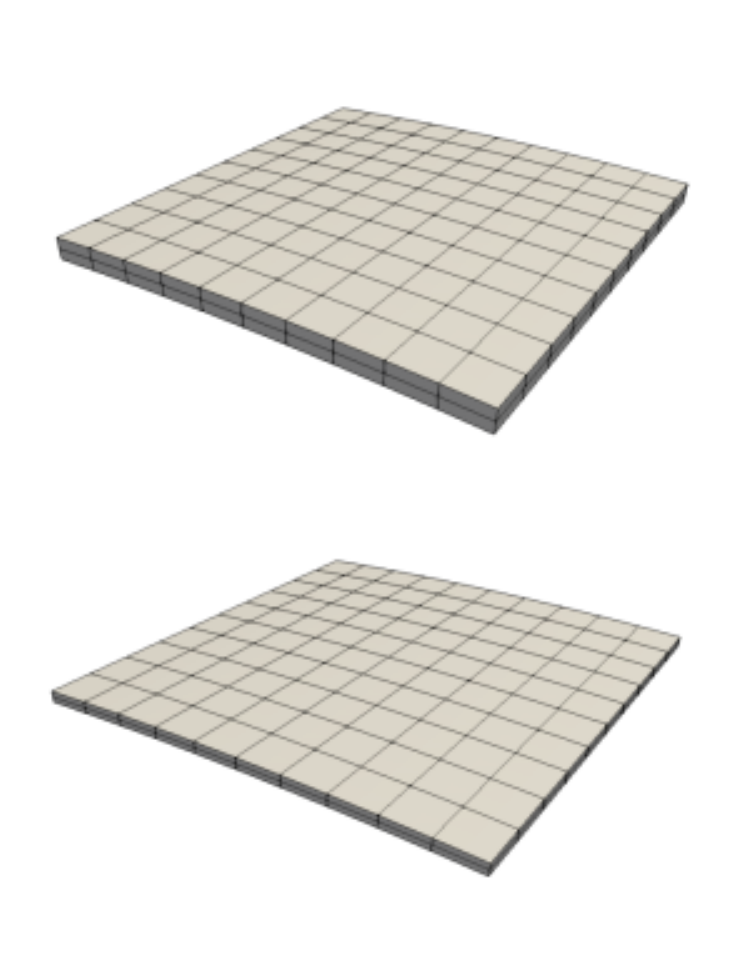} \label{fig-hingedpanel-meshes}}
\subfloat[]{\includegraphics[clip, scale=0.4]{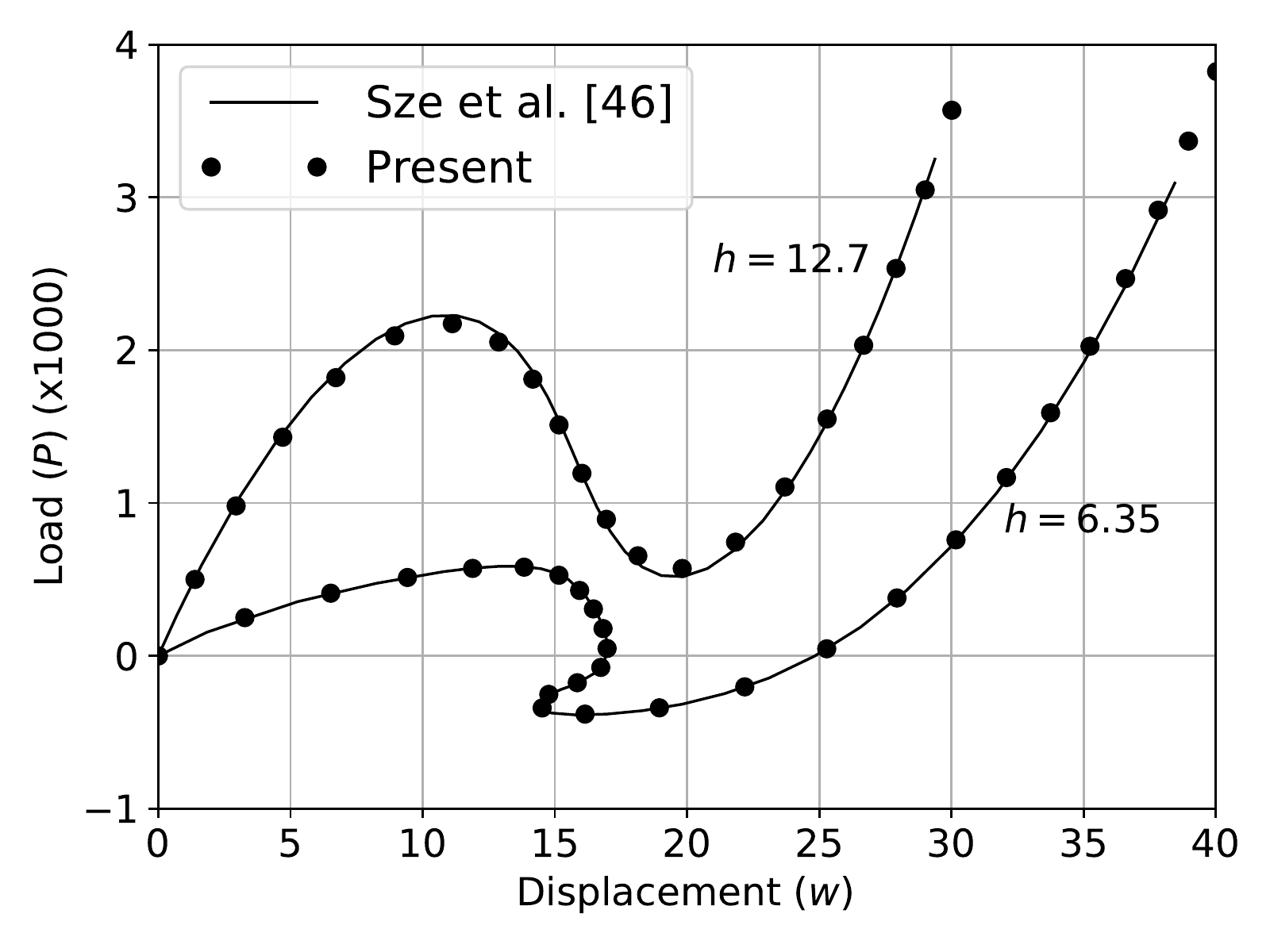} \label{fig-hingedpanel-graphs}}
\caption{Hinged cylindrical panel: (a) geometry and boundary conditions. $R = 2540$ mm, $L = 254$ mm and $\theta = 0.1$ rad, (b) $10 \times 10 \times 2$ mesh for the thick and thin models, and (c) load-displacement curves.}
\label{fig-hingedpanel}
\end{figure}

A hinged cylindrical panel, which is one of the popular benchmarks for nonlinear shell models \cite{SzeFEAD2004} is considered. The setup of the problem is as shown in Fig. \ref{fig-hingedpanel-geom}. Two different cases with thickness $h=12.7$ mm and $h=6.35$ mm are considered. The Young's modulus and Poisson's ratio, respectively, are $E = 3102.75$ N/mm$^2$ and $\nu = 0.3$. The material model is assumed to be Neo-Hookean. This problem is solved using continuum finite elements by adapting the mixed formulation recently proposed in Kadapa and Mokarram \cite{KadapaMAMS2020}. Due to the symmetry, only a quarter portion of the domain is considered. For the spatial discretisation, the Q2/Q1 element is used. For this element, the displacement and pressure field are discretised, respectively, using tri-quadratic (27-noded) and tri-linear (8-noded) hexahedron elements. The load ($P$) is distributed to all the nodes that are radially inline with its direction. Graph of load ($P$) versus the vertical displacement ($w$) of the node on the mid-surface shown in Fig. \ref{fig-hingedpanel-graphs} demonstrate that the numerical solutions obtained with the present arc-length implementation match well with the reference solution. It is worth highlighting that the proposed technique does not require significantly large number load steps for successful completion of the simulation.

\subsection{Pullout of an open-ended cylindrical shell}
\begin{figure}[H]
\centering
\subfloat[]{\includegraphics[trim=10mm 0mm 20mm 0mm, clip, scale=0.4]{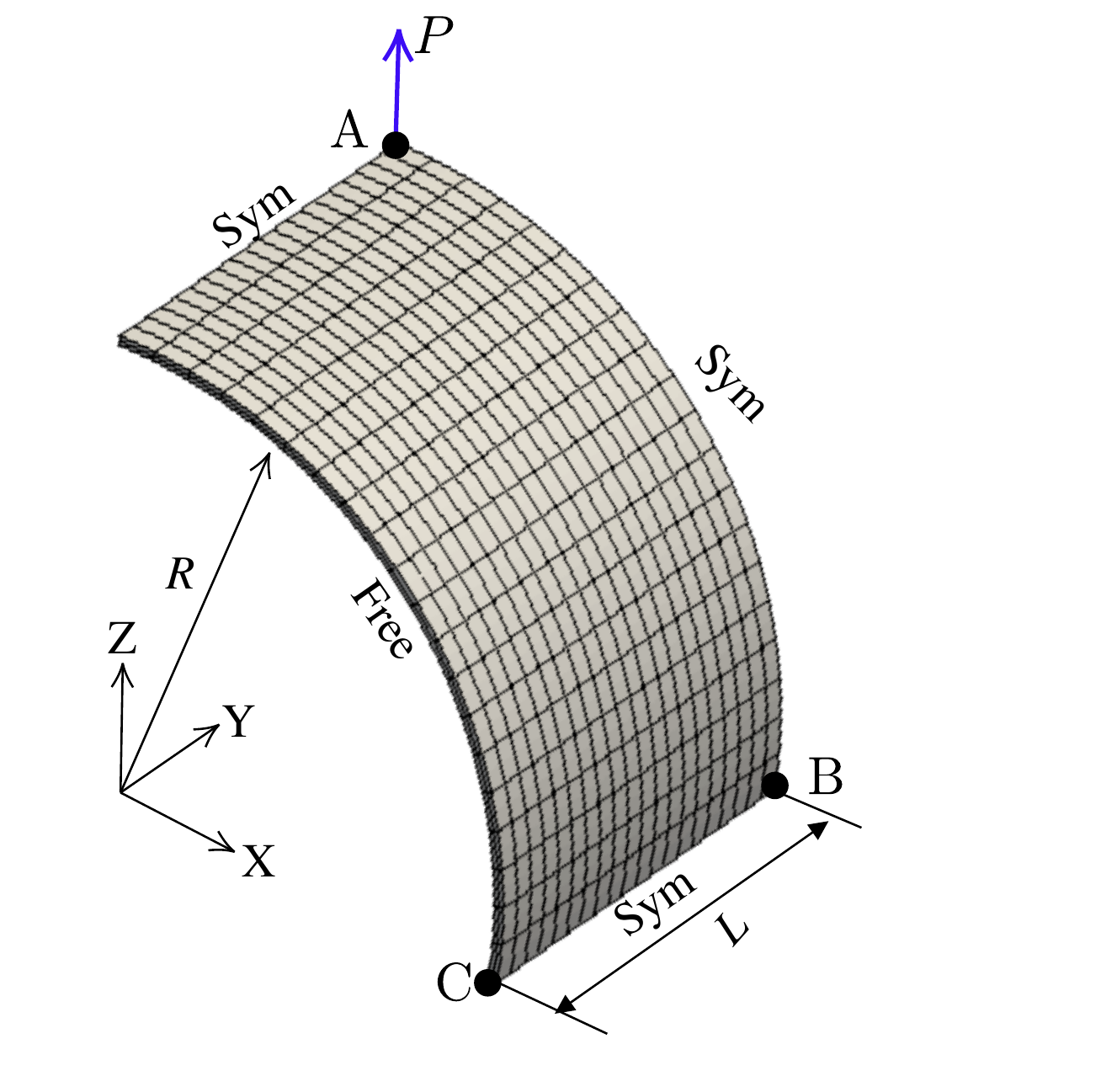} \label{fig-openededcylnshell-geom} }
\subfloat[]{\includegraphics[clip, scale=0.4]{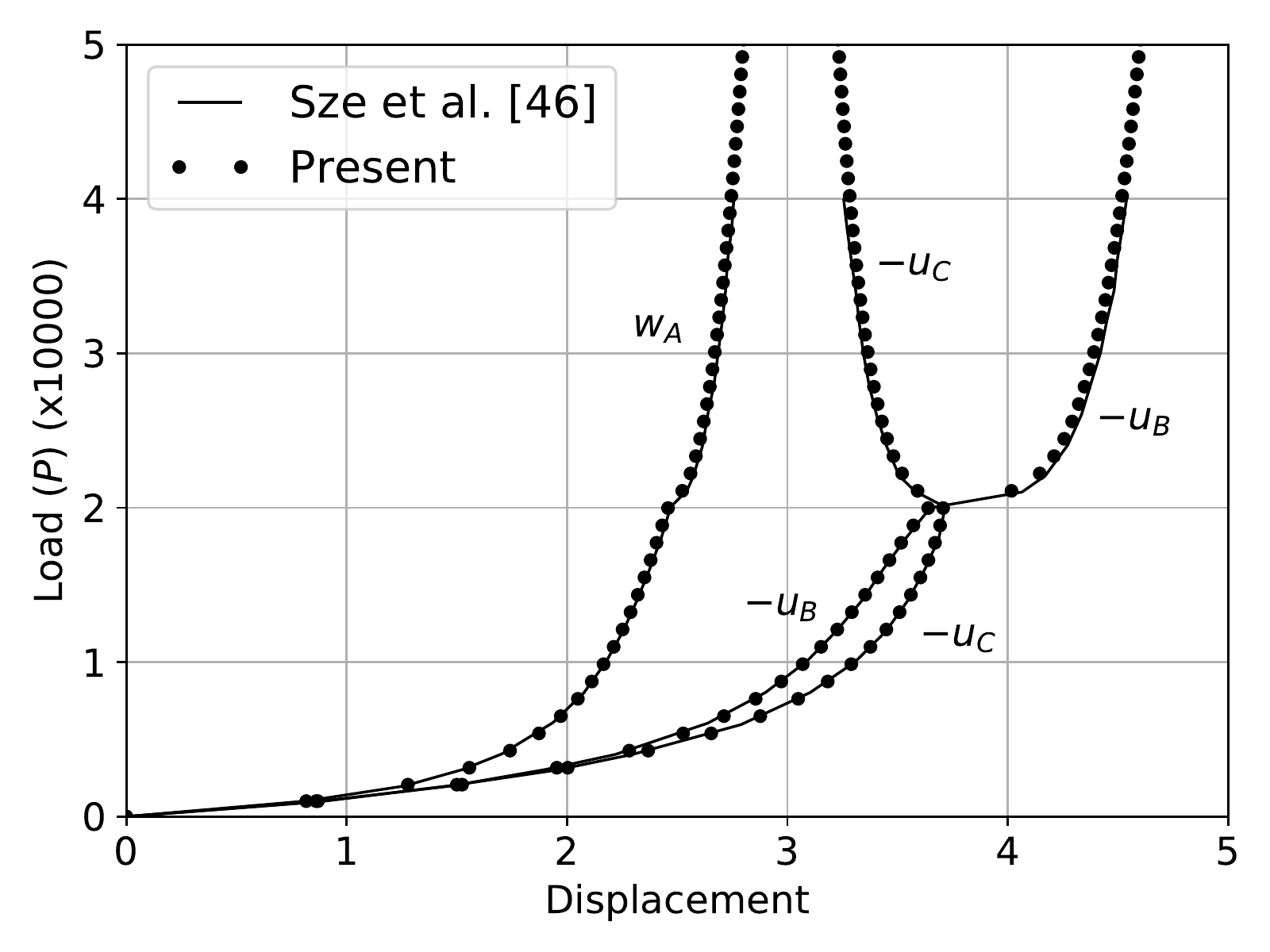} \label{fig-openededcylnshell-graph}}
\subfloat[]{\includegraphics[clip, scale=0.3]{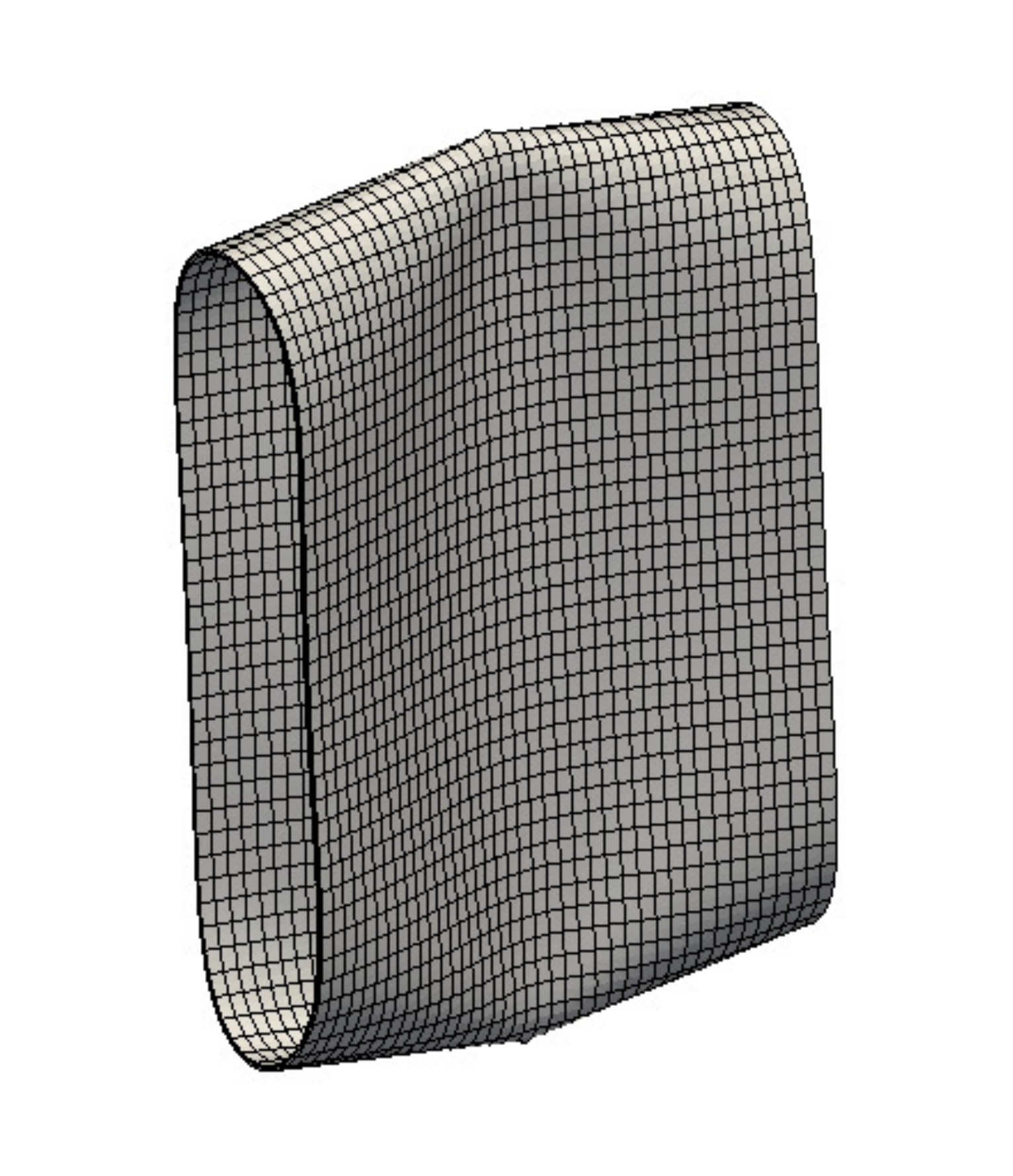} \label{fig-openededcylnshell-defshape}}
\caption{Open-ended cylindrical shell: (a) geometry, boundary conditions and finite element mesh, (b) load-displacement curve, and (c) deformed shape at $P=4.470279 \times 10^4$ units.}
\label{fig-openededcylnshell-figs}
\end{figure}

As the last example, another shell benchmark from Sze et al. \cite{SzeFEAD2004} is considered. The problem consists of a thin open-ended circular cylinder pulled outwards by a pair of radial forces. The half-length of the cylinder is, $L=5.175$ units; its mean radius is, $R=4.953$ units and the thickness is $h=0.094$ units. The material properties are: $E=10.5 \times 10^6$ units and $\nu=0.3125$. Due to symmetry, only 1/8th of the domain is modelled using $20 \times 20 \times 2$ Q2/Q1 elements, as shown in Fig. \ref{fig-openededcylnshell-geom}. Points A, B and C lie on the mid-surface. The load-displacement curves obtained with the proposed scheme, see Fig. \ref{fig-openededcylnshell-defshape}, match well with the reference solutions from Sze et al. \cite{SzeFEAD2004}.

\section{Summary and conclusions} \label{sec-conclusion}
This contribution presents a simple extrapolation operator for overcoming the well-known issues associated with the predictor step as well as with identifying the correct direction along the equilibrium path using the arc-length method. The proposed technique computes the predictor as a linear combination of two solutions at the previously converged load steps, which makes it simple and inexpensive when compared with the other techniques proposed in the literature on the arc-length method. The proposed approach is also applicable to the adaptive load stepping strategy. Another attractive feature of the present scheme is that it is free from ad-hoc parameters. The simplistic nature of the proposed technique renders it suitable for adaption in the existing computer implementations of the arc-length method with some minor modifications.

The ability of the proposed scheme in successfully computing complex equilibrium paths consisting of limit points as well as complex loops is demonstrated using seven benchmark examples in nonlinear structural mechanics. The presented numerical results show excellent agreement between the results obtained with the proposed approach and the analytical/reference solution. It is also worth pointing out that the present approach does not require prohibitively small increments for its success. The capability of the proposed seemingly simple and economical predictor to capture the complex response of the structure is quite remarkable especially considering that it does not involve any sophisticated computations and comparisons commonly employed for tracking the forward movement along equilibrium path in the classical implementation of arc-length methods.

In the present work, the standard Newton-Raphson method is employed. As the future work, the proposed technique can be explored with the modified Newton-Raphson method either for $\delta \bm{u}^{I}$ only or for the whole iteration. Other possibilities include combining the proposed technique with the existing strategies for improving the convergence in the vicinity of limit points towards enhancing the computational efficiency further. The ongoing work focuses on the adaption of the proposed technique for applications in soft and smart materials.

\section*{Supplementary material}
The computer implementation of the arc-length method using the proposed technique is available as GNU Octave scripts at the GitHub repository \url{https://github.com/chennachaos/ArcLengthMethod}. It is possible to use these scripts in MATLAB with some minor modifications.

\appendix

\renewcommand{\theequation}{A.\arabic{equation}}
\setcounter{equation}{0}
\section{Nonlinear truss element} \label{sec-appendixA}
For the two-noded space truss element, the length of the element in the original and the deformed configurations, $L_0$ and $L$, respectively, are given by
\begin{align}
L_{0} &= \sqrt{(X_2-X_1)^2 + (Y_2-Y_1)^2 + (Z_2-Z_1)^2}, \\
L     &= \sqrt{(x_2-x_1)^2 + (y_2-y_1)^2 + (z_2-z_1)^2}.
\end{align}
where, $(X_1,Y_1,Z_1)$ and $(X_2,Y_2,Z_2)$, respectively, are the coordinates of nodes 1 and 2 in the original configuration, and $(x_1,y_1,z_1)$ and $(x_2,y_2,z_2)$ are the coordinates in the current configuration. Nodal coordinates in the current configuration are related to their respective values in original configuration via the relations
\begin{align}
(x_1,y_1,z_1) &= (X_1,Y_1,Z_1) + (u_1,v_1,w_1) \\
(x_2,y_2,z_2) &= (X_2,Y_2,Z_2) + (u_2,v_2,w_2)
\end{align}
where, $(u_1,v_1,w_1)$ and $(u_2,v_2,w_2)$ are the displacement of nodes 1 and 2, respectively.

For the truss model based on the engineering strain ($\varepsilon_E$), which is defined as
\begin{align}
\varepsilon_E := \frac{L-L_0}{L_0},
\end{align}
the internal force vector ($\mathbf{F}^{\mathrm{int}}_{e}$) and the stiffness matrix ($\mathbf{K}_{e}$) for an element are given by,
\begin{align}
\mathbf{F}^{\mathrm{int}}_{e} &= \frac{E \, A \, \varepsilon_E}{L} \, \mathbf{B}^{\T}, \qquad
\mathbf{K}_{e} = \frac{E \, A}{L^3} \, \mathbf{B}^{\T} \, \mathbf{B} + \frac{A \, E \, \varepsilon_E}{L} \, \mathbf{H},
\end{align}
where, $A$ is the area of the element and $E$ is the Young's modulus,
\begin{align}
\mathbf{B} = [ x_1-x_2, \; y_1-y_2, \; z_1-z_2, \; x_2-x_1, \; y_2-y_1, \; z_2-z_1]^{\T},
\end{align}
\begin{align}
\mathbf{H} = \begin{bmatrix}
 1  &  0  &  0  & \tm 1  &  0  &  0  \\
 0  &  1  &  0  &  0  & \tm 1  &  0  \\
 0  &  0  &  1  &  0  &  0  & \tm 1  \\
\tm 1  &  0  &  0  &  1  &  0  &  0  \\
 0  & \tm 1  &  0  &  0  &  1  &  0  \\
 0  &  0  & \tm 1  &  0  &  0  &  1  \\
\end{bmatrix}.
\end{align}

For the truss model based on the Green-Lagrange strain ($\varepsilon_G$), which is defined as
\begin{align}
\varepsilon_G := \frac{L^2-L_0^2}{2 \, L_0^2},
\end{align}
the internal force vector and the stiffness matrix for an element are given by,
\begin{align}
\mathbf{F}^{\mathrm{int}}_{e} &= \frac{E \, A \, \varepsilon_G}{L_0} \, \mathbf{B}^{\T}, \qquad
\mathbf{K}_{e} = \frac{E \, A}{L_0^3} \, \mathbf{B}^{\T} \, \mathbf{B} + \frac{A \, E \, \varepsilon_G}{L_0} \, \mathbf{H}.
\end{align}

\section*{References}

\end{document}